%% file: main.tex
\documentclass[onecolumn,amsmath,amssymb,nofootinbib,12pt]{article}
\usepackage{jheppub} \usepackage{ifpdf}

\usepackage{graphicx,subcaption} 
\usepackage{amsfonts} 
\usepackage{amsmath} 
\usepackage{amssymb}
\usepackage{mathtools} 
\usepackage{epsfig}
\usepackage{pdfpages} \usepackage{bbm} \usepackage{graphicx,epstopdf}
\usepackage[numbers]{natbib} \usepackage[makeroom]{cancel} \usepackage{hyperref}
\usepackage{tensor} 
\hypersetup{pdftex,colorlinks=true,allcolors=blue}
\usepackage{array} 
\usepackage{cleveref} 

\usepackage[normalem]{ulem} 

\newcommand{\RN}[1]{%
	\textup{\uppercase\expandafter{\romannumeral#1}}%
}  

\newcommand{\bulk}{\text{bulk}}

\definecolor{limegreen}{rgb}{0.2, 0.8, 0.2}
\definecolor{deepcarrotorange}{rgb}{0.91, 0.41, 0.17}
\definecolor{darkviolet}{rgb}{0.58, 0.0, 0.83} \definecolor{cyan}{rgb}{0.0,
  0.72, 0.92} \definecolor{plum}{rgb}{0.67, 0.0, 0.55}

\newcommand{\eg}{{\it e.g.,}\ } \newcommand{\ie}{{\it i.e.,}\ }
\newcommand{\reef}[1]{(\ref{#1})} 
\newcommand{\mt}[1]{\textrm{\tiny #1}}

\newcommand{\beq}{\begin{equation}} \newcommand{\eeq}{\end{equation}}
\newcommand{\beqa}{\begin{eqnarray}} \newcommand{\eeqa}{\end{eqnarray}}

\newcommand{\bea}{\begin{eqnarray}} \newcommand{\eea}{\end{eqnarray}}

\newcommand{\phinew}{\hat\phi}

\renewcommand{\(}{\left(} \renewcommand{\)}{\right)} \renewcommand{\[}{\left[}
    \renewcommand{\]}{\right]}

\newcommand{\tR}{{\tilde R}} \newcommand{\veps}{\varepsilon}
\newcommand{\islands}{\mathrm{islands}}

\newcommand{\RNum}[1]{\uppercase\expandafter{\romannumeral #1\relax}}

\input{vincentsMacros}

\newcommand{\rg}{\tilde r}
\newcommand{\rr}{r}

\usepackage[thinlines]{easytable}
\usepackage{textcomp}

\title{\boldmath Quantum Extremal Islands Made Easy, Part II:\\
  Black Holes on the Brane}

\author[a,b]{Hong Zhe Chen,} \author[a]{Robert C. Myers,} \author[a]{Dominik
  Neuenfeld,} \author[c]{Ignacio A. Reyes} \author[a]{and Joshua Sandor}

\affiliation[a]{Perimeter Institute for Theoretical Physics, Waterloo, ON N2L
  2Y5, Canada} \affiliation[b]{Dept.~of Physics $\&$ Astronomy, University of
  Waterloo, Waterloo, ON N2L 3G1, Canada} \affiliation[c]{Max-Planck-Institut
  f\"ur Gravitationsphysik, Am M\"uhlenberg 1, 14476 Potsdam, Germany}

\emailAdd{hchen2@pitp.ca} \emailAdd{rmyers@pitp.ca}
\emailAdd{dneuenfeld@pitp.ca} \emailAdd{ignacio.reyes@aei.mpg.de}
\emailAdd{jsandor@stanford.edu}

\abstract{We discuss holographic models of extremal and non-extremal black holes
  in contact with a bath in $d$ dimensions, based on a brane world model
  introduced in \cite{Chen:2020uac}. The main benefit of our setup is that it
  allows for a high degree of analytic control as compared to previous work in
  higher dimensions. We show that the appearance of quantum extremal islands in
  those models is a consequence of the well-understood phase transition of RT
  surfaces, and does not make any direct reference to ensemble averaging. For
  non-extremal black holes the appearance of quantum extremal islands has the
  right behaviour to avoid the information paradox in any dimension. We further
  show that for these models the calculation of the full Page curve is possible
  in any dimension. The calculation reduces to numerically solving two ODEs. In
  the case of extremal black holes in higher dimensions, we find no quantum
  extremal islands for a wide range of parameters. In two dimensions, our
  results agree with \cite{Almheiri:2019yqk} at leading order; however a
  finite UV cutoff introduced by the brane results in subleading corrections.
  For example, these corrections result in the quantum extremal surfaces moving further outward from the
  horizon, and shifting the Page transition
  to a slightly earlier time.
}

\begin{document}
\maketitle
\flushbottom

\section{Introduction}\label{sec:intro}
\input{sections/01_intro}

\section{Braneworld framework}\label{sec:RS}
\input{sections/02_setup}

\section{Black hole in equilibrium with an external bath}\label{sec:nonextremal}
\input{sections/03_non_extremal_bh}

\section{Numerical results}\label{sec:numerics}
\input{sections/04_numerics}

\section{Extremal horizon in equilibrium with $T=0$ bath}\label{sec:extremal}
\input{sections/05_extremal_bh}

\section{Two dimensions revisited}\label{app:Page2d}
\input{sections/06_JT}

\section{Discussion}\label{sec:discuss}
\input{sections/07_discussion}

\section*{Acknowledgments}
We would like to thank Ahmed Almheiri, Raphael Bousso, Xi Dong, Roberto Emparan,
Netta Engelhardt, Zach Fisher, Greg Gabadadze, Juan Hernandez, Don Marolf,
Shan-Ming Ruan, Edgar Shaghoulian, Antony Speranza and Raman Sundrum for useful
comments and discussions. Research at Perimeter Institute is supported in part
by the Government of Canada through the Department of Innovation, Science and
Economic Development Canada and by the Province of Ontario through the Ministry
of Colleges and Universities. RCM is supported in part by a Discovery Grant from
the Natural Sciences and Engineering Research Council of Canada, and by the BMO
Financial Group. HZC is supported by the Province of Ontario and the University
of Waterloo through an Ontario Graduate Scholarship. RCM and DN also received
funding from the Simons Foundation through the ``It from Qubit'' collaboration.
The work of IR is funded by the Gravity, Quantum Fields and Information group at
AEI, which is generously supported by the Alexander von Humboldt Foundation and
the Federal Ministry for Education and Research through the Sofja Kovalevskaja
Award. IR also acknowledges the hospitality of Perimeter Institute, where part
of this work was done.

\appendix

\bibliography{references} \bibliographystyle{utphys}

\end{document}

%% file: vincentsMacros.tex
\usepackage{amsmath}
\usepackage{amssymb}
\usepackage{mathtools}
\usepackage{cancel}
\usepackage{tikzsymbols}

\newcommand{\AdS}{\mathrm{AdS}}
\newcommand{\CFT}{\mathrm{CFT}}

\newcommand{\eff}{\mathrm{eff}}
\newcommand{\QES}{\mathrm{QES}}
\newcommand{\brane}{\mathrm{brane}}
\newcommand{\bath}{\mathrm{bath}}
\newcommand{\island}{\mathrm{isl.}}
\newcommand{\noisland}{\cancel{\mathrm{isl.}}}
\newcommand{\therm}{\mathrm{th}}
\newcommand{\but}{\mathrm{but}}
\newcommand{\ent}{\mathrm{ent}}
\newcommand{\induced}{\mathrm{induced}}
\newcommand{\UV}{\mathrm{UV}}
\newcommand{\BH}{\mathrm{BH}}

\newcommand{\bdyReg}{\mathbf{R}}
\newcommand{\RTbdy}{\Sigma_{\CFT}}
\newcommand{\RTbrn}{\sigma_{\bdyReg}}
\newcommand{\blkSurf}{\mathbf{V}}
\newcommand{\RT}{\Sigma_{\bdyReg}}

\newcommand{\LAdS}{L}
\newcommand{\Lbrn}{\ell_{\mt{B}}}
\newcommand{\LJT}{\ell_{\mt{JT}}}
\newcommand{\cutoffIR}{\ell_\mt{IR}} 
\newcommand{\cutoffbdy}{\delta}
\newcommand{\cutoffbrn}{\tilde{\delta}}
\newcommand{\volPerp}{\vol_{d-2}^\perp}
\newcommand{\DGPRatio}{\lambda_{\mathrm{b}}}
\newcommand{\braneAngle}{\theta_{\mt{B}}}
\newcommand{\RTDiam}{D}
\newcommand{\beltHalfWidth}{b}
\newcommand{\bByzs}{\mathcal{F}}
\newcommand{\xDom}{\tilde{x}}
\newcommand{\tDom}{\tilde{t}}
\newcommand{\zDom}{\tilde{z}}
\newcommand{\yAlm}{y}
\newcommand{\aAlm}{a}
\newcommand{\bAlm}{b}

\newcommand{\rhoDom}{\varrho}
\newcommand{\phibare}{\phi}
\newcommand{\phiren}{\tilde{\phi}}
\newcommand{\Gbr}{G_\mt{brane}}
\newcommand{\Gbk}{G_\mt{bulk}}
\newcommand{\Geff}{G_\mt{eff}}

\newcommand{\area}{A}
\newcommand{\vol}{\mathrm{vol}}
\newcommand{\hyperF}{\tensor[_2]{F}{_1}}


%% file: sections/01_intro.tex
Understanding the quantum description of black holes remains a central question in theoretical physics. One unresolved question is the fate of information during black hole evaporation. In his seminal work, Hawking argued that in a quantum theory black holes evaporate into a mixed state of radiation, independently of how the black hole was formed \cite{Haw74,Haw75,Haw76a}. Of course, this is in tension with the assumption that to an outside observer, the black hole looks like an ordinary, unitary quantum mechanical system, \eg as suggested by the AdS/CFT correspondence \cite{Polchinski:2016hrw,Harlow:2014yka}. This tension is colloquially known as the black hole information paradox \cite{Mat09}.

One way of sharpening the paradox is to consider the von Neumann entropy of the Hawking radiation produced during black hole evaporation. Assuming the gravitational system begins in a pure state, this entropy gives a measure of the amount of entanglement between the radiation and the black hole. According to Hawking's original calculation, the entanglement increases monotonically throughout the evaporation process since the radiation is thermal. 
On the other hand, unitary evolution would require that the thermodynamic entropy of the black hole, which is proportional to its horizon area \cite{Bek72,Bek73,Haw76}, set an upper bound on the
the entanglement entropy of the radiation. Since the former decreases as the black hole radiates, at some time -- known as the Page time -- the thermodynamic entropy of the black hole will equal the entropy of the radiation, and the latter entropy must then decrease in the subsequent evolution, reaching zero when the black hole has disappeared. That is, subtle correlations between the quanta emitted at early and late times must produce a purification of the final state, in a unitary evolution of the full system. This qualitative behaviour of the radiation's entropy as a function of time is known as the Page curve \cite{Page:1993wv} -- see also \cite{Harlow:2014yka}.

While reconciling Hawkings calculation with the idea that quantum gravity is
unitary was a longstanding puzzle, recently progress has made it possible to
compute the Page curve in a controlled manner \cite{Almheiri:2019psf,
  Penington:2019npb, Almheiri:2019hni}. The new approach builds on insights
coming from holographic entanglement entropy
\cite{Ryu:2006ef,Ryu:2006bv,Hubeny:2007xt,Rangamani:2016dms} and its extension
to include quantum contributions \cite{Faulkner:2013ana,Engelhardt:2014gca}. It
is best understood in a setting where a black hole is coupled to an auxiliary,
non-gravitational reservoir -- referred to as the {\it bath} -- which captures
the Hawking radiation.\footnote{This approach has now also been applied in a variety of different situations involving black holes
  \cite{Gautason:2020tmk,Sully:2020pza,Chen:2019iro,Anegawa:2020ezn,Balasubramanian:2020hfs,Hartman:2020swn,Hollowood:2020cou,Alishahiha:2020qza,Almheiri:2019psy,Rozali:2019day,Hashimoto:2020cas,Geng:2020qvw,Bak:2020enw,Li:2020ceg,Chandrasekaran:2020qtn,Almheiri:2020cfm,Hollowood:2020kvk,Almheiri:2019qdq,Bousso:2019ykv,Penington:2019kki,Akers:2019nfi,Chen:2020wiq,Kim:2020cds,Verlinde:2020upt,Liu:2020gnp,Bousso:2020kmy, Balasubramanian:2020coy, Chen:2020jvn, Stanford:2020wkf} and cosmology \cite{Cooper:2018cmb,Marolf:2020xie,Hartman:2020khs,Giddings:2020yes,Chen:2020tes,VanRaamsdonk:2020tlr, Sybesma:2020fxg, Balasubramanian:2020xqf}.} This setup can be interpreted as a idealized picture, where we split the spacetime into two regions: The first, in which gravity is important, is close to the black hole while the second region is far away, where gravitational effects are negligible, at least semi-classically. In this situation, it was argued  that instead of using Hawking's calculation, the true entropy of  the Hawking radiation captured in a region $\bdyReg$ of the bath should be calculated using the so-called \emph{island rule} \cite{Almheiri:2019hni} 
\beq\label{eq:islandformula}
 S_\mt{EE}(\bdyReg) = \text{min} \left\{\underset{\islands}{\text{ext}}  \left(S_\mt{QFT}( \bdyReg \cup \islands ) +\frac{A\!\left( \partial( \islands )\right)}{4G_N}   \right)  \right\}\,.
\eeq
This formula instructs us to evaluate the (semiclassical) entanglement entropy of the quantum fields in the bath region $\bdyReg$ combined with any codimension-two -- and possibly disconnected -- subregions in the gravitating region. The boundary of the candidate islands also contributes a gravitational term in the form of the usual Bekenstein-Hawking entropy. One extremizes the right-hand side of eq.~\eqref{eq:islandformula} over all such choices, and if the latter yields  multiple extrema, the correct choice is the one that yields the smallest entropy for $\bdyReg$. If this procedure yields a solution with a nontrivial region `$\islands$', the latter is called a \emph{quantum extremal island} -- see \cite{Almheiri:2020cfm} for a recent review.

For an evaporating black hole, an obvious choice for the island region which extremizes the entropy functional is the empty set, in which case the result of eq.~\eqref{eq:islandformula} agrees with Hawking's calculation. However, if radiation in the region $\bdyReg$ shares a large amount of entanglement with the quantum fields behind the horizon, new quantum extremal islands can appear. In particular, this occurs for an old evaporating black hole, and in this case a quantum extremal island appears just behind the horizon \cite{Penington:2019npb}. It turns out that after the Page time, this configuration yields the minimal entropy in eq.~\eqref{eq:islandformula}. As time evolves further, the entropy of $\bdyReg$ is controlled by the horizon area of the black hole which enters through the second term in eq.~\eqref{eq:islandformula}. Hence as the black hole evaporates, the latter shrinks to zero size and the island rule \eqref{eq:islandformula} gives a unitary Page curve.

\begin{figure}[t]
	\def\svgwidth{0.9\linewidth}
	\centering{
		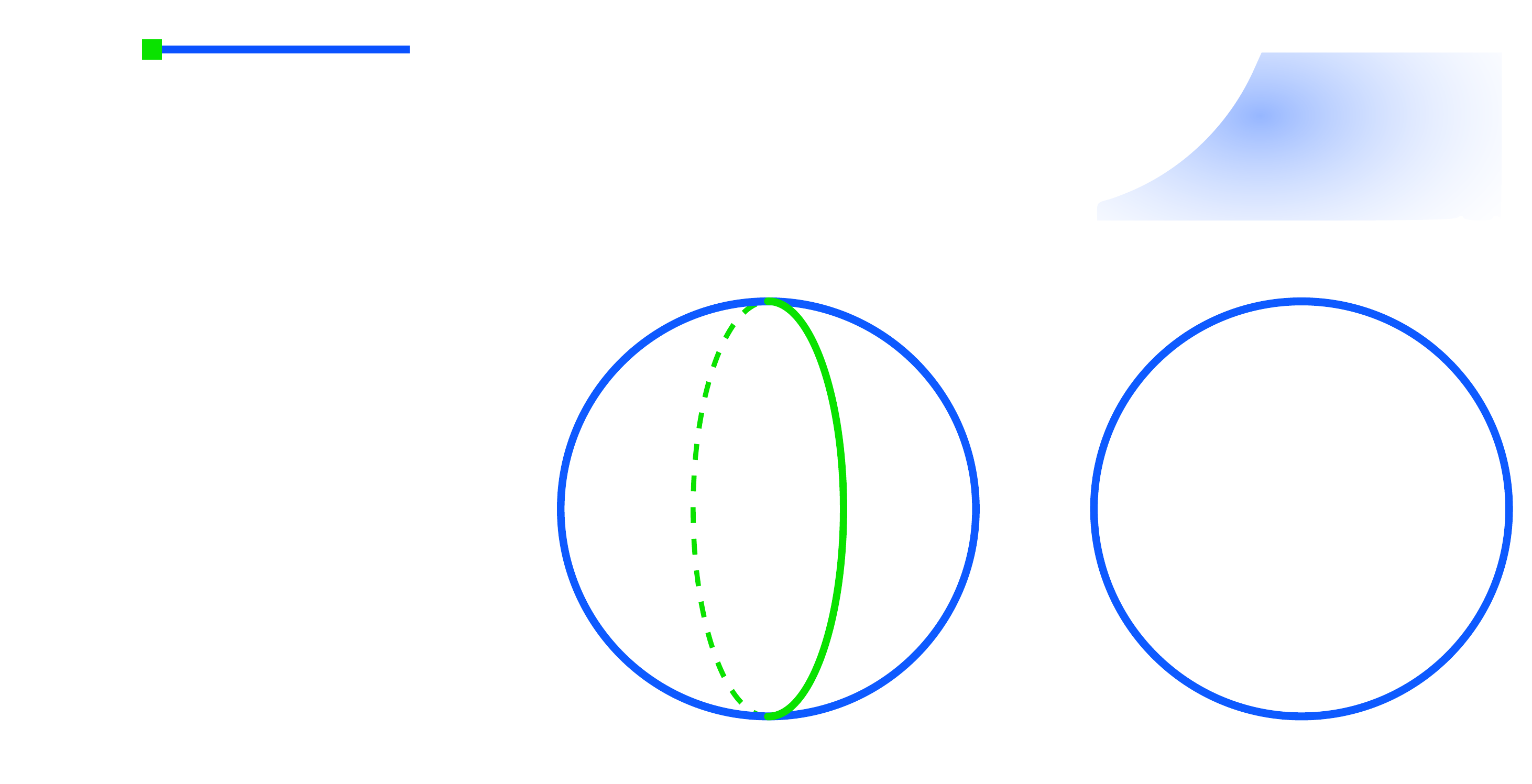
		\caption{Illustration of doubly-holographic models: The top row illustrates (a time slice of) the three perspectives of the model in \cite{Almheiri:2019hni}, while the bottom row displays the analogous descriptions of our construction in higher dimensions \cite{Chen:2020uac}. In the latter, we are using the global conformal frame where the boundary CFT lives on $R\times S^{d-1}$ and the conformal defect appears on the equator of the ($d-1$)-sphere -- see discussion in section \ref{sec:RS} and \cite{Chen:2020uac}. The bottom row reduces to the top upon setting $d=2$ and taking a $\mathbb{Z} _2$ quotient across the defect in the boundary or the brane in the bulk. The boundary, brane and bulk gravity perspectives correspond to panels a \& d, b \& e, and c \& f,  respectively.	}
		\label{threetalesX}
	}
\end{figure}

Eq.~\eqref{eq:islandformula} was motivated in part by analyzing a ``doubly-holographic'' model in \cite{Almheiri:2019hni}. This model provides three different descriptions of the physical phenomena: First, from the {\it boundary perspective}, the system consists of two (one-dimensional) quantum mechanical systems, which are entangled in a thermofield double state. Further, one of the quantum mechanical systems is coupled to a two-dimensional holographic CFT, which plays the role of the bath -- see figure \ref{threetalesX}a. With the {\it brane perspective}, the quantum mechanical systems are replaced by their holographic dual, a two-dimensional black hole in JT gravity. The latter has an AdS$_2$ geometry, which also supports another copy of the two-dimensional holographic CFT -- see figure \ref{threetalesX}b. Finally, with the {\it bulk perspective}, the holographic CFT is replaced everywhere with three-dimensional Einstein gravity in an asymptotically AdS$_3$ geometry. The latter effectively has two boundaries: the standard asymptotically AdS boundary and the region where JT gravity is supported, 
which is referred to as the Planck brane -- see figure \ref{threetalesX}c. An advantage of working in the bulk perspective is that entanglement entropies of subregions in the bath can be computed geometrically using the usual rules of holographic entanglement entropy \cite{Ryu:2006ef,Ryu:2006bv,HubRan07}, taking into account that the RT surfaces that can also end on the Planck brane \cite{Tak11,FujTak11}. 

One direction for progress is to understand the Page curve and quantum extremal islands in higher dimensions. While limited results have been obtained on this front \cite{Penington:2019npb,Almheiri:2019psy,Rozali:2019day,Hashimoto:2020cas,Geng:2020qvw,Bak:2020enw}, we focus here on the holographic model which we introduced in \cite{Chen:2020uac}. Our model allows us to obtain analytic results, while being powerful enough to do calculations in the regime where the gravitational theory on the brane is well-approximated by Einstein gravity.  In our previous paper, we showed that quantum extremal islands can appear in any spacetime dimension, and clarified several of the properties of the doubly-holographic model in \cite{Almheiri:2019hni}. Here, we will extend our earlier work and discuss the presence of quantum extremal islands for black holes coupled to bath at a finite temperature. That is, our analysis provides a higher dimensional extension of the two-dimensional scenario considered in \cite{Almheiri:2019yqk}.

The key feature of our holographic model \cite{Chen:2020uac} is that it reproduces the three descriptions of the underlying physics discussed above for the doubly-holographic model of \cite{Almheiri:2019hni}.  From the boundary perspective, our system consists of a $d$-dimensional holographic CFT coupled to codimension-one conformal defect, as shown in figure \ref{threetalesX}d. Using the standard AdS/CFT dictionary, this description is translated to the {bulk gravity perspective}. The latter describes the system in terms of ($d+1$)-dimensional Einstein gravity in an asymptotically AdS$_{d+1}$ geometry coupled to a $d$-dimensonal brane, which intersects the boundary at the location of the conformal defect -- see figure \ref{threetalesX}f. According to the Randall-Sundrum (RS) scenario \cite{Randall:1999ee,Randall:1999vf,Karch:2000ct}, the gravitational backreaction of the brane warps the bulk geometry creating new localized graviton modes in its vicinity. This mechanism allows for the brane perspective, shown in figure \ref{threetalesX}e, where the system is described by an effective theory of Einstein gravity coupled to (two copies of) the holographic CFT on the brane, all coupled to the boundary CFT. In \cite{Chen:2020uac}, we also considered introducing an intrinsic Einstein term to the brane action, analogous to the construction of Dvali, Gabadadze and Porrati (DGP) \cite{Dvali:2000hr}. 

Hence our construction \cite{Chen:2020uac} provides a natural generalization to higher dimensions of the two-dimensional doubly-holographic setup considered in \cite{Almheiri:2019hni}. Let us also note that our model resembles the setup in \cite{Almheiri:2019hni} even more closely upon taking a $\mathbb Z_2$ orbifold quotient across the brane. Further, we emphasize that while the three different perspectives were presented on a more or less equal footing, the fact that the RS gravity on the brane has a finite UV cutoff \cite{Randall:1999ee,Randall:1999vf} singles out the brane prespective as an effective low-energy description, in contrast to the boundary and bulk descriptions.\footnote{This does not mean that the bulk description in terms of a(n infinitely thin) brane in AdS${}_{d+1}$ is UV complete. However, it is reasonable to expect that the bulk description can be completed in the UV by a more complicated configuration which can be obtained within string theory, \eg see \cite{DHoker:2007hhe,Chiodaroli:2009yw,Chiodaroli:2011nr}. In contrast, the brane theory has a fundamental cutoff.}

Again, the bulk gravity perspective allows us to calculate entanglement entropies of boundary regions geometrically with the usual rules of holographic entanglement entropy \cite{Ryu:2006ef,Ryu:2006bv,HubRan07}. From the brane perspective then, quantum extremal islands simply arise when the minimal RT surfaces in the bulk extend across the brane for certain configurations. 

In this case, the entanglement entropy of the corresponding boundary region $\bdyReg$ is given by
\begin{align}\label{eq:island2}
 S_\text{EE}(\bdyReg)= \text{min} \left\{\underset{\RT}{\text{ext}}  \left( \frac{\area(\RT)}{4 G_\bulk}
  + \frac{\area(\RTbrn)}{4 G_\brane}  \right)  \right\}
\end{align}
where $\RT$ is the usual bulk RT surface, \ie an extremal codimension-two  surface in the bulk homologous to $\bdyReg$. As argued in \cite{Chen:2020uac}, when the brane supports an intrinsic gravitational action, we must also include a Bekenstein-Hawking area contribution for the brane region $\RTbrn=\RT\cap \text{brane}$. This intersection of the RT surface with the brane becomes the boundary of the islands seen in the brane prespective. 

\begin{figure}[t]
	\def\svgwidth{.7\linewidth}
	\centering
		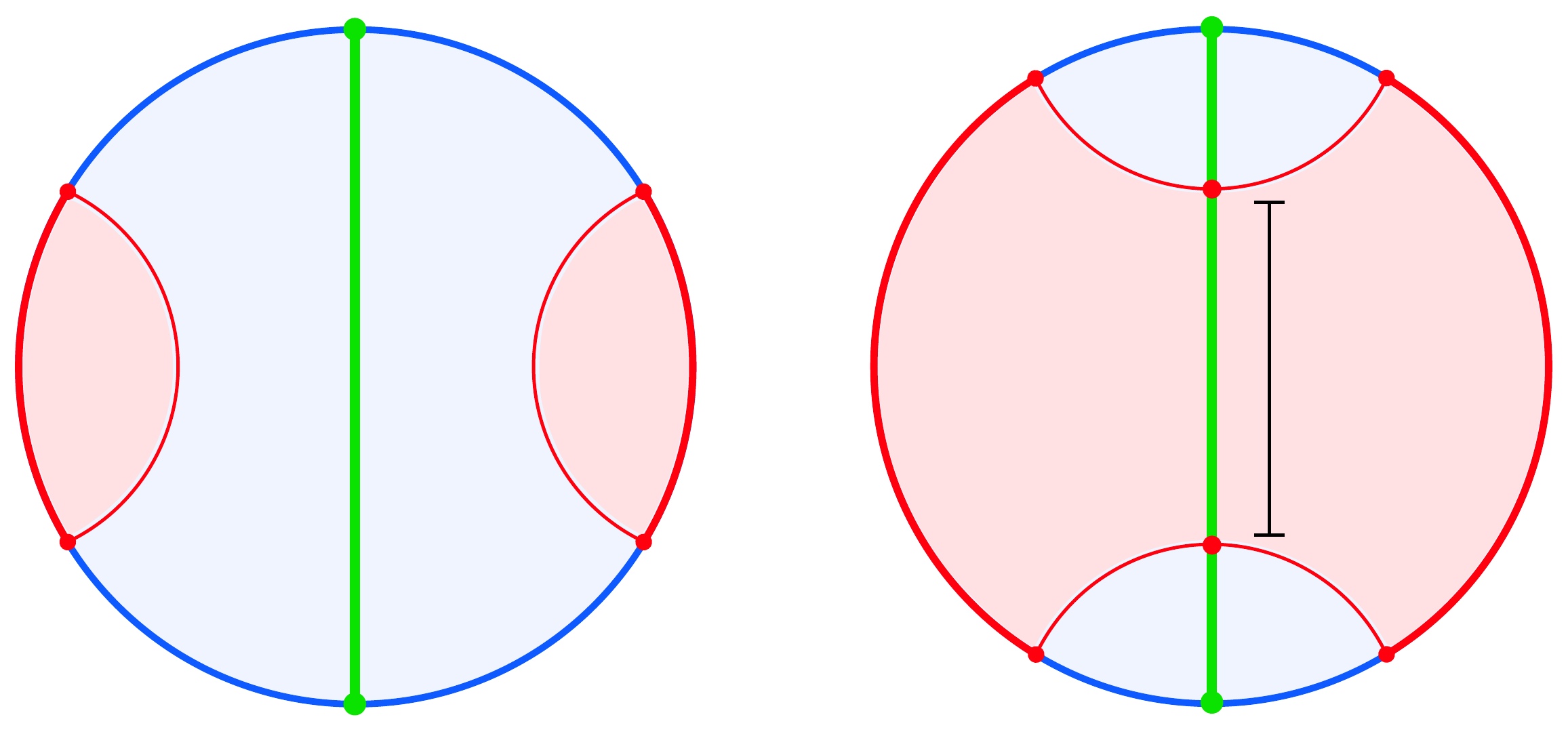
		\caption{The choice of RT surfaces on a constant time slice in the presence of the brane (coloured green), showing the different ingredients involved in eq.~\eqref{eq:island2}. }
\label{fig:RTPhases_intro}		
\end{figure}

The equivalence between eqs.~\eqref{eq:islandformula} and \eqref{eq:island2} can be easily understood as follows: The bulk term in eq.~\eqref{eq:island2} describes the leading planar contributions of the entanglement entropy of the boundary CFT, and so matches the first term in eq.~\eqref{eq:islandformula}. However, expanding this geometric contribution near the brane also reveals an Bekenstein-Hawking term that matches the induced Einstein term in the effective gravitational action on the brane \cite{Chen:2020uac}. This contribution combines with the brane term  in eq.~\eqref{eq:island2} to produce the expected gravitational contribution appearing in eq.~\eqref{eq:islandformula}. In fact, the RT contribution also captures higher derivative contributions matching the Wald-Dong entropy \cite{Wald:1993nt,Iyer:1994ys,Jacobson:1993vj,Dong:2013qoa} of the higher curvature terms appearing in the effective gravitational action \cite{Chen:2020uac}. Further, as discussed in \cite{Chen:2020uac}, the competition between candidate quantum extremal islands, denoted by the `min' in eq.~\eqref{eq:islandformula} simply becomes the usual competition between different possible RT surfaces in the holographic formula \eqref{eq:island2}, \eg see figure \ref{fig:RTPhases_intro}.

In the following, we will study the question of quantum extremal islands for black holes in arbitrary dimensions using the purely geometric description \reef{eq:island2} of the bulk gravity perspective. As emphasized in \cite{Chen:2020uac}, the transition between the phase without an island and that with the island is nothing more than the usual transition between different classes of RT surfaces \cite{Hea10,Faulkner:2013yia,Hartman:2013mia} -- see figure \ref{fig:RTPhases_intro}. In particular, in the island phase, the RT surface crosses the brane so that a portion of the latter, \ie the island, is included in the corresponding entanglement wedge. Thus the appearance of quantum extremal islands is simply decribed by a well understood feature of holographic entanglement entropy in a new setting.  The main advantage of our construction here and in \cite{Chen:2020uac} lies in its simplicity. As we will show, our framework allows us to carry the calculations remarkably far analytically, complementing previous approaches which heavily relied on numerics \cite{Almheiri:2019psy}. In our case, the numerics required to extract quantitative results are limited to solving few ODEs. 

The remainder of this paper is organised as follows: In section \ref{sec:RS}, we review the bulk geometry and effective action of our model presented in \cite{Chen:2020uac}, which is based on the Karch-Randall setup \cite{Karch:2000ct,Porrati:2001gx,Karch:2001jb,Porrati:2001db,Aharony:2003qf,Miemiec:2000eq,Schwartz:2000ip} for branes embedded in AdS. We also discuss the addition of a DGP term \cite{Dvali:2000hr} to the brane action. For the two dimensional bulk gravity case, we summarize the setup of \cite{Almheiri:2019yqk}, describe the connection to our model and introduce eternal black holes. In section \ref{sec:nonextremal}, we construct eternal black holes on the brane in higher dimensions. As in the $d=2$ case, these black holes are in equilibrium with the bath at finite temperature and so they do not evaporate. Nonetheless, there is a continuous exchange of radiation between the black hole and the bath, which has the potential to create an information paradox \cite{Almheiri:2019yqk}. Hence, we use eq.~\eqref{eq:island2} to investigate under which conditions islands appear. We present the general analysis for the time dependence of the entropy, exploring the island and no-island phases. In section \ref{sec:numerics}, we develop the numerics associated to some integral equations found in the previous section and explicitly evaluate the Page curve for $d=3$, 4 and 5. Section \ref{sec:extremal} examines an extremal horizon with a vanishing temperature, and find that in contrast to two dimensions \cite{Almheiri:2019yqk}, generally islands do not form in higher dimensions. However, this is not problematic, since at zero temperature the black hole and bath are not actually exchanging radiation and thus no information paradox arises. Details for the special case $d=2$ appear in section \ref{app:Page2d}. We review the induced action on the two-dimensional brane, including the introduction of JT gravity terms, given in \cite{Chen:2020uac}. We also evaluate the corresponding quantum extremal surfaces and the Page curve, and show that the brane cutoff produces subleading corrections compared to the results in \cite{Almheiri:2019yqk}. Finally in section \ref{sec:discuss}, we discuss our results and point towards some future directions.

%% file: images/ThreeTales_EditcopyJS5.pdf_tex
\begingroup%
  \makeatletter%
  \providecommand\color[2][]{%
    \errmessage{(Inkscape) Color is used for the text in Inkscape, but the package 'color.sty' is not loaded}%
    \renewcommand\color[2][]{}%
  }%
  \providecommand\transparent[1]{%
    \errmessage{(Inkscape) Transparency is used (non-zero) for the text in Inkscape, but the package 'transparent.sty' is not loaded}%
    \renewcommand\transparent[1]{}%
  }%
  \providecommand\rotatebox[2]{#2}%
  \newcommand*\fsize{\dimexpr\f@size pt\relax}%
  \newcommand*\lineheight[1]{\fontsize{\fsize}{#1\fsize}\selectfont}%
  \ifx\svgwidth\undefined%
    \setlength{\unitlength}{841.88976378bp}%
    \ifx\svgscale\undefined%
      \relax%
    \else%
      \setlength{\unitlength}{\unitlength * \real{\svgscale}}%
    \fi%
  \else%
    \setlength{\unitlength}{\svgwidth}%
  \fi%
  \global\let\svgwidth\undefined%
  \global\let\svgscale\undefined%
  \makeatother%
  \begin{picture}(1,0.51178451)%
    \lineheight{1}%
    \setlength\tabcolsep{0pt}%
    \put(0.01744583,0.33564636){\color[rgb]{0,0,0}\makebox(0,0)[lt]{\lineheight{1.25}\smash{\begin{tabular}[t]{l}a.\end{tabular}}}}%
    \put(0.01744583,0.02926547){\color[rgb]{0,0,0}\makebox(0,0)[lt]{\lineheight{1.25}\smash{\begin{tabular}[t]{l}d.\end{tabular}}}}%
    \put(0.36558402,0.33564636){\color[rgb]{0,0,0}\makebox(0,0)[lt]{\lineheight{1.25}\smash{\begin{tabular}[t]{l}b.\end{tabular}}}}%
    \put(0.36558402,0.02926547){\color[rgb]{0,0,0}\makebox(0,0)[lt]{\lineheight{1.25}\smash{\begin{tabular}[t]{l}e.\end{tabular}}}}%
    \put(0,0){\includegraphics[width=\unitlength,page=1]{images/ThreeTales_EditcopyJS5.pdf}}%
    \put(0.71482202,0.02967891){\color[rgb]{0,0,0}\makebox(0,0)[lt]{\lineheight{1.25}\smash{\begin{tabular}[t]{l}f.\end{tabular}}}}%
    \put(0,0){\includegraphics[width=\unitlength,page=2]{images/ThreeTales_EditcopyJS5.pdf}}%
    \put(0.71482161,0.33564561){\color[rgb]{0,0,0}\makebox(0,0)[lt]{\lineheight{1.25}\smash{\begin{tabular}[t]{l}c.\end{tabular}}}}%
    \put(0,0){\includegraphics[width=\unitlength,page=3]{images/ThreeTales_EditcopyJS5.pdf}}%
  \end{picture}%
\endgroup%

%% file: images/RTPhases_updated.pdf_tex
\begingroup%
  \makeatletter%
  \providecommand\color[2][]{%
    \errmessage{(Inkscape) Color is used for the text in Inkscape, but the package 'color.sty' is not loaded}%
    \renewcommand\color[2][]{}%
  }%
  \providecommand\transparent[1]{%
    \errmessage{(Inkscape) Transparency is used (non-zero) for the text in Inkscape, but the package 'transparent.sty' is not loaded}%
    \renewcommand\transparent[1]{}%
  }%
  \providecommand\rotatebox[2]{#2}%
  \newcommand*\fsize{\dimexpr\f@size pt\relax}%
  \newcommand*\lineheight[1]{\fontsize{\fsize}{#1\fsize}\selectfont}%
  \ifx\svgwidth\undefined%
    \setlength{\unitlength}{666.14173228bp}%
    \ifx\svgscale\undefined%
      \relax%
    \else%
      \setlength{\unitlength}{\unitlength * \real{\svgscale}}%
    \fi%
  \else%
    \setlength{\unitlength}{\svgwidth}%
  \fi%
  \global\let\svgwidth\undefined%
  \global\let\svgscale\undefined%
  \makeatother%
  \begin{picture}(1,0.46808511)%
    \lineheight{1}%
    \setlength\tabcolsep{0pt}%
    \put(0,0){\includegraphics[width=\unitlength,page=1]{images/RTPhases_updated.pdf}}%
    \put(-0.01836025,0.14149102){\color[rgb]{0,0,0}\makebox(0,0)[lt]{\lineheight{1.25}\smash{\begin{tabular}[t]{l}$\bdyReg$\end{tabular}}}}%
    \put(0.08292115,0.32291755){\color[rgb]{0,0,0}\makebox(0,0)[lt]{\lineheight{1.25}\smash{\begin{tabular}[t]{l}$\RT$\end{tabular}}}}%
    \put(0.57051508,0.06604643){\color[rgb]{0,0,0}\makebox(0,0)[lt]{\lineheight{1.25}\smash{\begin{tabular}[t]{l}$\bdyReg$\end{tabular}}}}%
    \put(0.65859472,0.33696537){\color[rgb]{0,0,0}\makebox(0,0)[lt]{\lineheight{1.25}\smash{\begin{tabular}[t]{l}$\RT$\end{tabular}}}}%
    \put(0.717236,0.13420884){\color[rgb]{0,0,0}\makebox(0,0)[lt]{\lineheight{1.25}\smash{\begin{tabular}[t]{l}$\RTbrn$\end{tabular}}}}%
    \put(0.82572333,0.22366794){\color[rgb]{0,0,0}\makebox(0,0)[lt]{\lineheight{1.25}\smash{\begin{tabular}[t]{l}island\end{tabular}}}}%
  \end{picture}%
\endgroup%

%% file: sections/02_setup.tex
\subsection{Braneworlds in higher dimensions}
Let us review the holographic model discussed in \cite{Chen:2020uac}. Beginning with the {\it bulk gravity perspective} our setup is described by ($d+1$)-dimensional Einstein gravity with a negative cosmological constant,\footnote{Throughout the paper, we ignore surface terms for the gravitational action, \eg see \cite{PhysRevLett.28.1082,PhysRevD.15.2752,Emparan:1999pm}.}
\beq
I_\mt{bulk} = \frac{1}{16 \pi G_\mt{bulk}}\int d^{d+1}x\sqrt{-g}
\[ {R}(g) + \frac{d(d-1)}{L^2} \]\,,
\label{act2}
\eeq
where $g_{ab}$ denotes the bulk metric. We also introduce a codimension-one brane in the bulk with action 
\beq
I_\mt{brane} = -(T_o-\Delta T)\int d^dx\sqrt{-\tilde{g}} + \frac{1}{16\pi G_\mt{brane}}\int d^dx\sqrt{-\tilde{g}}\, \tilde{R}(\tilde g)\,,
\label{act1}
\eeq
where $\tilde{g}_{ij}$ is the induced metric on the brane. As well as the usual tension term, we have also introduced an intrinsic Einstein-Hilbert term in the brane action, in a manner analogous to Dvali-Gabadadze-Porrati (DGP) braneworld gravity \cite{Dvali:2000hr}.
We have separated the brane tension into $T_o$ and $\Delta T$, and will tune $\Delta T\propto 1/G_\mt{brane}$ so that the brane position is determined entirely by $T_o$. Adding the DGP term is a natural generalisation to higher dimensions of having JT gravity on a  two-dimensional brane \cite{Almheiri:2019hni} -- see section \ref{app:Page2d}. 

\begin{figure}[t]
	\def\svgwidth{0.9\linewidth}
	\centering{
		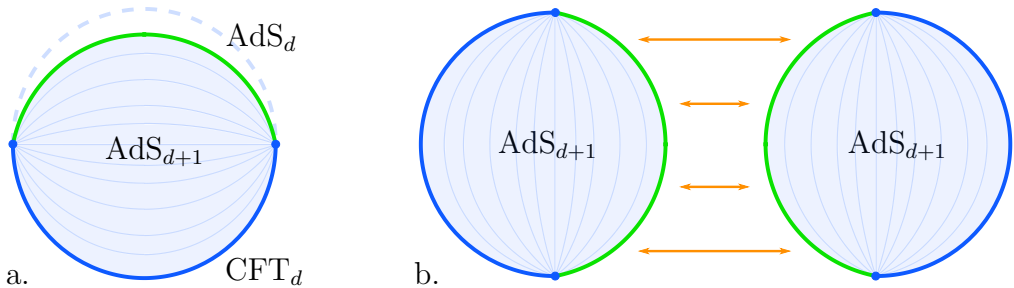
		\caption{A timeslice of our Randall-Sundrum setup. In panel (a), we cut off the AdS$_{d+1}$ spacetime along an AdS$_d$ slice near the asymptotic boundary $\theta=0$, in the metric \reef{metric33}. Two of these spaces are glued together in panel (b) and the brane is realized as the interface between the two geomeries.}
		\label{fig:brane2}
	}
\end{figure}

Since the brane is codimension-one, the bulk geometry away from the brane locally takes the form of AdS$_{d+1}$ with the curvature scale set by $L$. We will work in a regime where the induced geometry on the brane will be that of AdS$_d$ space -- see \cite{Chen:2020uac} for details -- and so it is useful to consider the following foliation of the AdS$_{d+1}$ geometry by AdS$_d$ slices
\beq\label{metric33}
ds_{\AdS_{d+1}}^2
= \frac{L^2}{\sin^2\theta}\,\(d\theta^2 + ds^2_{\AdS_{d}}\)\,.
\eeq
The AdS$_d$ metric is dimensionless with unit curvature. This metric would cover the entire AdS$_{d+1}$ vacuum spacetime if we take $0\le\theta\le\pi$. The solution for the backreacting brane is constructed by first cutting off the spacetime along an AdS$_d$ slice near the asymptotic boundary $\theta=0$, \ie at $\theta=\theta_\mt{B}\ll1$ where $\theta_\mt{B}$ is determined by the brane tension $T_o$ -- see below. Then, two such spaces are joined together along this surface, and the brane is realized as the interface between the two geometries. With this construction, the brane divides the bulk spacetime in half, but the backreaction of the brane has enlarged the geometry -- see figure \ref{fig:brane2}. In this case, the metric \reef{metric33} can be used to cover a coordinate patch with $\theta_\mt{B}\le\theta\le\pi$ on either side of the brane.

With the above construction, the induced geometry on the brane is simply AdS$_d$ and using the Israel junction conditions \cite{israel1966singular,Chen:2020uac}, one finds the curvature scale to be
\beq
\frac{1}{\ell_\mt{B}^2}=\frac{\sin^2\theta_\mt{B}}{L^2}=
 \frac{2}{L^2}\,\veps\(1-\veps/2\),\qquad{\rm where}\ \ \veps\equiv\(1-\frac{4 \pi G_\mt{bulk} L T_o}{d-1}\)\,.
\label{curve2}
\eeq
For the most part, we will be interested in the regime where $L^2/\ell_\mt{B}^2\ll1$ or $\veps\ll1$. As we will explain below, this ensures that the gravitational theory on the brane is essentially Einstein gravity. Implicitly in eq.~\reef{curve2}, we have tuned  the ``shift" $\Delta T$ to produce an embedding of the brane that is independent of the DGP coupling $\Gbr$, \ie the brane location remains unchanged when we vary $\Gbr$. This is achieved by setting
\beq
\Delta T=\frac{(d-1)(d-2)}{16\pi G_\mt{brane}\, \ell_\mt{B}^2}\,.
\label{tense1}
\eeq

The {\it boundary perspective} simply considers the dual description of the above gravitational system using the standard rules of the AdS/CFT correspondence. As described in \cite{Chen:2020uac}, when considered in ``global" coordinates, the dual solution is naturally the boundary CFT on a spherical cylinder $R\times S^{d-1}$ (where the $R$ is the time direction). Further there is also a codimension-one conformal defect positioned on the equator of the sphere, where the brane reaches the asymptotic boundary. The central charge of the boundary CFT is given by the standard expression $c_T\sim L^{d-1}/G_\mt{bulk}$, \eg see \cite{Buchel:2009sk}, whereas the ($d-1$)-dimensional CFT of the conformal defect has $\tilde{c}_T\sim \ell_\mt{eff}^{d-2}/G_\mt{eff}\gg c_T$. Similarly, one can consider the ratio of the couplings in the defect and bath CFTs: ${\tilde \lambda}/{\lambda}\sim{\ell_\mt{eff}}/{L}\gg1$.

We arrive at the {\it brane perspective} by replacing the conformal defect in the boundary perspective by its gravitational dual. 
Hence this description includes the boundary CFT on the asymptotic AdS$_{d+1}$ boundary, but also two copies of the boundary CFT on the brane, as dictated by the usual Randall-Sundrum (RS) scenario. Of course, the latter is an effective theory with a finite UV cutoff set by the position of the brane, \eg see \cite{Gubser:1999vj} and references therein.\footnote{In fact, working with the induced metric on the brane (as we do in the following), the short-distance cutoff on the brane is $\tilde \delta \simeq L$ -- see \cite{Chen:2020uac} for further details.}
Further, new (nearly) massless graviton modes localized in the vicinty of the brane also appear and so the brane also supports a gravitational theory. We can think that integrating out the brane CFT (or the bulk gravity) induces an effective gravitational action on the brane of the form \cite{Chen:2020uac}
\beqa\label{inducedAction}
I_\mt{induced}&=&\frac{1}{16 \pi G_\mt{eff}}\int d^{d}x\sqrt{-\tilde{g}}
\[\frac{(d-1)(d-2)}{\ell_\mt{eff}^2} + \tilde{R}(\tilde{g})\]
\label{act3}\\
&&+
\frac{1}{16 \pi G_\mt{RS}}\int d^{d}x\sqrt{-\tilde{g}}
\[\frac{L^2}{(d-4)(d-2)}\(\tilde{R}^{ij}\tilde{R}_{ij}-
\frac{d}{4(d-1)} \tilde{R}^2\)+\cdots\]\,,
\nonumber
\eeqa
where
\beqa
\begin{split}
\frac{1}{G_\mt{eff}}&\equiv&\frac{1}{G_\mt{RS}}(1+\DGPRatio)\qquad
{\rm with} \ \ \ \DGPRatio \equiv \frac{G_\mt{RS}}{G_\mt{brane}}\,,
\\
\frac{1}{G_\mt{RS}}&=&\frac{2L}{(d-2)G_\mt{bulk}}\,,\qquad\qquad
\frac{1}{\ell_\mt{eff}^2}=
 \frac{2}{L^2}\,\veps\,,
\end{split}
\label{Newton3}
\eeqa
and $\veps$ is given in eq.~\reef{curve2}. Note that in the regime of interest
(\ie $\veps\ll1$), we have $\ell_\mt{eff}\simeq\ell_\mt{B}$. Hence to leading
order, the above gravitational theory \reef{act3} corresponds to Einstein
gravity coupled to a negative cosmological constant. In the second line of
eq.~\reef{act3}, we show the first of a(n infinite) sequence of higher curvature
corrections, involving powers of $L^2\ \times$~curvature. Since the
gravitational equations of motion set the curvatures to be roughly
$1/\ell_\mt{eff}^2$ (at least for the background of interest), the contribution of these terms is highly suppressed since we work in the regime where $L^2/\ell_\mt{eff}^2\ll1$.\footnote{A more careful examination in \cite{Chen:2020uac} showed that the gravitational theory on the brane was well approximated as semiclassical Einstein gravity with $L^2/\ell_\mt{eff}^2\ll1$ for $\DGPRatio>0$, but required $L^2/\ell_\mt{eff}^2\ll1+\DGPRatio$ for $\DGPRatio<0$. However, the latter constraint is replaced by $L^2/\ell_\mt{eff}^2\ll(1+\DGPRatio)^2$ for the special case of $\DGPRatio<0$ and $d=3$.  \label{bigtoe}} Lastly, let us note that $1/G_\mt{RS}$ is the standard RS gravitational coupling induced in the absence of a DGP term, \ie $\DGPRatio=0$.

It turns out that in the case of a brane theory with negative cosmological constant, like the one we are considering here, the graviton acquires a mass \cite{Karch:2000ct,Porrati:2001gx,Karch:2001jb,Porrati:2001db,Aharony:2003qf,Miemiec:2000eq,Schwartz:2000ip}. For small brane angles, the graviton mass is proportional so some power of the brane angle \cite{Miemiec:2000eq,Schwartz:2000ip} and thus vanishes as we take the zero-angle limit. 
It was suggested in \cite{Geng:2020qvw} that this mass is a crucial ingredient for islands to exist, since the limit of vanishing graviton mass coincides with a limit in which islands cannot be created since their area becomes infinite. Alternatively, it is possible that in the Karch-Randall model, the graviton mass simply depends on the effective gravitational coupling on the brane, and is thus correlated with the island size, but not responsible for the island.

\subsection{Two dimensions and black holes} \label{twod}

\begin{figure}[t]
	\def\svgwidth{.7\linewidth} \centering{ 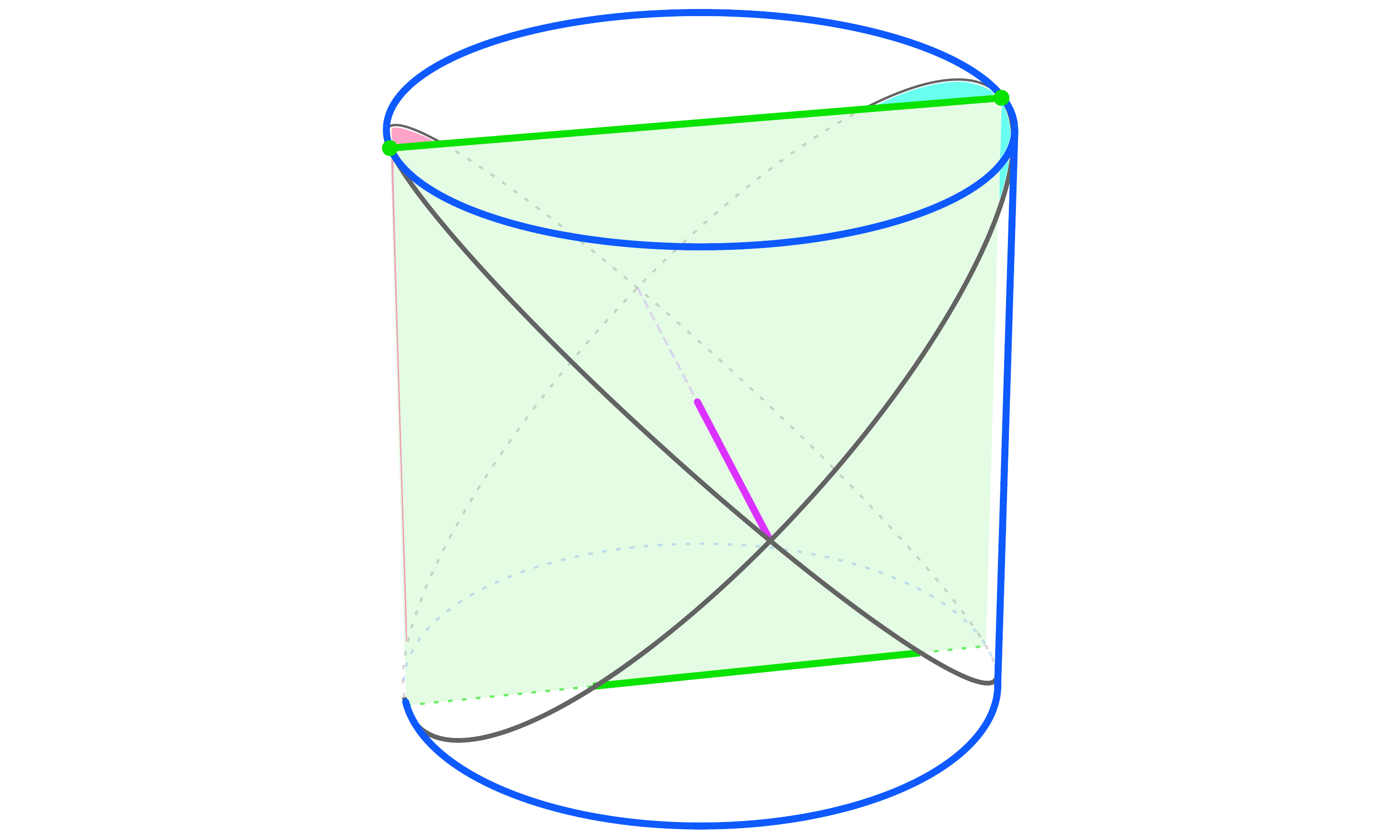
		\caption{Our eternal black hole coupled to the CFT bath, as seen from the
      \textit{bulk perspective}.}
		\label{fig:bulkpenrose}
	}
\end{figure}

In two dimensions, we need to revisit our setup for an accurate effective brane action
and to make connection to \cite{Almheiri:2019yqk}. First, there are factors of
$1/(d-2)$ appearing in eq.~\reef{act3},\footnote{Similar factors of $1/(d-4)$
  appear and are problematic in $d=4$, but we work in a regime such that the
  curvature squared terms of eq. (\ref{inducedAction}) are irrelevant.} which
indicate that the bulk integration analysis leading to this result must be
reconsidered for $d=2$. As reviewed in section \ref{app:Page2d}, we find that
the induced brane action is non-local, a signature of the trace anomaly. In
addition, the two-dimensional analogue of the DGP brane action is a JT
gravitational action localized on the brane. Having accounted for these changes,
we may relate our setup directly to that of \cite{Almheiri:2019yqk}, which we
now briefly review.

Ref. \cite{Almheiri:2019yqk} interprets the two Rindler patches of AdS$_2$ as
exteriors of an eternal non-zero temperature black hole and subsequently
considers coupling each exterior to a flat half-space, consitituting a bath region. A matter $\CFT_2$ theory
spans both the bath and $\AdS_2$ regions and JT gravity is placed on the
$\AdS_2$ region. Invoking $\AdS_2/\CFT_1$, this setup is alternatively described
by the thermofield double (TFD) state of a BCFT living on
two half-lines (the bath regions) coupled to quantum mechanics (dual to the $\AdS_2$
spacetime) on the boundaries of the half-lines.
The authors then compute the
entanglement entropy of a region consisting of intervals on both sides of the
TFD including the defect and with endpoints in the bath regions. From the
$\AdS_2$ perspective, this entropy is obtained using
eq.~\eqref{eq:islandformula}, allowing for the possibility of islands in the
$\AdS_2$ spacetime. In particular, this gives rise to a competition between a
no-island phase and an island phase, with the former dominating at early times
and the latter at late times. In the island phase, quantum extremal surfaces
(QESs) appear in the $\AdS_2$ spacetime just outside the horizon, marking the
boundaries of an island, stretching through the $\AdS_2$ wormhole,
which now belongs to the entanglement wedge of the bath complements to
the intervals.

\begin{figure}[t]
	\def\svgwidth{.5\linewidth} \centering{ 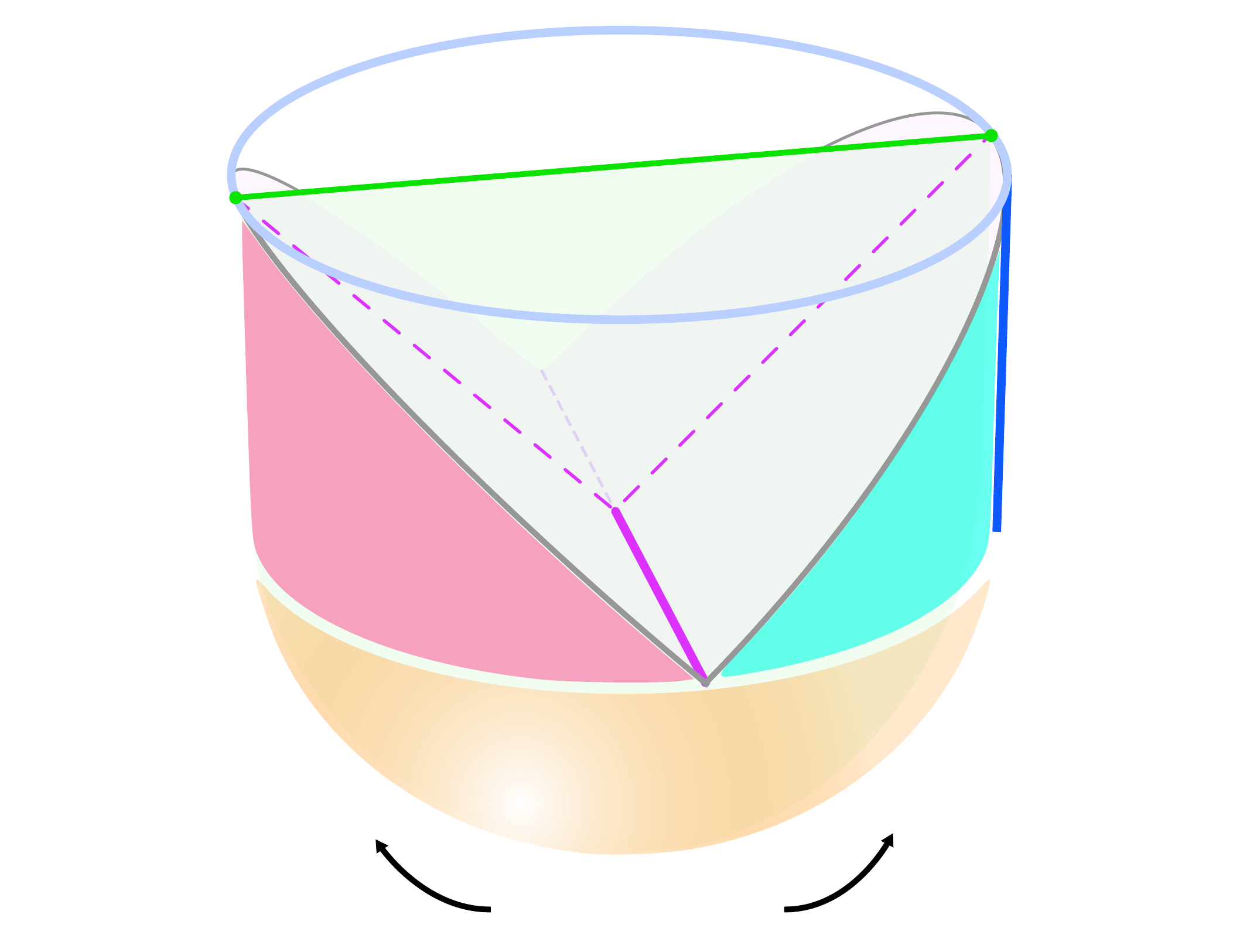
		\caption{The Euclidean path integral (orange region) prepares the
      Hawking-Hartle state. The black hole temperature $T=1/(2\pi R)$ is derived
      in section \ref{sec:nonextremal}.}
		\label{fig:HH}
	}
\end{figure}

\begin{figure}[t]
	\def\svgwidth{.6\linewidth} \centering{
    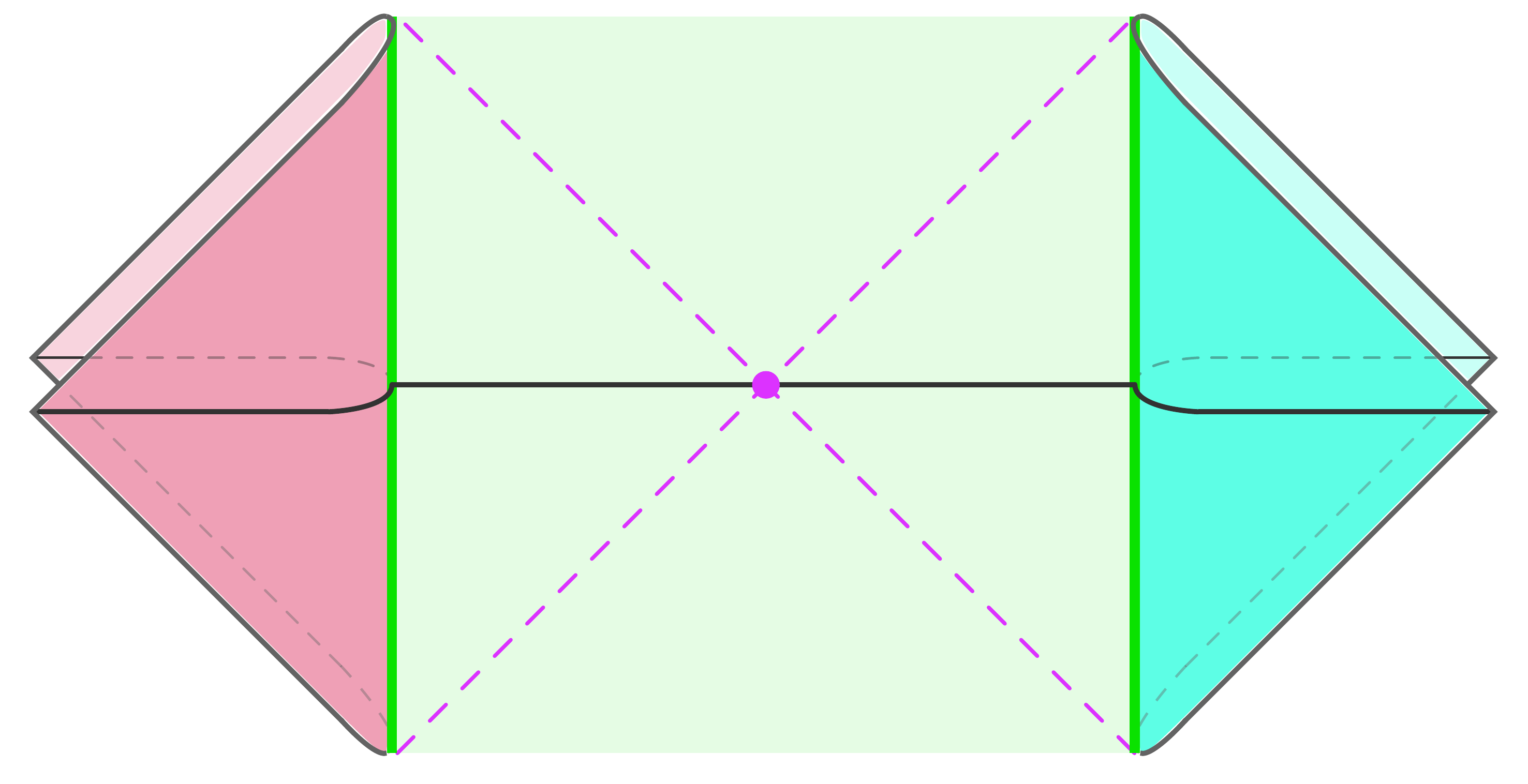
		\caption{Our eternal black hole coupled to the CFT bath, as seen from the
      effective \textit{brane perspective}. Each point in the Penrose diagram
      represents a hyperbolic space $H_{d-2}$. For $d=2$ this is simply a
      point.}
		\label{fig:Eternal}
	}
\end{figure}

\begin{figure}[t]
	\def\svgwidth{.6\linewidth} \centering{
    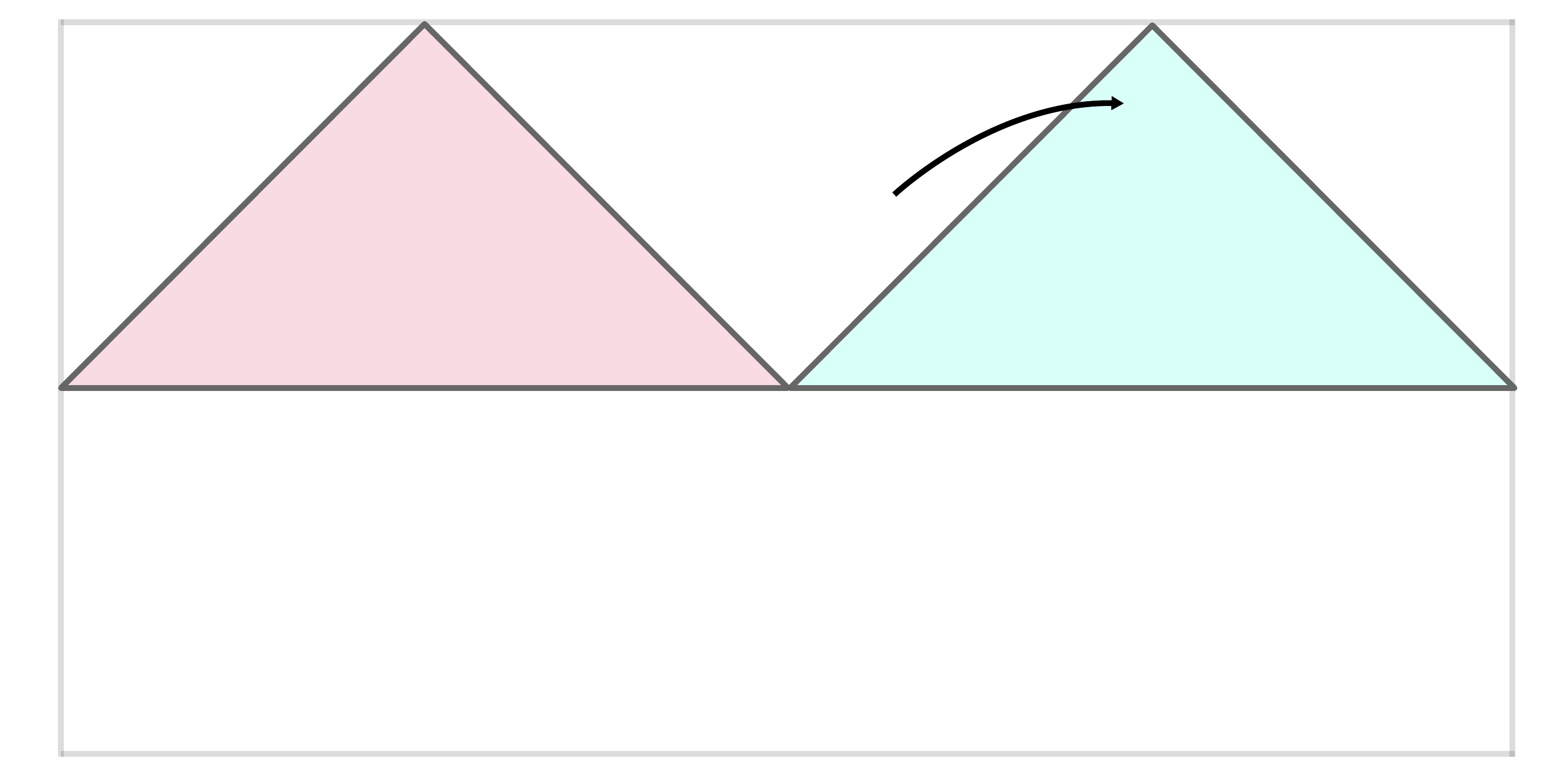
		\caption{Conformal defects along a CFT bath in the \textit{boundary
        perspective}.}
		\label{fig:bndrypenrose}
	}
\end{figure}

Let us return to our braneworld to see how our setup mimics that of
\cite{Almheiri:2019yqk} described above. From the bulk perspective,
we have an AdS$_3$ spacetime with a brane lying along an AdS$_2$ slice (fig.
\ref{fig:bulkpenrose}). We may reproduce the $\AdS_2$ black hole on the brane
by taking Rindler-AdS coordinates in the $\AdS_3$ bulk --- this equips the
$\AdS_3$ bulk with a horizon and `left' and `right' exterior regions. The resulting picture is that of a
Hartle-Hawking state prepared by the Euclidean path integral drawn in
fig.~\ref{fig:HH}. The Rinder $\AdS_3$ coordinates also induce a horizon on the
brane. In fact, the geometry of the brane is itself Rindler-$\AdS$,
\beq
ds^2 = \Lbrn^2\left[-(\rho^2 - 1)d\tau^2 +
  \frac{d\tilde{\rho}^2}{\rho^2 -1} \right] \,,
\eeq
supporting a dilaton profile $\Phi\propto \rho$.
In the brane perspective, we then have a CFT spanning the left and right
asymptotic boundary regions
-- the baths -- and the Rindler-$\AdS_2$ brane, which also
supports a theory of JT gravity. Illustrated in figure \ref{fig:Eternal}, this
is essentially the same setup as in \cite{Almheiri:2019yqk}, up to a
$\mathbb{Z}_2$-quotient across the brane. We may alternatively take the boundary
perspective, wherein the bulk $\AdS_3$ plus brane theory is dual to a $\CFT_2$
plus defect theory. More precisely, the Euclidean path integral preparing the
Hartle-Hawking bulk is equated to a thermal path integral
preparing a TFD state of two copies of a $\CFT_2$ with a defect running through
its middle. We are thus led to the boundary picture drawn in figure
\ref{fig:bndrypenrose}. Taking a $\mathbb{Z}_2$ quotient across the defect,
this, of course, is the alternative description of the setup in
\cite{Almheiri:2019yqk} as a thermal BCFT coupled to quantum mechanics.

With our setup in place, we can then consider subregions of the boundary CFT and
use the RT formula \eqref{eq:island2} to compute the corresponding entanglement
entropies. Analogous to \cite{Almheiri:2019yqk}, we choose `belt' subregions consisting of intervals symmetric about
the defect. The details of the resulting entropy calculation in two dimensions are provided in section
\ref{app:Page2d}. The upshot is that we find a competition between a no-island
phase and an island phase, as
sketched in fig.~\ref{fig:RTPhases_intro}, with the former dominating at early
times and the latter past a Page time. Notice that these phases are analogous
to the no-island and island phases of \cite{Almheiri:2019yqk}, with now the QESs demarked by the intersection between our bulk RT
surface and the brane. Namely, it is clear from the bulk picture shown in the
right panel of 
fig.~\ref{fig:RTPhases_intro} that the island region between these intersection
points belongs to the entanglement wedge of the bath region complementary to the
belt.

In section \ref{app:Page2d}, we also explicitly demonstrate that our bulk RT calculation using
eq.~\eqref{eq:island2} precisely reproduces
the results of \cite{Almheiri:2019yqk}, in the limit where the brane approaches
the would-be $\AdS_3$ boundary by slicing through the bulk at a small brane
angle $\braneAngle$ (that is, the high-tension limit of higher
dimensions). For early times,
we find that the entanglement entropy grows linearly in the no-island phase\footnote{Recall our setup is related to that of \cite{Almheiri:2019yqk}
by a $\mathbb{Z}_2$-orbifold, hence factors of $2$ must be accounted for when
comparing results.} as $4\pi c
t/(3\beta)$ (see eq.~\eqref{eq:hawking2d2}), whereas for late times it is dominated by the island and given by a
constant, $\frac{1}{2\Gbr}\left( \tilde\Phi_0+ \Phi_r\right)$ (see
eq.~\eqref{eq:ofTheirDeathMarkdLove}). Thus, as in \cite{Almheiri:2019yqk}, the
appearance of an island caps off the entropy growth at the expected
course-grained entropy of two copies of the black hole on the brane, rescuing the system from a
potential information paradox (the resulting Page curve is shown in fig.~\ref{fig:Page-curve}). While we find
perfect agreement with \cite{Almheiri:2019yqk} at leading order in
$\braneAngle$, we also find corrections to these results due to the brane
imposing a UV cutoff at finite $\braneAngle$. The result is
$O(\braneAngle^2)$ corrections which, for instance, push the QES
further from the horizon, lower the entropy of the island phase, and lead to a
hastened Page transition. (Note that, in the no-island phase, no such corrections appear as
the bulk RT surface does not intersect the brane.)

It would be straightforward to use our setup to perform the zero-temperature analysis also
covered in \cite{Almheiri:2019yqk} for $d=2$. Here one would instead take
Poincar\'e coordinates which would equip the $\AdS_3$ bulk and $\AdS_2$ brane
with an extremal horizon. We then expect entanglement entropy of large regions
in the bath to
require the inclusion of islands on the gravitating brane. In
particular, intervals stretching from some location in the bath out to infinity
require the inclusion of an island localized around the horizon. (This is to be
contrasted with our findings in $d\ge 3$, where islands are lacking in the
extremal case at small brane angle $\braneAngle$.)

The benefit of our Randall-Sundrum setup is that it allows great flexibility in generalizing the construction of
\cite{Almheiri:2019yqk} to higher dimensions. Indeed, it is straigtforward to
re-interpret figures \ref{fig:bulkpenrose},
\ref{fig:Eternal}, and \ref{fig:bndrypenrose} with a suppressed hyperbolic
$H_{d-2}$ direction. In the following sections,
we shall apply our setup to extend the results mentioned here to higher dimensions.

%% file: images/Gluing2.pdf_tex
\begingroup%
  \makeatletter%
  \providecommand\color[2][]{%
    \errmessage{(Inkscape) Color is used for the text in Inkscape, but the package 'color.sty' is not loaded}%
    \renewcommand\color[2][]{}%
  }%
  \providecommand\transparent[1]{%
    \errmessage{(Inkscape) Transparency is used (non-zero) for the text in Inkscape, but the package 'transparent.sty' is not loaded}%
    \renewcommand\transparent[1]{}%
  }%
  \providecommand\rotatebox[2]{#2}%
  \newcommand*\fsize{\dimexpr\f@size pt\relax}%
  \newcommand*\lineheight[1]{\fontsize{\fsize}{#1\fsize}\selectfont}%
  \ifx\svgwidth\undefined%
    \setlength{\unitlength}{609.4488189bp}%
    \ifx\svgscale\undefined%
      \relax%
    \else%
      \setlength{\unitlength}{\unitlength * \real{\svgscale}}%
    \fi%
  \else%
    \setlength{\unitlength}{\svgwidth}%
  \fi%
  \global\let\svgwidth\undefined%
  \global\let\svgscale\undefined%
  \makeatother%
  \begin{picture}(1,0.31162791)%
    \lineheight{1}%
    \setlength\tabcolsep{0pt}%
    \put(0,0){\includegraphics[width=\unitlength,page=1]{images/Gluing2.pdf}}%
    \put(0.48494745,0.15246383){\color[rgb]{0,0,0}\makebox(0,0)[lt]{\lineheight{1.25}\smash{\begin{tabular}[t]{l}AdS$_{d+1}$\end{tabular}}}}%
    \put(0.81719978,0.15246383){\color[rgb]{0,0,0}\makebox(0,0)[lt]{\lineheight{1.25}\smash{\begin{tabular}[t]{l}AdS$_{d+1}$\end{tabular}}}}%
    \put(0,0){\includegraphics[width=\unitlength,page=2]{images/Gluing2.pdf}}%
    \put(0.11141953,0.14697592){\color[rgb]{0,0,0}\makebox(0,0)[lt]{\lineheight{1.25}\smash{\begin{tabular}[t]{l}AdS$_{d+1}$\end{tabular}}}}%
    \put(0,0){\includegraphics[width=\unitlength,page=3]{images/Gluing2.pdf}}%
    \put(0.2262286,0.25431299){\color[rgb]{0,0,0}\makebox(0,0)[lt]{\lineheight{1.25}\smash{\begin{tabular}[t]{l}AdS$_d$\end{tabular}}}}%
    \put(0,0){\includegraphics[width=\unitlength,page=4]{images/Gluing2.pdf}}%
    \put(0.22559529,0.03217163){\color[rgb]{0,0,0}\makebox(0,0)[lt]{\lineheight{1.25}\smash{\begin{tabular}[t]{l}CFT$_d$\end{tabular}}}}%
    \put(0.01709566,0.02819796){\color[rgb]{0,0,0}\makebox(0,0)[lt]{\lineheight{1.25}\smash{\begin{tabular}[t]{l}a.\end{tabular}}}}%
    \put(0.40515807,0.0281964){\color[rgb]{0,0,0}\makebox(0,0)[lt]{\lineheight{1.25}\smash{\begin{tabular}[t]{l}b.\end{tabular}}}}%
  \end{picture}%
\endgroup%

%% file: images/BraneBH_4.pdf_tex
\begingroup%
  \makeatletter%
  \providecommand\color[2][]{%
    \errmessage{(Inkscape) Color is used for the text in Inkscape, but the package 'color.sty' is not loaded}%
    \renewcommand\color[2][]{}%
  }%
  \providecommand\transparent[1]{%
    \errmessage{(Inkscape) Transparency is used (non-zero) for the text in Inkscape, but the package 'transparent.sty' is not loaded}%
    \renewcommand\transparent[1]{}%
  }%
  \providecommand\rotatebox[2]{#2}%
  \newcommand*\fsize{\dimexpr\f@size pt\relax}%
  \newcommand*\lineheight[1]{\fontsize{\fsize}{#1\fsize}\selectfont}%
  \ifx\svgwidth\undefined%
    \setlength{\unitlength}{850.39370079bp}%
    \ifx\svgscale\undefined%
      \relax%
    \else%
      \setlength{\unitlength}{\unitlength * \real{\svgscale}}%
    \fi%
  \else%
    \setlength{\unitlength}{\svgwidth}%
  \fi%
  \global\let\svgwidth\undefined%
  \global\let\svgscale\undefined%
  \makeatother%
  \begin{picture}(1,0.6)%
    \lineheight{1}%
    \setlength\tabcolsep{0pt}%
    \put(0,0){\includegraphics[width=\unitlength,page=1]{images/BraneBH_4.pdf}}%
    \put(0.15095189,0.39913967){\color[rgb]{0,0,0}\makebox(0,0)[lt]{\lineheight{1.25}\smash{\begin{tabular}[t]{l}defect\end{tabular}}}}%
    \put(0,0){\includegraphics[width=\unitlength,page=2]{images/BraneBH_4.pdf}}%
    \put(0.05701361,0.21160838){\color[rgb]{0,0,0}\makebox(0,0)[lt]{\lineheight{1.25}\smash{\begin{tabular}[t]{l}Rindler Left\end{tabular}}}}%
    \put(0.76818634,0.44630191){\color[rgb]{0,0,0}\makebox(0,0)[lt]{\lineheight{1.25}\smash{\begin{tabular}[t]{l}Rindler Right\end{tabular}}}}%
    \put(0.76458679,0.28726339){\color[rgb]{0,0,0}\makebox(0,0)[lt]{\lineheight{1.25}\smash{\begin{tabular}[t]{l}$\tau=t=0$\end{tabular}}}}%
    \put(0,0){\includegraphics[width=\unitlength,page=3]{images/BraneBH_4.pdf}}%
  \end{picture}%
\endgroup%

%% file: images/HawkingHartle.pdf_tex
\begingroup%
  \makeatletter%
  \providecommand\color[2][]{%
    \errmessage{(Inkscape) Color is used for the text in Inkscape, but the package 'color.sty' is not loaded}%
    \renewcommand\color[2][]{}%
  }%
  \providecommand\transparent[1]{%
    \errmessage{(Inkscape) Transparency is used (non-zero) for the text in Inkscape, but the package 'transparent.sty' is not loaded}%
    \renewcommand\transparent[1]{}%
  }%
  \providecommand\rotatebox[2]{#2}%
  \newcommand*\fsize{\dimexpr\f@size pt\relax}%
  \newcommand*\lineheight[1]{\fontsize{\fsize}{#1\fsize}\selectfont}%
  \ifx\svgwidth\undefined%
    \setlength{\unitlength}{617.95275591bp}%
    \ifx\svgscale\undefined%
      \relax%
    \else%
      \setlength{\unitlength}{\unitlength * \real{\svgscale}}%
    \fi%
  \else%
    \setlength{\unitlength}{\svgwidth}%
  \fi%
  \global\let\svgwidth\undefined%
  \global\let\svgscale\undefined%
  \makeatother%
  \begin{picture}(1,0.75688073)%
    \lineheight{1}%
    \setlength\tabcolsep{0pt}%
    \put(0,0){\includegraphics[width=\unitlength,page=1]{images/HawkingHartle.pdf}}%
    \put(0.42897062,0.01350375){\color[rgb]{0,0,0}\makebox(0,0)[lt]{\lineheight{1.25}\smash{\begin{tabular}[t]{l}$\frac{\beta}{2}=\pi R$\end{tabular}}}}%
    \put(0,0){\includegraphics[width=\unitlength,page=2]{images/HawkingHartle.pdf}}%
    \put(0.01224146,0.54431984){\color[rgb]{0,0,0}\makebox(0,0)[lt]{\lineheight{1.25}\smash{\begin{tabular}[t]{l}defect\end{tabular}}}}%
    \put(0,0){\includegraphics[width=\unitlength,page=3]{images/HawkingHartle.pdf}}%
  \end{picture}%
\endgroup%

%% file: images/Penrose_braneperspective.pdf_tex
\begingroup%
  \makeatletter%
  \providecommand\color[2][]{%
    \errmessage{(Inkscape) Color is used for the text in Inkscape, but the package 'color.sty' is not loaded}%
    \renewcommand\color[2][]{}%
  }%
  \providecommand\transparent[1]{%
    \errmessage{(Inkscape) Transparency is used (non-zero) for the text in Inkscape, but the package 'transparent.sty' is not loaded}%
    \renewcommand\transparent[1]{}%
  }%
  \providecommand\rotatebox[2]{#2}%
  \newcommand*\fsize{\dimexpr\f@size pt\relax}%
  \newcommand*\lineheight[1]{\fontsize{\fsize}{#1\fsize}\selectfont}%
  \ifx\svgwidth\undefined%
    \setlength{\unitlength}{850.39370079bp}%
    \ifx\svgscale\undefined%
      \relax%
    \else%
      \setlength{\unitlength}{\unitlength * \real{\svgscale}}%
    \fi%
  \else%
    \setlength{\unitlength}{\svgwidth}%
  \fi%
  \global\let\svgwidth\undefined%
  \global\let\svgscale\undefined%
  \makeatother%
  \begin{picture}(1,0.51333333)%
    \lineheight{1}%
    \setlength\tabcolsep{0pt}%
    \put(0.83864797,0.0580426){\color[rgb]{0,0,0}\makebox(0,0)[lt]{\lineheight{1.25}\smash{\begin{tabular}[t]{l}CFT$_\mt{R}$\end{tabular}}}}%
    \put(0.06629712,0.05787043){\color[rgb]{0,0,0}\makebox(0,0)[lt]{\lineheight{1.25}\smash{\begin{tabular}[t]{l}CFT$_\mt{L}$\end{tabular}}}}%
    \put(0,0){\includegraphics[width=\unitlength,page=1]{images/Penrose_braneperspective.pdf}}%
    \put(0.26859103,0.27281995){\color[rgb]{0,0,0}\makebox(0,0)[lt]{\lineheight{1.25}\smash{\begin{tabular}[t]{l}$\tau=t=0$\end{tabular}}}}%
    \put(0.43330414,0.4534251){\color[rgb]{0,0,0}\makebox(0,0)[lt]{\lineheight{1.25}\smash{\begin{tabular}[t]{l}horizon\end{tabular}}}}%
    \put(0,0){\includegraphics[width=\unitlength,page=2]{images/Penrose_braneperspective.pdf}}%
    \put(0.31041003,0.03626119){\color[rgb]{0,0,0}\makebox(0,0)[lt]{\lineheight{1.25}\smash{\begin{tabular}[t]{l}defect\end{tabular}}}}%
    \put(0,0){\includegraphics[width=\unitlength,page=3]{images/Penrose_braneperspective.pdf}}%
  \end{picture}%
\endgroup%

%% file: images/Penrose_boundaryperspective.pdf_tex
\begingroup%
  \makeatletter%
  \providecommand\color[2][]{%
    \errmessage{(Inkscape) Color is used for the text in Inkscape, but the package 'color.sty' is not loaded}%
    \renewcommand\color[2][]{}%
  }%
  \providecommand\transparent[1]{%
    \errmessage{(Inkscape) Transparency is used (non-zero) for the text in Inkscape, but the package 'transparent.sty' is not loaded}%
    \renewcommand\transparent[1]{}%
  }%
  \providecommand\rotatebox[2]{#2}%
  \newcommand*\fsize{\dimexpr\f@size pt\relax}%
  \newcommand*\lineheight[1]{\fontsize{\fsize}{#1\fsize}\selectfont}%
  \ifx\svgwidth\undefined%
    \setlength{\unitlength}{771.02362205bp}%
    \ifx\svgscale\undefined%
      \relax%
    \else%
      \setlength{\unitlength}{\unitlength * \real{\svgscale}}%
    \fi%
  \else%
    \setlength{\unitlength}{\svgwidth}%
  \fi%
  \global\let\svgwidth\undefined%
  \global\let\svgscale\undefined%
  \makeatother%
  \begin{picture}(1,0.49264706)%
    \lineheight{1}%
    \setlength\tabcolsep{0pt}%
    \put(0,0){\includegraphics[width=\unitlength,page=1]{images/Penrose_boundaryperspective.pdf}}%
    \put(0.44431165,0.34025389){\color[rgb]{0,0,0}\makebox(0,0)[lt]{\lineheight{1.25}\smash{\begin{tabular}[t]{l}defect\end{tabular}}}}%
    \put(0,0){\includegraphics[width=\unitlength,page=2]{images/Penrose_boundaryperspective.pdf}}%
    \put(0.10541946,0.26344482){\color[rgb]{0,0,0}\makebox(0,0)[lt]{\lineheight{1.25}\smash{\begin{tabular}[t]{l}CFT$_\mt{L}$\end{tabular}}}}%
    \put(0.79448711,0.26344546){\color[rgb]{0,0,0}\makebox(0,0)[lt]{\lineheight{1.25}\smash{\begin{tabular}[t]{l}CFT$_\mt{R}$\end{tabular}}}}%
    \put(0,0){\includegraphics[width=\unitlength,page=3]{images/Penrose_boundaryperspective.pdf}}%
  \end{picture}%
\endgroup%

%% file: sections/03_non_extremal_bh.tex
%
In this section, we discuss how islands arise in the presence of certain topological, non-extremal black holes in higher-dimensional brane-world models. Topological black holes are characterized as having nontrivial horizon topology, and we will be interested in the case of neutral black holes with a hyperbolic horizon \cite{Mann:1996gj, Birmingham:1998nr}. The general metric is given by
\begin{equation}\label{eq:topo_bh_metric}
d s^{2} = - f(r)\,\frac{L^2}{R^2 }
\, d t^{2} + \frac{d r^{2}}{f(r)} + r^{2}\, d H^{2}_{d-1}\,,
\end{equation}
with the blackening factor
\begin{align}\label{BFactor}
f(r) = \frac{r^2}{\LAdS^2}-1 -
\frac{\omega^{d-2}}{r^{d-2}}\,.
\end{align}
Here, $\LAdS$ denotes the AdS curvature scale and $d H^{2}_{d-1}$ denotes the line element on a ($d-1$)-dimensional hyperbolic plane with unit curvature. This bulk geometry \reef{eq:topo_bh_metric} is dual to a thermofield double (TFD) state for two copies of the boundary CFT \cite{Mal01}, where each resides on a spatial geometry
$H_{d-1}$. After an appropriate Weyl rescaling, the boundary metric for each CFT reads 
\beq
ds^2_\mt{CFT}=-dt^2 + R^2 \, d H^{2}_{d-1}\,,
\label{bmetric}
\eeq
and hence the scale $R$ (introduced in eq.~\reef{eq:topo_bh_metric}) corresponds to the curvature scale of the spatial geometry. The full boundary geometry is then two copies of $\mathbb R \times H_{d-1}$, where the $\mathbb R$ corresponds to the time direction in each of the CFTs.
 
Turning back to eq.~\reef{eq:topo_bh_metric}, the relation between the position of the horizon $r_h$, the black hole mass $M$, and the `mass' parameter $\omega$ is \cite{Myers:1999psa,Emparan:1999pm,Chapman:2016hwi}
\begin{align}
\omega^{d-2}= r_h^{d-2}\left(\frac{r_h^2}{\LAdS^2}-1\right) = \frac{16 \pi \, G_N}{(d-1) \, \vol_{H_{d-1}}}\,\frac R {\LAdS} M.
\label{horiz}
\end{align}
Here and in the following, we use $\vol_{H_{d-1}}$ to denote the dimensionless volume of the spatial boundary geometry, \ie the volume measured by the metric $dH^2_{d-1}$. Of course, this volume is infinite and we must introduce an infrared regulator -- see below.

In the following, we will consider the special case of a topological black hole with vanishing mass $M = \omega = 0$. Note that despite the fact that $\omega=0$, we still find a horizon at $r_h=L$ from eq.~\reef{BFactor}. In fact, the bulk geometry corresponds to the AdS vacuum (as expected for $M=0$), but we are describing this geometry with the AdS-Rindler coordinates where the metric resembles that of black hole \cite{Casini:2011kv}. In this case, it is straightforward to evaluate the entropy and the temperature of the black hole 
\begin{align}\label{eq:ST}
S =  \frac{\vol_{H_{d-1}} L^{d-1} }{4 G_N}\,,\qquad
T = \frac{1}{2\pi R}\,.
\end{align}

In terms of the dual CFT, we are considering a pure state (\ie the vacuum) in the conformal frame where the boundary geometry corresponds to $\mathbb R \times S^{d-1}$. However, with an appropriate conformal transformation, we produce the TFD state on two copies of $\mathbb R \times H_{d-1}$ with temperature $T=1/(2\pi R)$ \cite{Casini:2011kv}. The entropy in eq.~\reef{eq:ST} corresponds to the entanglement entropy between the two copies of the CFT -- and alternatively, can be interpreted as the entanglement entropy between two halves of the sphere in the original conformal frame. From the point of view of the CFT, masslessness of the black hole corresponds to a fine tuning of the temperature to $T = \frac{1}{2\pi R}$. 

Following the brane world construction outlined in the previous section, we locate a codimension-one defect at the center of each CFT. By the holographic dictionary, this corresponds to a brane which cuts through the bulk and orthogonally intersects the horizon -- see figure \ref{fig:bulkpenrose}. Since with $\omega=0$ , the bulk geometry is just the AdS vacuum, our previous discussion of the brane geometry (above and in \cite{Chen:2020uac}) is still applicable. Hence, the brane position in the bulk is determined precisely as described above in terms of the brane tension $T_o$. In fact, this bulk geometry provides a higher dimensional generalization of the construction discussed in section \ref{twod}, and we will see that the brane  inherits a black hole metric with temperature $T = 1/(2\pi R)$, from the AdS-Rindler coordinates in the bulk. 

Our aim will be to use eq.~\eqref{eq:island2} to investigate the appearance of quantum extremal islands, from the brane perspective, where (two copies of) the boundary CFT are supported in this black hole geometry on the brane. Further, we will compute the entanglement entropy associated to symmetric regions $\bdyReg$ on each side of the defect as a function of time -- see figure \ref{fig:RTPhases_intro}. The regions $\bdyReg$ of interest consist of those points on a CFT timeslice which are further than a distance $\chi=\chi_\Sigma$  away form the defect.\footnote{The coordinate $\chi$ is introduced in eq.~\reef{eq:metric_bdry_slicing} below. Of course, since the global state which we are considering is pure, we could equivalently discuss the entanglement entropy of the belt regions $-\chi_\Sigma <\chi<\chi_\Sigma$ in both CFTs, including the conformal defects.} The entanglement entropy is evaluated using the holographic prescription of the bulk perspective and as described in the introduction, the corresponding RT surfaces  can be in one of two phases. Either they connect through the horizon, which we will call the no-island phase, or they connect through the brane, which we will call the island phase. The reason for those names is apparent from the $d$-dimensional effective gravity on the brane, \ie the region bounded by the intersection of the RT surface and the brane is a quantum entremal island, which now contributes to the entropy of $\bdyReg$. This also implies that from the $(d+1)$-dimensional bulk perspective, the appearance of islands is simply explained as a standard phase transition of an RT surface. We will see in the remainder of this section that at early times, the RT surfaces starts out in the no-island phase, \ie connects throught the horizon. As is well known \cite{Hartman:2013qma}, the volume of the corresponding surfaces grows linearly with time. At some point its volume will have grown so large, that the RT surface in the island phase has smaller area and gives the correct entanglement entropy. 

The calculation of the time-dependence of the area of RT surfaces will proceed in two steps: In sections \ref{sec:island_phase_non_extremal} to \ref{sec:no_island_phase_2_non_extremal}, we will derive expressions for the area of three special cases of extremal surfaces. The first one will be RT surfaces in the island phase anchored at Rindler time $\tau_\Sigma = 0$. The second and third special cases will be RT surfaces $\RT$ in the no-island phase which either end on entangling surfaces $\partial \bdyReg$ at $\chi=\pm \chi_\Sigma$  
and $\tau_\Sigma = 0$, or end on entangling surfaces located at the defect ($\chi_\Sigma = 0$) and arbitrary $\tau_\Sigma$. While these special cases naively might seem not to contain enough information to completely reconstruct the time-evolution of the entanglement entropy, we will argue in section \ref{sec:time_evolution_non_extremal} that the time-evolution of any symmetric RT surface in the no-island phase can always be reduced to one of those three cases.

We remind the reader that as described in section \ref{twod}, we are considering eternal black holes which do not evaporate. Nonetheless, from the effective brane point of view, the black hole on the brane and the fields on the asymptotic boundary are in contact, and can therefore continuously exchange radiation. If island are not accounted for appropriately, this leads to information loss \cite{Almheiri:2019hni}. In section \ref{sec:islands_for_non_extremal} we will argue, using results obtained below, that also in higher dimensions the presence of islands makes the entanglement dynamics of the joint system of black holes and radiation compatible with unitarity.


\subsection{Geometry on the brane}
\label{sec:geometry_on_brane_non_extremal}
To set the stage for the following calculations, we will start by discussing the bulk and brane geometry. As noted above, the bulk metric is described by AdS-Rindler coordinates
\begin{align}
\label{eq:metric_rindler_bulk}
ds^2 = \LAdS^2 \left( - (r^2-1) d\tau^2 + \frac{dr^2}{r^2-1}  + r^2\, dH_{d-1}^2\right)\,,
\end{align}
which is obtained from eq.~\eqref{eq:topo_bh_metric} by taking the massless limit $\omega,M \to 0$ and rescaling the coordinates $t \to R\, \tau$ and $r \to L\, r$, such that the coordinates in eq.~\eqref{eq:metric_rindler_bulk} are dimensionless. Although the underlying geometry is simply the AdS vacuum, the metric \reef{eq:metric_rindler_bulk} resembles a black hole metric with  horizons at $r=\pm1$ and an apparent singularity at $r=0$.   We can also extend the spacetime at a fixed time-slice through the bifurcation surface and arrive at a second Rindler wedge. The bulk spacetime thus has two asymptotic regions, located at $r \to \infty$, each of which hosts one copy of the boundary CFT on the $\mathbb R \times H_{d-1}$ geometry. As noted above (in terms of the dimensionful coordinates), the corresponding TFD state has a (dimensionful) temperature $T = 1/(2\pi R)$, which is tuned in relation to the curvature scale $R$ of the hyperbolic geometries \eqref{bmetric}. Lastly, note that since the Rindler wedges are simply a reparametrization of pure AdS, it is clear that the singularity at $r=0$ is only a coordinate singularity.\footnote{This is in contrast to the general metric \reef{eq:topo_bh_metric} where $r\to0$ does yields a curvature singularity.} In fact, we can extend the coordinates smoothly through the interior to negative $r$ where we can exit the region behind the (inner) horizon at $r=-1$ and enter a new set of Rindler wedges. 

For each CFT, we introduce a codimension-one conformal defect (with zero extrinsic curvature) at the center of the hyperbolic spatial geometry. It is convenient to choose slicing coordinates for the hyperbolic boundaries, such that
\begin{align}
	\label{eq:metric_bdry_slicing}
  dH_{d-1}^2 = d\chi^2 + \cosh^2 \chi \,dH^2_{d-2}.  
\end{align}
In these coordinates, the location of the conformal defect is $\chi = 0$. 

From the bulk perspective, the CFT defects are dual to a co-dimension one brane, which spans a slice of constant extrinsic curvature of the bulk spacetime and intersects the asymptotic boundary at the location of the CFT defect. In order to describe its trajectory, it is convenient to write the bulk metric in terms of the slicing coordinates in eq.~\eqref{metric33}.
The brane is located at constant $\theta = \theta_\mt{B}$, which is determined by the tension $T_o$ through eq.~\eqref{curve2} with
\begin{align}
\label{eq:brane_curvature_scale}
\ell_\mt{B} = \frac{L }{\sin\theta_\mt{B}} \,. 
\end{align}
The trajectory of a hypersurfaces of constant $\theta_\mt{B}$ in the bulk spacetime is then given by
\begin{align}
	\label{eq:brane_trajectory}
  r^2 \sinh^2 \chi
  =&\cot^2\theta_\mt{B}
	=\left(\frac{\ell_\mt{B}}{L}\right)^2 - 1
  \,.
\end{align}
As noted in \cite{Chen:2020uac}, this means that a brane with positive tension (\ie $T_o\ge0$) creates additional geometry by its backreaction. Of course, the backreaction of a negative-tension brane would remove geometry. However, let us add that there is no (nearly) massless graviton induced on a negative-tension brane\footnote{We thank Raman Sundrum for explaining this point to us.} and therefore we will only consider positive tensions in the following, \ie $0\le\theta_\mt{B}\le\frac{\pi}{2}$.

For such a (positive-tension) brane, the bulk geometry to one side of the brane can be described by eq.~\eqref{eq:metric_rindler_bulk}, with $r \sinh \chi\leq\cot\theta_\mt{B}$, while the geometry to the other side of the brane is given by the same metric with $r \sinh \chi\geq - \cot\theta_\mt{B}$. We can therefore treat either side of the brane as an AdS-Rindler geometry which is cut off by the brane.

Using eq.~\eqref{eq:brane_trajectory}, we can determine the induced metric on the brane. After a short calculation, one finds
\begin{align}
\label{eq:metric_rindler_brane}
ds^2 = \ell_\mt{B}^2 \left( - (\rho^2-1)\, d\tau^2 + \frac{d\rho^2}{\rho^2-1}  + \rho^2\, dH_{d-2}^2\right)\,,
\end{align}
where we have changed the radial coordinate with
\begin{align}
  \ell_\mt{B}^2(\rho^2-1) = L^2 (r^2 - 1) \,.
  \label{eq:donTTellMe}
\end{align}
This brane metric again takes the form of an AdS-Rindler metric, c.f. eq.~\eqref{eq:metric_rindler_bulk}. Further, this demonstrates that the Rindler horizon in the bulk (at $r=1$) induces a Rindler horizon on the brane (at $\rho=1$), as one would expect from the bulk perspective.\footnote{However, it is interesting to note that $r=0$ corresponds to $\rho=\cos\theta_\mt{B}=1-(L/\ell_\mt{B})^2$, and hence one cannot reach $\rho=0$ in the $r$-coordinate system (unless $\theta_\mt{B}=\pi/2$).} From the boundary perspective, this behavior is readily explained by the fact that the conformal defect is in thermal equilibrium with the surrounding CFT. In the effective Randall-Sundrum description of the brane perspective, this behaviour arises because the region of dynamical gravity is coupled to the bath CFT along an accelerated trajectory,  so that the temperature felt by the accelerated boundary agrees with the temperature of the CFT, \eg see \cite{Deser:1997ri,Jacobson:1997ux,Parikh:2012kg}. As already mentioned, this setup generalizes the two-dimensional framework presented in \cite{Almheiri:2019yqk} to higher dimensions. 

All calculations below will be done for the case of positive tension branes. However, when it comes to interpretation, we will be particularly interested in the case where $1 \gg \theta_\mt{B} \simeq \frac{L}{\ell_\mt{B}}$, for which the brane theory is well described as Einstein gravity coupled to two copies of the boundary CFT (with a high cutoff). The reason is that in this limit, we can interpret the intersection of the brane and the RT surface as bounding an island in this effective gravitational theory.


\subsection{Island phase at $\tau_\Sigma = 0$}
\label{sec:island_phase_non_extremal}
\begin{figure}[t]
	\def\svgwidth{0.9\linewidth}
	\centering{
		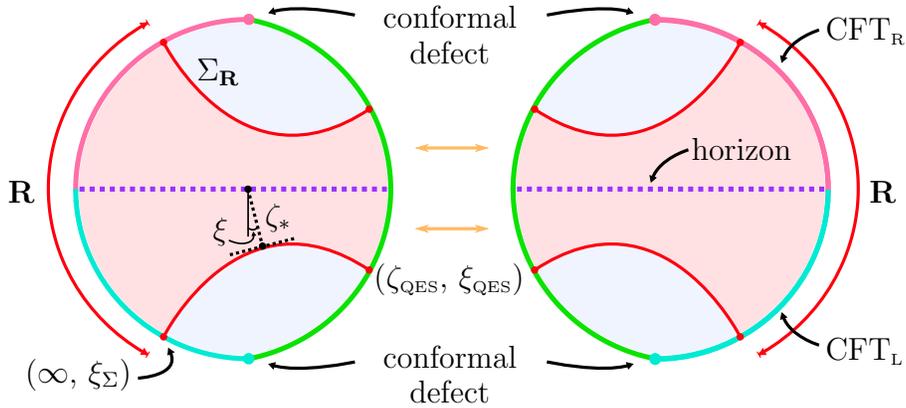
		\caption{This figure shows the RT surface and various quantities defined in the text for the RT surface in the connected phase. The entangling region $\bdyReg$ in the boundary is composed of the two regions $|\chi| >\chi_\Sigma$ (where $\tan\xi_\Sigma=\sinh\chi_\Sigma$) in both the left and right CFTs. Note that the right (left) CFT occupies the region on the asymptotic boundary marked in pink (aqua). The conformal defects (\ie $\chi=0$ or $\xi=0$ and $\pi$) are positioned where the brane (green) reaches these boundary regions. }
		\label{fig:connected_phase_non_extremal}
	}
\end{figure}

We will start our analysis by calculating the area of the RT surface for an entangling surface lying in the $\tau = \tau_\Sigma = 0$ plane and crossing the Planck brane. 
In other words, the RT surface is in the connected phase -- see figure \ref{fig:connected_phase_non_extremal}. We are interested in the entanglement entropy of $\bdyReg$ comprised of the combined regions $\chi >\chi_\Sigma$ and $\chi <-\chi_\Sigma$ in both the left and right CFTs. Hence the entangling surfaces of interest have two components (in each CFT) sitting a constant distance away from the defect at $\chi=\pm\chi_\Sigma$. We note that the induced metric on the latter surfaces is proportional to $\cosh^{d-2} \!\chi_{\Sigma}$. 

In two dimensions, the analysis of the RT surfaces is simplified because the metric \eqref{eq:metric_rindler_bulk} has a shift symmetry $\chi \to \chi + \text{const}$, but the latter is absent in higher dimensions. However, we can find a similar simplification by going to a different coordinate system defined via \cite{Chen:2020uac, Krtous:2014pva}
\begin{align}
\label{eq:new_coords_non_extremal}
(1 + \zeta^2) = r^2\, \cosh^2 \chi\,, && \tan \xi = \frac{r}{\sqrt{r^2-1}}\, \sinh \chi\,,	
\end{align}
such that the horizon is located at $\xi = \pm \frac \pi 2$. By time-translation invariance, we know that the RT surface lies on a constant Rindler time slice and hence we consider the metric on the $\tau = 0$ slice in the new coordinates,\footnote{Note that the full metric takes the form $ds^2 = \LAdS^2\,\zeta^2\cos^2\!\xi\,d\tau^2 + ds_\mt{E}^2$, and hence the shift symmetry does not extend to the full spacetime metric. \label{carrot3}} which reads
\begin{align}
\label{eq:new_coords_non_extremal_metric}
ds_\mt{E}^2 = \LAdS^2 \left( \frac{d \zeta^2}{ 1 + \zeta^2} + \zeta^2 d \xi^2 + (1 + \zeta^2) dH_{d-2}^2\right)\,. 
\end{align}
Hence the geometry of this spatial slice (or any constant $\tau$ slice) is invariant under $\xi \to \xi + \text{const}$, which will simplify the following.  

Making the ansatz $\zeta = \zeta(\xi)$ for the profile of the RT surface, the induced metric on these surfaces takes the form
\begin{align}
ds_\text{ind}^2 = \LAdS^2 \left[ \left(\left(\frac{\partial \zeta}{\partial \xi}\right)^2 + \zeta^2 (1 + \zeta^2)\right) \frac{d \xi^2}{(1 + \zeta^2)} + (1 + \zeta^2) dH_{d-2}^2\right]\,,
\end{align}
with metric determinant
\begin{align}
\det(\gamma) = \LAdS^{2(d-1)} (1 + \zeta^2)^{d-3} \left(\left(\frac{\partial \zeta}{\partial \xi}\right)^2 + \zeta^2 (1 + \zeta^2) \right)\, .
\end{align}
To obtain the correct RT surface, we now need to extremize the area functional
\begin{align}
\label{carrot1}
 \area(\RT) = \int_{\RT} \sqrt{\det(\gamma)}\,,
\end{align}
subject to the correct boundary conditions. Here, a few observations are in order. The boundary condition is determined by the RT surface ending at the entangling surface on both sides of the defect. Alternatively, since our setup is reflection-symmetric across the brane, we can also consider a family of bulk extremal surfaces which end on the brane and vary with respect to the point of intersection of the brane and the RT surface \cite{Chen:2020uac}. Even in higher dimensions, this variation takes a fairly simple form (see eq.~\eqref{eq:var_RT_area_varrho} below), since extremizing the RT surface can be cast as an effectively two-dimensional problem with metric
\begin{align}
\label{eq:effective_2d_nonextremal_metric}
ds^2_{2D} = \LAdS^{2(d-1)} \vol_{H_{d-2}}  (1 + \zeta^2)^{d-2} \left( \frac{d\zeta^2 }{1 + \zeta^2} + \zeta^2 d\xi^2 \right) .
\end{align}

\begin{figure}[t]
	\def\svgwidth{0.9\linewidth}
	\centering{
		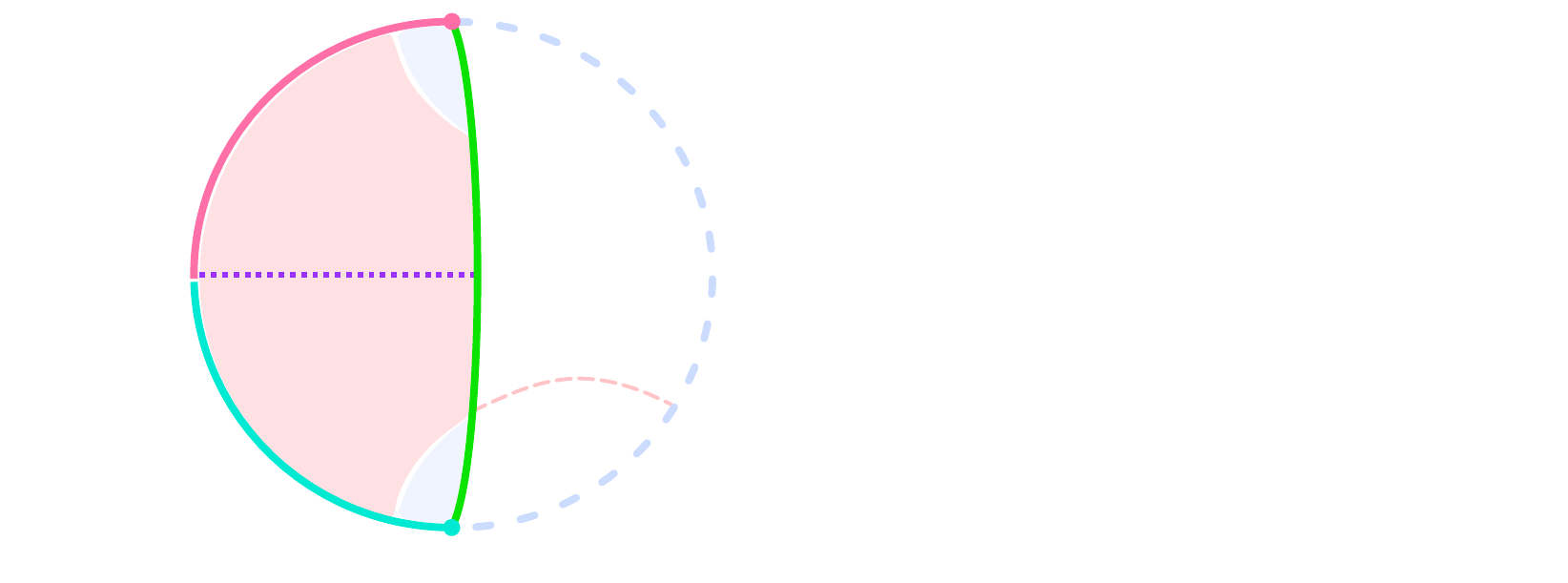
	}
	\caption{This figure shows how RT surfaces can intersect the brane before reaching the turnaround point $\zeta_*$, with relatively small brane tension $T_o$, \ie $\theta_\mt{B} \sim O(1)$, and positive DGP coupling.
	}
	\label{fig:island_phase_other_sign}
\end{figure}

Note that the area functional does not explicitly depend on $\xi$. Rather, $\xi$ plays the role of an angular coordinate and its associated Hamiltonian is conserved. This allows us to turn the second order equation which determines extremal surfaces into a first order expression,
\begin{align}
\label{eq:first_order_RT_surface}
\frac{d \zeta}{d \xi} = \pm \sqrt{ \zeta^2 (1 + \zeta^2) \left( \frac{\zeta^2 ( 1 + \zeta^2)^{d-2}}{\zeta_*^2 ( 1 + \zeta_*^2)^{d-2}}-1\right) },
\end{align}
where we have introduced $\zeta_*$ which is the turn-around point for $\zeta$ as a function of $\xi$ -- see figure \ref{fig:connected_phase_non_extremal}. The sign depends on whether $\zeta$ is going towards ($+$) or away ($-$) from the boundary as $\xi$ increases. In the latter case, where the RT surface does not turn around before it intersects the brane we have to think of $\zeta_*$ as a coordinate of vacuum AdS extended past the brane, as shown in figure \ref{fig:island_phase_other_sign}. More generallly, the sign starts out negative and generally flips after $\zeta = \zeta_*$ has been reached.

The area functional for the RT surfaces satisfying eq.~\reef{eq:first_order_RT_surface} then becomes
\begin{align}
\label{eq:RT_area}
\area(\RT) = 4 \LAdS^{d-1} \vol_{H_{d-2}} \left( \int^\infty_{\zeta_*} \pm  \int_{\zeta_\QES}^{\zeta_*}\right) d\zeta \frac{\zeta (1 + \zeta^2)^{d-\frac 5 2}}{\sqrt{\zeta^2 (1 + \zeta^2)^{d-2} - \zeta^2_* (1 + \zeta_*^2)^{d-2}}},
\end{align}
where here and below, we use the subscript $\QES$ to mark coordinates of the intersection between RT surface and brane, which corresponds to a quantum extremal surface in the brane theory. The upper limit of integration indicated as $\infty$ must be regulated, since the area of the RT surface is infinite. The sign here is the same sign as in eq.~\eqref{eq:first_order_RT_surface}. We have also included a factor of four, since there is one RT surface to each side of the defect and considering both CFTs, we need to multiply the result by another factor of two.

Eq.~\reef{eq:first_order_RT_surface} yields a family of RT surfaces (parameterized by $\zeta_*$) which are locally extremal in the bulk away from brane. However, fully extremizing the area functional \reef{carrot1} requires that we also extremize over the possible locations where these candidate surfaces intersect the brane.
That is, we consider the extremization condition of the RT surfaces' area (plus possibly the area of the QES, should there be extra DGP gravity) with respect to the position of the intersection $\RTbrn$,
\begin{align}
	0 =&  \frac{\partial }{\partial \rho_{\QES}}\left( \frac{\area(\RT)}{4 G_\bulk} + \frac{\area(\RTbrn)}{4 G_\brane} \right)\,, \label{eq:full_variation_RT_surface_intersection}
\end{align}
where the two contributions reflect the two contributions in eq.~\eqref{eq:island2}. Here, $\rho_{\QES}$ denotes the location of $\RTbrn$ in coordinates along the brane in eq.~\eqref{eq:metric_rindler_brane}. 

As described in \cite{Chen:2020uac}, this extremization leads to a boundary condition restricting the angle at which the RT surface meets the brane. Normally, this would be a difficult problem in higher dimensions. However, here we are leveraging the hyperbolic symmetry along the transverse directions, which reduces the present case to a two-dimensional problem. That is, we need only extremize a one-dimensional profile $\zeta(\xi)$ of the RT surface in the effective two-dimensional geometry given by eq.~\eqref{eq:effective_2d_nonextremal_metric}. Assuming that we consider an extremal bulk surface which is anchored at the asymptotic boundary, the variation of the surface's area with respect to its intersection point with the brane is given by
\begin{align}
  \delta_{\RTbrn} \area(\RT) = & h_{ij}\, T^i X^j|_{\text{end-point}}, \label{eq:variation_RT_surface_intersection}
\end{align}
where $h_{ij}$ is the two-dimensional metric \eqref{eq:effective_2d_nonextremal_metric} and $T^i$ is a normalized (w.r.t.~$h_{ij}$) tangent
vector to the RT surface, which can be obtained from
eq.~\eqref{eq:brane_trajectory}. The vector $X^i$ determines the variation along the brane.

In the absence of a DGP gravity term in the action, this variation must vanish for $X^j$ along the brane; hence we have a boundary condition which sets the RT surface perpendicular to the brane. More
generally, we must balance the above variation against the variation of the entropy contribution intrinsic to the brane, as can be seen from eq.~\eqref{eq:full_variation_RT_surface_intersection}.

The first contribution to eq.~\eqref{eq:full_variation_RT_surface_intersection} is then calculated using eq.~\eqref{eq:variation_RT_surface_intersection} and yields
\begin{align}
\label{eq:var_RT_area_varrho}
\begin{split}
\partial_\rho \area(\RT) = 4 \LAdS^{d-1} \vol_{H_{d-2}} \frac{\zeta_* ( 1 + \zeta_*^2)^\frac{d-2}2}{\zeta^2 \sin \theta_\mt{B}} \left( \sqrt{\frac{\zeta^2 + 1}{\tan^2 \theta_\mt{B}\, \zeta^2 - 1}} \pm \sqrt{\frac{\zeta^2 ( 1 + \zeta^2)^{d-2}}{\zeta_*^2 ( 1 + \zeta_*^2)^{d-2}} - 1}  \right),
\end{split}
\end{align}
which is evaluated at $\zeta = \zeta_\QES$. Here we have used the brane angle $\theta_\mt{B}$ defined in eq.~\eqref{eq:brane_curvature_scale}.

If the brane DGP coupling is turned on, the variation of the area also obtains a contribution from the second term in eq.~\eqref{eq:full_variation_RT_surface_intersection},
\begin{align}
\label{eq:var_brane_area_varrho}
\partial_\rho \area(\RTbrn) = 2 \LAdS^{d-2} \vol_{H_{d-2}} \frac{\zeta_* ( 1 + \zeta_*^2)^\frac{d-2}2}{\zeta^2\sqrt{ \zeta^2 \sin^2 \theta_\mt{B} - \cos^2 \theta_\mt{B}}} (d-2) (\zeta^2 + 1)^\frac{d-2}{2}.
\end{align}
Substituting eqs.~\eqref{eq:var_RT_area_varrho} and \eqref{eq:var_brane_area_varrho} into eq.~\eqref{eq:full_variation_RT_surface_intersection}, we obtain the following relation between the QES position $\zeta_\QES$ and the deepest point $\zeta_*$ reached by the RT surface:
\begin{align}
	\begin{split}
	&\zeta_* (1+\zeta_*^2)^{\frac{d-2}{2}} = (\zeta_\QES^2 + 1)^{\frac{d-3}{2}} \sqrt{\zeta_\QES^2 \sin^2 \theta_\mt{B} - \cos^2 \theta_\mt{B}} 
	\\
	&\times\left[ \DGPRatio \cos (\theta_\mt{B}) \sqrt{1+\zeta_\QES^2} 	+ \sqrt{1+\zeta_\QES^2-\DGPRatio^2\left( \zeta_\QES^2 \sin^2 \theta_\mt{B} - \cos^2 \theta_\mt{B} \right) }
	\right]\,,
	\end{split}
\label{eq:condition_1}
\end{align}
where $\DGPRatio$ was defined in eq.~\reef{Newton3}.

A final relation associating $\zeta_\QES$ and the belt width $\xi_\Sigma$ comes from integrating eq.~\eqref{eq:first_order_RT_surface} from the boundary to the brane,
\begin{align}
	\xi_\QES
	=& \xi_\Sigma +
	\int_{\zeta_*}^\infty d\zeta\; \left|\frac{d\zeta}{d\xi}\right|^{-1}
	\pm \int_{\zeta_*}^{\zeta_\QES} d\zeta\; \left|\frac{d\zeta}{d\xi}\right|^{-1}\,.
	\label{eq:condition_2}
\end{align}
After using eq.~\eqref{eq:condition_1}, this can then be rewritten as a relation between the location of the entangling surface $\xi_\Sigma$ and the QES $\zeta_\QES$ only, if we further use eq.~\eqref{eq:brane_trajectory} together with eq.~\eqref{eq:new_coords_non_extremal} to find the brane trajectory in $\zeta, \xi$ coordinates and determine the relationship between $\xi$ and $\zeta$ on the brane
\begin{align}
\label{eq:brane_trajectory_2}
\zeta^2 \, \sin^2 \xi = \cot^2 \theta_\mt{B}\,.
\end{align}
In section \ref{sec:numerics}, we will use eqs.~\eqref{eq:RT_area}, \eqref{eq:condition_1} and \eqref{eq:condition_2} to produce the late-time part of the Page curve for a topological black hole coupled to a bath in higher dimensions.



\subsection{No-island phase for $\tau_\Sigma = 0$}\label{sec:no_island_phase_1_non_extremal}
We can use the result of the previous subsection to obtain a solution for the no-island phase. The first order equation \eqref{eq:first_order_RT_surface} (where we choose the minus sign) again determines the shape of extremal surface. By symmetry, we know that $\zeta_*$ must lie on the bifurcate horizon and is thus determined by solving
\begin{align}
\label{eq:zeta_star_no_island}
\int_\infty^{\zeta_*} \left| \frac{d\zeta}{d\xi}\right|^{-1} d\zeta = - \frac \pi 2 - \xi_\Sigma.
\end{align}
Here we have implicitly chosen to perform the calculation in the asymptotic CFT which sits at negative $\xi$, \ie to a particular side of the brane. By symmetry the calculation on the other side of the brane yields the same result.
The total area of the two RT surfaces which connect both CFTs through the horizon is then given by 
\begin{align}
\label{eq:RT_area_no_island}
\area(\RT) = 4 \LAdS^{d-1} \vol_{H_{d-2}}  \int^\infty_{\zeta_*}  d\zeta \frac{\zeta (1 + \zeta^2)^{d-\frac 5 2}}{\sqrt{\zeta^2 (1 + \zeta^2)^{d-2} - \zeta^2_* (1 + \zeta_*^2)^{d-2}}},
\end{align}
with $\zeta_*$ given by eq.~\eqref{eq:zeta_star_no_island}. 
In the case of small brane angle $\theta_\mt{B}$ this phase always dominates at early times. The reason is that the the RT surface in the competing phase, \ie the phase where the RT surface crosses the brane, has to travel a large distance to the brane before it can return to the asymptotic boundary across the brane. This additional distance can be made arbitrarily small by choosing a small enough brane angle. We will furthermore see in section \ref{sec:time_evolution_non_extremal} how the time evolution of an RT surface at early times can be mapped to this case.


\subsection{No-island phase for $\chi_\Sigma = 0$}\label{sec:no_island_phase_2_non_extremal}
\begin{figure}[t]
	\def\svgwidth{0.8\linewidth}
	\centering{
		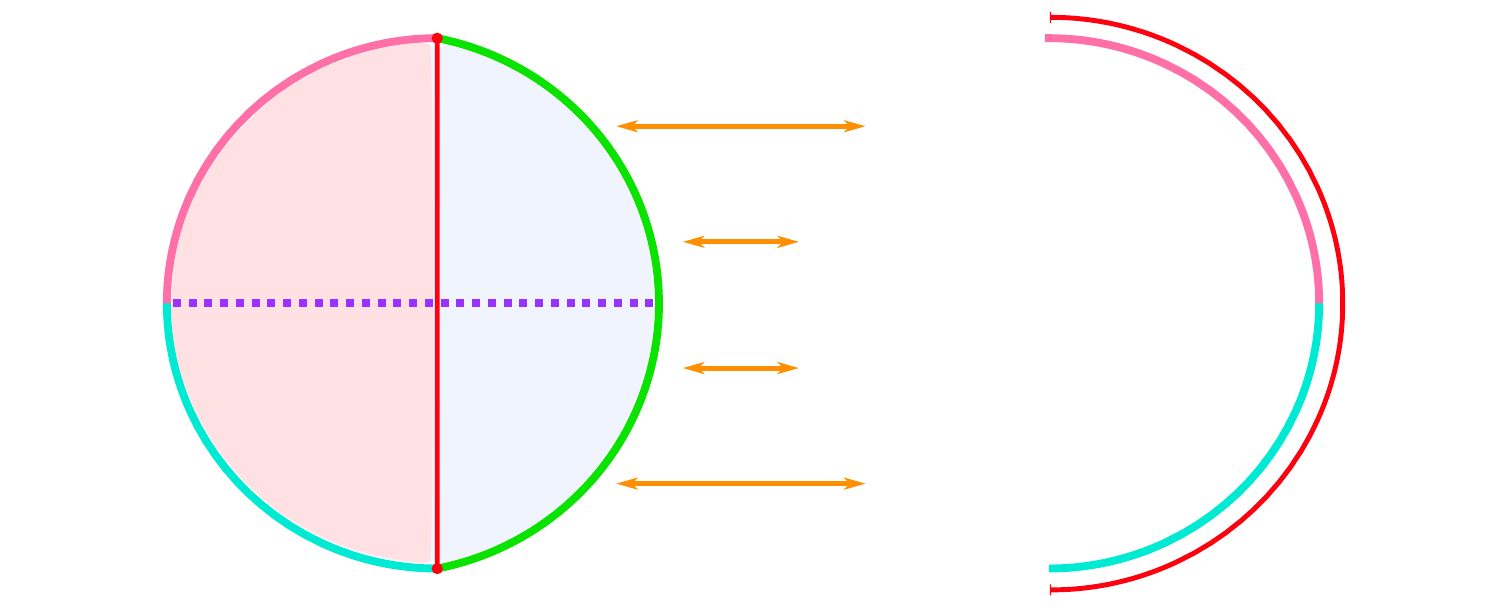
		\caption{The RT surface of an entangling surface located at the defect in the no-island phase.
		}
		\label{fig:no_island_2}
	}
\end{figure}

Lastly, we will consider the case of a zero-width belt, \ie the case where the location of the entangling surface is taken towards the defect, so that the RT surface falls straight through the bulk along constant boundary slicing coordinate $\chi = \chi_\Sigma$, c.f. figure \ref{fig:no_island_2}. Note that this setup is essentially the same as considered in \cite{Hartman:2013qma}, which studied entanglement entropy of identical half-spaces in the two sides of a time-evolved TFD.

Due to symmetry, the trajectory of the RT surface is determined by its radial coordinate $r$ as a function of time $\tau$.
However, it is convenient to introduce Eddington-Finkelstein coordinates to avoid the coordinate singularity at $r=1$. Hence, describing ingoing null rays, we have
\beq
v=\tau+r_\mt{tor}(r)\qquad{\rm where}\quad r_\mt{tor}(r)=\frac12\,
\log\!\(\frac{|r-1|}{r+1}\)\,,
\label{EFcoord}
\eeq
where $r_\mt{tor}(r)$ denotes the usual tortoise coordinate.\footnote{We extend our defintion of $r_\mt{tor}(r)$ across the horizon using the standard prescription given in \cite{Chapman:2016hwi}.}
Note that with the above definitions, $r_\mt{tor}(r\to\infty)\to0$ and hence $v=\tau$ at the asymptotic AdS boundary. Then the metric becomes
\beq
\label{EFcoord2}
ds^2 = \LAdS^2 \left( - (r^2-1)\, dv^2 + 2\,dv\,dr + r^2\, dH_{d-1}^2\right)\,.
\eeq

Now the extremal surface will fall from the asymptotic boundary, through the exterior, across the Rindler horizon, reaching a minimal radius at $r_*$, within the interior. Then the surface will continue emerging into the second exterior region. Due to reflection symmetry, we need only track the trajectory of the RT surface until it reaches $r_*$. Using eq.~\reef{EFcoord2}, the area functional can be written as 
\begin{align}
\label{eq:RT_area_no_island_tau}
\area(\RT) = 4\, \vol_{H_{d-2}} L^{d-1} \int^{\lambda_\mt{UV}}_{\lambda_*} d\lambda\, r^{d-2}\sqrt{-(r^2-1) \dot v^2 + 2\dot v\dot r}\,,
\end{align}
where $\lambda$ is a radial coordinate intrinsic to the surface, which increases along the surface moving from the left asymptotic AdS boundary to the right boundary. The limits of integration here correspond to $\lambda_*$, the value at the minimal radius $r_*$, and $\lambda_\mt{UV}$, the value at the UV cutoff near the right boundary -- see figure \ref{fig:no_island_2}. We have also included a factor of 4 to account for the fact that we only integrate from the $\mathbb Z_2$ symmetric point $\lambda_*$ out to the right boundary, and the fact that there are two such RT surfaces, one on either side of the brane. Of course, we have also integrated out the directions along the belt, \ie along the $H^{d-2}$. Now, we fix the reparametrization symmetry of the area functional with the following convenient gauge choice
\beq
\sqrt{-(r^2-1) \dot v^2 + 2\dot v\dot r} = r^{d-2}\,.
\label{gaugefix}
\eeq
 
The integrand in eq.~\reef{eq:RT_area_no_island_tau} is independent of $v$ and so we have a conserved `$v$-momentum' 
\beq
P_v=\frac{\partial {\cal L}}{\partial \dot v}=\frac{r^{d-2}(\dot r-(r^2-1) \dot v)}{\sqrt{-(r^2-1) \dot v^2 + 2\dot v\dot r}} = \dot r-(r^2-1) \dot v\,,
\label{pv}
\eeq
where the second expression results from substituting in the gauge choice \reef{gaugefix}. Using  eqs.~\eqref{gaugefix} and \eqref{pv} to solve for $\dot r$ and $\dot v$, we find
\beqa
\dot r\left[P_v,r\right]  &= & \sqrt{(r^2-1)\,	 r^{2(d-2)} + P_v^2}\,,\nonumber \\
\dot v\left[P_v,r\right]  &= & \frac{\dot r-P_v}{r^2-1 } = \frac{1}{r^2-1 } \left( -P_v+ \sqrt{(r^2-1)\,	 r^{2(d-2)} + P_v^2} \right)\, .\label{onedot}
\eeqa
Note that we have implicitly chosen a positive sign for $\dot r$ indicating that $r$ is increasing as we move along the surface out towards the asymptotic boundary.

An intuitive picture of the dynamics of the extremal surfaces is given by recasting the $\dot r$ equation above as a Hamiltonian constraint,
\begin{equation}
\dot  r^2 + U(r)   =  P_v^2\,,
\label{yarnSho}
\end{equation}
where the effective potential is given by
\begin{equation}\label{PotentialShocks}
U(r)=-(r^2-1)\, r^{2(d-2)}\, .
\end{equation}
In this framework, $P_v^2$ plays the role of the conserved energy and the minimum radius $r_*$ corresponds to the turning point where $\dot r =0$, \ie
\beq
(1-r_*^2)\, r_*^{2(d-2)}=  P_v^2\,.
\label{rmin}
\eeq
The area \reef{eq:RT_area_no_island_tau} of the extremal surface becomes
\begin{align}
\label{area22}
\area(\RT) 
=4\, \vol_{H_{d-2}} L^{d-1} \int_{r_*}^{r_\mt{UV}} dr\, \frac{r^{2(d-2)}}{\sqrt{(r^2-1)\,	 r^{2(d-2)}+P_v^2}}\,,
\end{align}
using eqs.~\reef{gaugefix} and \reef{onedot}. Note that $r_\mt{UV}$ denotes the position of the UV cutoff surface near the asymptotic AdS boundary.

With eq.~\reef{rmin}, the extremal surface can be specified by the  integration
constant $P_v$ or the boundary condition $r_*$. However, we want to examine the
time evolution of the entanglement entropy and so we must determine a relation
between these constants and the boundary time. In particular, using
eq.~\reef{onedot}, we can integrate out to the right boundary to determine 
\beq
v_\mt{bound}-v_*=\int_{r_*}^{r_\mt{UV}} dr\, \frac{\dot v}{\dot r}
=\int_{r_*}^{r_\mt{UV}} dr\, \frac{1}{r^2-1}\[1-\frac{P_v}{\sqrt{(r^2-1)\,	 r^{2(d-2)} + P_v^2}}\]\,,
\label{vbound1}
\eeq
where $v_*$ denotes the value of the Eddington-Finklestein time at the turning point. However, because of the $\mathbb Z_2$ symmetry of the extremal surface, we know that the turning point lies on the surface $t=0$, and so we may use eq.~\reef{EFcoord} to write
\beq
v_*=r_\mt{tor}(r_*)=\frac12\,
\log\!\(\frac{1-r_*}{1+r_*}\)\,.
\label{vbound2}
\eeq
Further, we know that $v_\mt{bound}=\tau[P_v]$ and hence we find
\beq
\tau [P_v]=\frac12\,
\log\!\(\frac{1-r_*}{1+r_*}\)+\int_{r_*}^{r_\mt{UV}} dr\, \frac{1}{r^2-1}\[1-\frac{P_v}{\sqrt{(r^2-1)\,	 r^{2(d-2)} + P_v^2}}\]\,.
\label{vbound3}
\eeq
Note that the integrand is nonsingular in the vicinity of the horizon, \ie near
$r=1$. 

The time derivative of the area \reef{eq:RT_area_no_island_tau} admits a very simple form
\begin{equation}\label{simple0}
\frac{d \area(\RT)}{d \tau_\Sigma} = 4\, \vol_{H_{d-2}} L^{d-1}\, P_v =
4\, \vol_{H_{d-2}}L^{d-1}\, r_*^{d-2}\,\sqrt{1-r_*^2}\,,
\end{equation}
where $\tau$ is the boundary time parameter.\footnote{
A quick derivation of this result follows by considering a small variation of the surface profile in eq.~\reef{eq:RT_area_no_island_tau}. The bulk contributions naturally vanishes by the equations of motion determining the extremal surface. However, deriving the latter requires an integration by parts which produces boundary terms. These are usually eliminated by fixing the boundary conditions at infinity. In the above result, we instead allow for a small variation in the boundary time.}
Further, we also observe that the critical radius where $\partial_r U=0$ is given by
\beq
r_c^2=\frac{d-2}{d-1}\,.
\label{rcrit}
\eeq
At late times, the turning point is very close to this critical radius, \ie, the critical surface lies near the surface $r=r_c$ for a long time, and so we can replace $r_* \to r_c$ into eq.~\reef{simple0}. Hence we expect the growth of the area is fixed at late times, \ie 
\begin{equation}\label{carrot2}
\frac{d \area(\RT)}{d \tau_\Sigma} = 
4\, \vol_{H_{d-2}}L^{d-1}\, \frac{(d-2)^{(d-2)/2}}{(d-1)^{(d-1)/2}}\,.
\end{equation}

As we will see momentarily, the late time behavior of the entropy of any subregion bounded by constant $\chi$ in the no-island phase is determined by a zero-belt width calculation. Thus, as in the two-dimensional case studied in \cite{Bak:2020enw} (as well as the higher dimensional case \cite{Almheiri:2019psy}), the entropy corresponding to the no-island phase grows without bound.


\subsection{Time-evolution for general $\chi_\Sigma,\tau_\Sigma \neq 0$}
\label{sec:time_evolution_non_extremal}

Given the region $\bdyReg$ of interest,\footnote{Recall that $\bdyReg$ consists of all points more than a distance $\chi_\Sigma$ away from the defect in both CFTs.} we can ask how the RT surface changes under time evolution. If we are in the island phase, the RT surface is completely contained inside the Rindler patch so that time translations are a symmetry and the entropy is a constant. On the other hand, in the no-island phase, the RT surface connects to both bath CFTs. Forward time evolution of both sides is not a symmetry and the area of the RT surface changes.

Obtaining RT surfaces in the no-island phase which are anchored on symmetric entangling surfaces of arbitrary width and at arbitrary times in higher dimensions is generally difficult. However, as we will now show, our choice of entangling surfaces with the hyperbolic symmetry of $H_{d-2}$ allows us to map the RT surface at any $(\chi_\Sigma,\tau_\Sigma)$ either to some RT surface in the $\tau=0$ slice, \ie with $(\chi'_\Sigma,\tau'_\Sigma=0)$ or to the case where the entangling surface is at $\chi=0$, \ie with $(\chi'_\Sigma=0,\tau'_\Sigma)$. In particular, this means that the solutions obtained in the last two subsections are sufficient to discuss the full time evolution of the symmetric entangling surfaces of interest.

The strategy we will employ in this chapter is the following. We will perform a coordinate change from Rindler space to a particular Poincar\'e coordinate system defined below. In the new coordinates, the entangling surfaces are straight lines. By exploiting the boost symmetry of the Poincar\'e patch and mapping back to Rindler space, the task of calculating entanglement entropy of a subregion with $\chi_\Sigma$ at time $\tau_\Sigma$ can be reduced to one of the cases discussed in sections \ref{sec:no_island_phase_1_non_extremal} and \ref{sec:no_island_phase_2_non_extremal}.

To understand the required coordinate changes it is convenient to embed AdS${}_{d+1}$ into $\mathbb R^{d,2}$, \ie we are looking for a parametrization of (parts of) the hyperboloid defined via 
\begin{align}
-T_1^2 - T_2^2 + X_1^2 + \dots + X_{d} = -L^2\,. 
\end{align}
Our original two Rindler patches correspond to the parametrization
\begin{align}
	T_1 &= \pm L \sqrt{r^2-1} \sinh \tau\,,  \qquad \qquad  \, T_2 = L r \cosh \chi \cosh \eta\,, \nonumber \\
	X_1 &= \pm L \sqrt{r^2-1} \cosh \tau\,,   \qquad \qquad  X_2  = \pm L r \sinh \chi\,, \label{eq:rindler_coords}  \\
	X_i &=  L r \cosh \chi \sinh \eta\, \mu_i \qquad  \text{ with } i  =3,4,\ldots,d\,, \nonumber
\end{align}
where $\mu_i$ denotes further angular coordinates, \eg
$\mu_3=\cos\phi_1,\ \mu_4=\sin\phi_1\cos\phi_2,\,\ldots$, which, together with $\eta$ parametrize the $H_{d-2}$ slice of the metric \eqref{eq:metric_bdry_slicing}. The AdS boundary is located at $r \to \infty$, and each sign corresponds to one of the two Rindler wedges. On a fixed $r$ slice, we can reach the boundary by taking $\chi \to \pm \infty$ or $\eta \to \pm\infty$.  For any constant Rindler time (\ie fixed $\tau$), the bifurcation surface reached with $r\to 1$. The defect in the CFT is located at $\chi = 0 = X_2$. The entangling surfaces are defined to be at $\chi=\pm\chi_\Sigma$ in both CFTs.

\begin{figure}[t]
	\def\svgwidth{0.6\linewidth}
	\centering{
		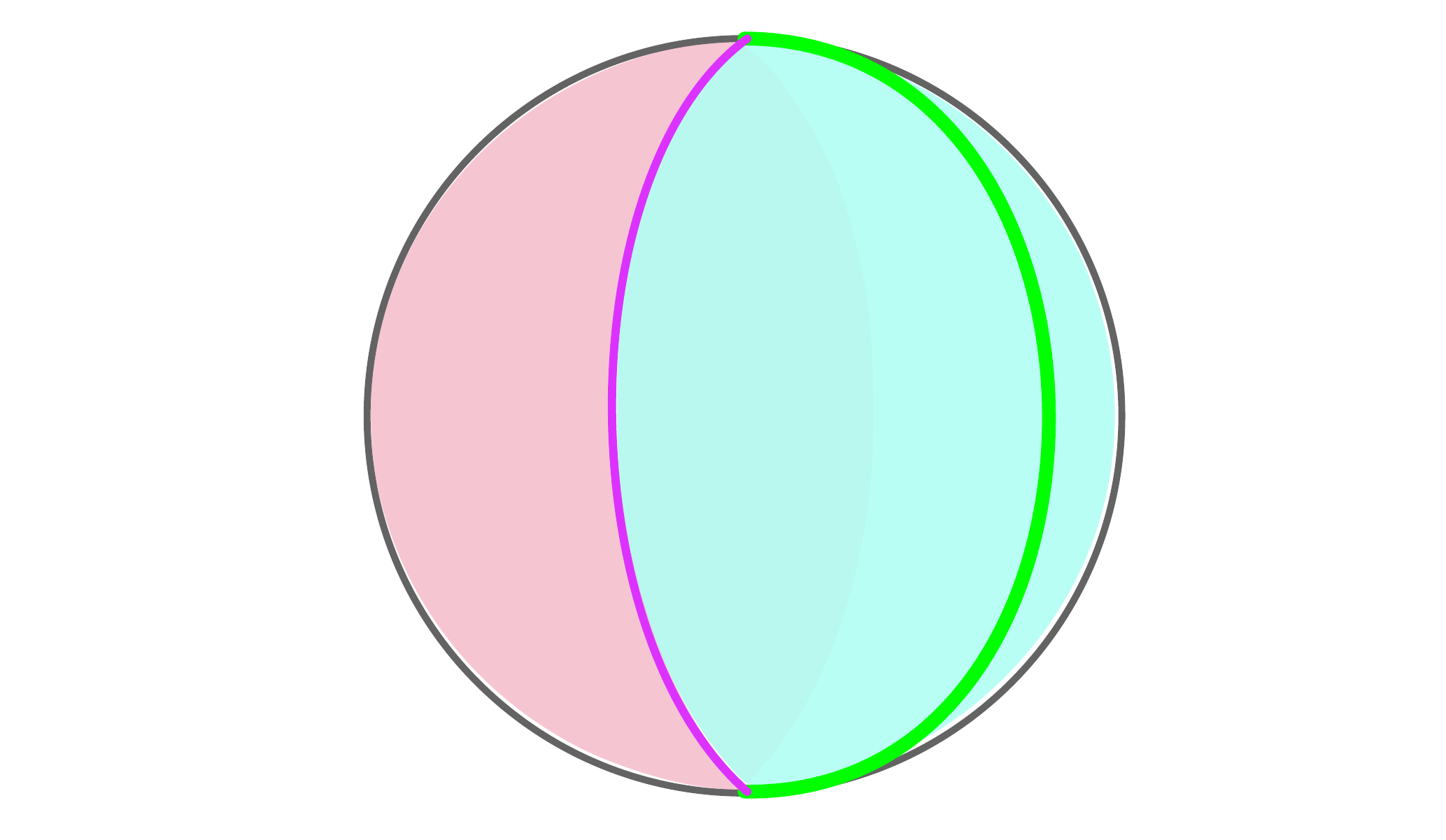
		\caption{A time-slice of our setup. The spatial boundary $S^{d-1}$ (in global coordinates) is split into two hyperbolic discs $H_{d-1}$, shown in pink and aqua, which are glued together at infinity. At the same location, the bifurcate horizon intersects the boundary. The CFT on either disc is dual to a Rindler wedge in the bulk. The defect (green) is a great circle on the global boundary. As indicated in the figure, the Poincar\'e coordinates introduced in this section cover the full sphere, with the point at infinity appearing on the south pole of the sphere. Entangling surfaces are the semi-circles shown in red. }
		\label{fig:coordinates_rindler_poincare}
	}
\end{figure}

We will now consider a particular Poincar\'e  coordinate system, which covers both Rindler wedges and is defined in terms of embedding coordinates as
\beqa
        T_1 &=& L \frac{\tDom}{\zDom}\,, \qquad
        X_1 = L \frac{\xDom_1}{\zDom}\,, \qquad
        X_2 = L \frac{\xDom_2}{\zDom}\,, \quad
        \cdots
        \nonumber \\
        X_d &=& \frac{\zDom^2 + \xDom^2- \tDom^2  - L^2}{2\zDom}\,, \qquad
        T_2 = \frac{\zDom^2  + \xDom^2- \tDom^2 + L^2}{2\zDom}\,,  	\label{eq:new_poincare_coords}
\eeqa
where $\xDom^2=\xDom_1^2+\xDom_2^2+\cdots+\xDom_{d-1}^2$. In these coordinates, the bifurcation surface intersects the boundary ($\zDom \to 0$) at $\xDom_1 = \tDom = 0$,\footnote{The full Rindler horizons reach the boundary along $\xDom_1^2-\tDom^2=0$.} while the defects are located at $\xDom_2 = 0$.  The two CFTs are mapped to the regions $\xDom_1 > 0$ and $\xDom_1 < 0$, respectively. We will denote the CFT at $\xDom_1 > 0$ as the right CFT, and the one at $\xDom_1 < 0$ as the left CFT.  Comparing eqs.~\eqref{eq:rindler_coords} and \eqref{eq:new_poincare_coords} in the boundary limit, it is easy to see that the entangling surfaces in the right CFT get mapped to
\beq
    \label{eq:location_entangling_surfaces}
        \xDom_2 = \pm \frac{\sinh \chi_\Sigma}{\cosh \tau_\Sigma} \cdot \xDom_1\,,\qquad
        \tDom = \tanh \tau_\Sigma \cdot \xDom_1\,.
\eeq
This shows the convenient property of the new Poincar\'e coordinates:  entangling surfaces lie along rays (\ie straight lines) in the positive half-space with $\xDom_1>0$, whose slope depends on the spatial location $\chi_\Sigma$ and the Rindler time $\tau_\Sigma$ at which the entangling surfaces are defined. 
Further, flipping the sign of $\xDom_1$ to $-\xDom_1$ in the above expressions yields the entangling surfaces in the left CFT.
The relation between the Rindler coordinate given in eq.~\eqref{eq:rindler_coords} and the new Poincar\'e coordinates of eq.~\eqref{eq:location_entangling_surfaces} is illustrated in figure \ref{fig:coordinates_rindler_poincare}.

We now need to choose cutoffs in order to regulate the area integrals of the RT surfaces. First, we need to regulate the UV divergence in the entanglement entropy by introducing maximum radius in both AdS-Rindler patches $r_\mt{UV}\gg1$. This translates to a $\zDom$-dependent cutoff in the new coordinates,
\begin{align}
	\label{eq:cutoff_z}
	\zDom^2 > \zDom^2_\mt{min} = \frac{\xDom_1^2-\tDom^2}{r_\mt{UV}^2-1} \sim \frac{\xDom_1^2-\tDom^2}{r_\mt{UV}^2} \,,
\end{align} 
where in the last step, we used that $r_\mt{UV}\gg 1$. 

Second, we need an IR cutoff which we impose in the transverse directions along the entangling surface. Since the solution is independent of shifts in all directions along the brane, the transverse directions should just contribute an overall volume factor. We choose $\eta_\mt{max} = \frac{\cutoffIR}{R}\gg1$, which translates to
\begin{align}
	\label{eq:cutoff_IR}
  \frac{\zDom^2 -\tDom^2 + \xDom^2 + \LAdS^2}{\sqrt{\zDom^2-\tDom^2 + \xDom_1^2 + \xDom_2^2}}
  < 2 \LAdS\, \cosh\!\frac{\cutoffIR}{R}\,.
\end{align}

\subsubsection{Island phase}

\begin{figure}[t]
	\def\svgwidth{0.9\linewidth}
	\centering{
		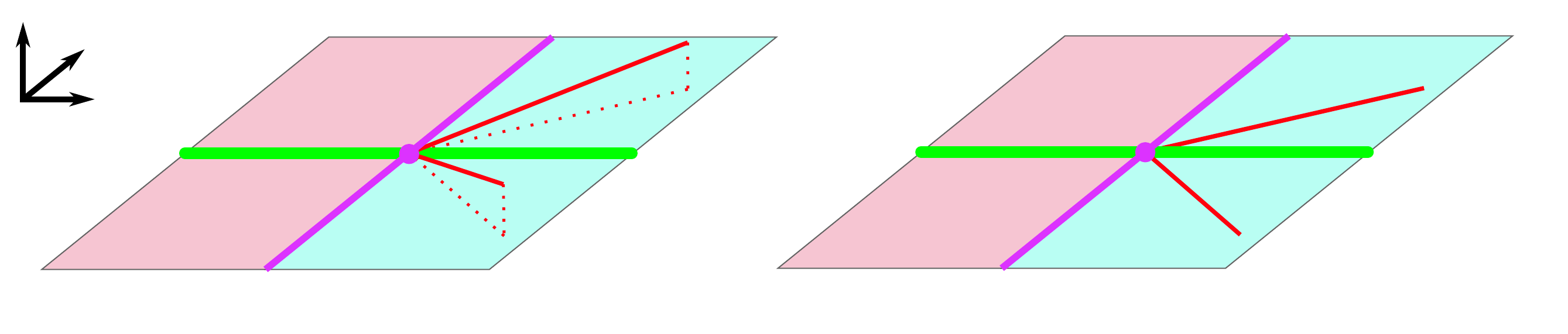
		\caption{The left panel shows two components of the entangling surface (red) at non-zero Rindler time $\tau_\Sigma$ in the right CFT in the Poincar\'e coordinates \reef{eq:new_poincare_coords}. In the island phase, these two rays in the boundary geometry are connected by an RT surface in the bulk. We can perform a boost in $\xDom_1$ direction to map this set of entangling surfaces to the $t = 0$ slice, which also corresponds to $\tau = 0$ slice of the hyperbolic boundary geometry. The boost is a symmetry of the defect (green).
		}
		\label{fig:island_pedagogy}
	}
\end{figure}

As a warm-up exercise, we will show that the entropy on the island phase is in fact invariant under time evolution. This is obviously true, since the RT surface is completely contained within one Rindler wedge and $\tau$ is a Killing coordinate for the corresponding metric \reef{eq:metric_rindler_brane}. Hence the corresponding time evolution of a single Rindler wedge is an isometry of that wedge. 
In this case, we are looking for an extremal surface which ends on the boundary at the location defined by eq.~\eqref{eq:location_entangling_surfaces} for either $\xDom_1>0$ or $\xDom_1<0$, depending on which Rindler wedge we are interested in. Here, we choose $\xDom_1 > 0$. We can express the problem in a boosted coordinate systems
\beq
\tDom' = \gamma(\tDom - \beta \xDom_1)\,,\qquad
\xDom_1' = \gamma(\xDom_1 - \beta \tDom)\,,
\eeq
with boost parameter $\beta = \tanh \tau_\Sigma$. This is depicted in figure \ref{fig:island_pedagogy}. This boost leaves the cutoffs given in eqs.~\eqref{eq:cutoff_z} and \eqref{eq:cutoff_IR} invariant, and changes the equation for the entangling surface to
\beq
 \xDom'_2 = \pm \xDom'_1 \sinh \chi_\Sigma\,,\qquad   \tDom' = 0\,.
\eeq
This is precisely the entangling surface of the same region at $\tau=\tau'_\Sigma = 0$ with the appropriate cutoffs. We may thus conclude that entropy of the region $\bdyReg$ remains constant in the island or connected phase, as anticipated. Again because we have a pure state globally, we can see that the entropy of the complementary region, \ie the two belts centered on the conformal defects in each of the two CFTs, is independent of $\tau_\Sigma$ in this connected phase.


\subsubsection{No-island phase}

\begin{figure}[t]
	\def\svgwidth{0.9\linewidth}
	\centering{
		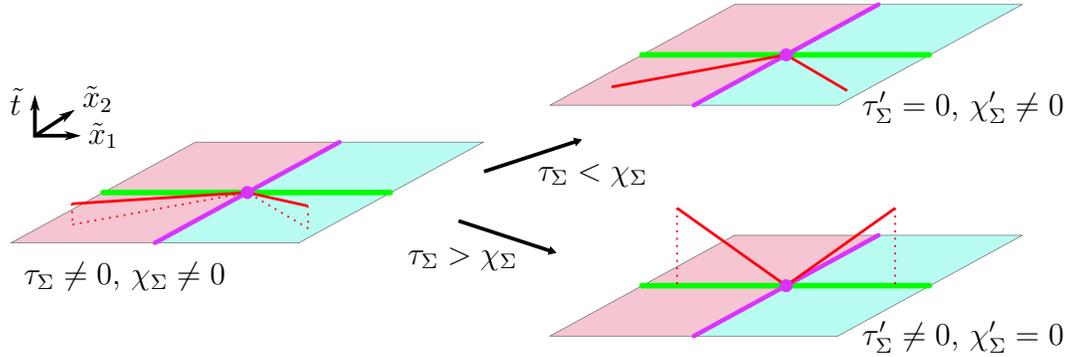
		\caption{
		The left panel shows two components of the entangling surface (red) at non-zero Rindler time $\tau_\Sigma$ in the right CFT in the Poincar\'e coordinates \reef{eq:new_poincare_coords}. 
These two rays are located in different CFTs so that in the no-island phase, they are joing by an RT surface in the bulk which passes through the Rindler horizon. In this case, we can now boost in $\xDom_2$ direction to map these two rays to $\tau_\Sigma'=0$ when $\tau_\Sigma<\chi_\Sigma$ or to $\chi_\Sigma'=0$ when $\tau_\Sigma>\chi_\Sigma$. 
		}
		\label{fig:no_island_pedagogy}
	}
\end{figure}

For the no-island phase, we focus on the case in which the RT surface connects entangling surfaces in the CFTs dual to different Rindler patches. The entangling surfaces are located at
\begin{align}
  \tDom = -\frac{\sinh \tau_\Sigma}{\sinh \chi_\Sigma}\, \xDom_2 , && \xDom_1 = \pm\frac{\cosh \tau_\Sigma}{\sinh \chi_\Sigma}\,  \xDom_2,
\end{align}
where we have chosen to focus on $\xDom_2 < 0$, \ie to the region on one side of the defect. Similarly to the island phase, we want to go to a new coordinate system in which the calculation becomes simpler. Now, however, we have to distinguish two cases.

\paragraph{Case 1:}
If $\tau_\Sigma < \chi_\Sigma$,\footnote{We are assuming that both $\tau_\Sigma$ and $\chi_\Sigma$ are positive (or zero). Let us also note here that $\tau_\Sigma = \chi_\Sigma$ is a special case, where the entangling surfaces lie in the null plane $\tDom = - \xDom_2$. Our approach of boosting in the $\xDom_2$ direction fails in this case, but the results for the time evolution are smooth across this point. }  we can boost this problem in $\xDom_2$ direction with boost parameter $\beta = -\frac {\sinh \tau_\Sigma}{\sinh \chi_\Sigma}$. This is depicted in the upper panel of figure \ref{fig:no_island_pedagogy}. The new entangling surfaces are then located at
\begin{align}
  \tDom' = 0, && \xDom_1' = \pm \xDom_2'\, \sqrt{\frac{\cosh^2 \tau_\Sigma}{\sinh^2 \chi_\Sigma - \sinh^2 \tau_\Sigma}},
                 \label{eq:fluoxetine}
\end{align}
where $\xDom_2' < 0$. Expressing the result in Rindler coordinates, we are dealing with the case of an entangling surface in the $\tau = \tau'_\Sigma= 0$ plane. The new location of the entangling surface $\chi_\Sigma'$ is given by
\begin{align}
\label{eq:b_prime}
\cosh \chi_\Sigma'= \frac{\cosh \chi_\Sigma}{\cosh \tau_\Sigma}.
\end{align}
Note that as $\cosh \tau_\Sigma \to \cosh \chi_\Sigma$ (and so as $|\frac {\sinh \tau_\Sigma}{\sinh \chi_\Sigma}|\to1$), the new entangling surface gets closer and closer to the defect, \ie $\chi_\Sigma'\to0$.

Importantly,  the cutoffs are not boost invariant in this case. The IR cutoff given in eq.~\eqref{eq:cutoff_IR} remains unchanged, but the UV cutoff in eq.~\eqref{eq:cutoff_z} changes along the trajectory of the entangling surface to 
\begin{align}
  r'_\mt{UV}= r_\mt{UV}\, \cosh\tau_\Sigma\,.
  \label{eq:cutoff_case_1}
\end{align}
We should caution the reader that we arrived at eq.~\eqref{eq:cutoff_case_1} by substituting the trajectory of the entangling surface into the boosted cutoff. This means that eq.~\eqref{eq:cutoff_case_1} is only correct for a small cutoff. Luckily, the corrections to the new cutoff only change the entanglement entropy at order $\mathcal O(1/{r_\mt{UV}})$.

In conclusion, we found that if $ \tau_\Sigma < \chi_\Sigma$, the entanglement entropy of the region $|\chi| > \chi_\Sigma$ at time $\tau=\tau_\Sigma$ is the same as that of a region $|\chi| > \chi'_\Sigma$ given in eq.~\eqref{eq:b_prime} at time $\tau=\tau_\Sigma ' = 0$ calculated with a different cutoff, given by eq.~\eqref{eq:cutoff_case_1}.

\paragraph{Case 2:}
The other case,  $\tau_\Sigma > \chi_\Sigma$, is shown in the lower panel of figure \ref{fig:no_island_pedagogy}. Now we can boost in the $\xDom_2$ direction again, but using $\tilde \beta = - \frac{\sinh \chi_\Sigma}{\sinh \tau_\Sigma}$. The new entangling surfaces are located at
\begin{align}
    \xDom_2' =0\,,\qquad \xDom_1' = \pm \tDom' \sqrt{\frac{\cosh^2 \tau_\Sigma}{\sinh^2 \tau_\Sigma - \sinh^2 \chi_\Sigma}}\,.
\end{align}
While this does not reduce to a surface lying in the $\tau' = 0$ plane, in Rindler coordinates it reduces to an entangling surface for a belt width $\chi_\Sigma' = 0$ and
\begin{align}
	\label{eq:new_time}
    	\cosh \tau_\Sigma' = \frac{\cosh \tau_\Sigma}{\cosh \chi_\Sigma}\,.
\end{align}
Again, the IR cutoff in eq.~\reef{eq:cutoff_IR} is unchanged, however, the UV cutoff changes to\footnote{Note that, like above, we have substituted the trajectory of the entangling surface into the boosted expression. Thus, this equation is only strictly correct in the $r_\mt{UV} \to \infty$ limit, but the corrections are subleading to the finite part of the entanglement entropy.} 
\begin{align}
  r^{\prime}_\mt{UV} =  r_\mt{UV}\, \cosh \chi_\Sigma\,.
  \label{eq:sertraline}
\end{align}
Let us note that the cutoff location still is continuous. In the previous case, the new cutoff was the old cutoff multiplied by $\cosh^2 \tau_\Sigma$. The latter was reliable as long as $ \tau_\Sigma < \chi_\Sigma$. However, we see here that once $\tau_\Sigma >  \chi_\Sigma$, the cutoff is no longer time-dependent.

\subsection{The information paradox}
\label{sec:islands_for_non_extremal}

Now the preceding results can be combined to give a qualitative description of the time evolution of the entanglement entropy. Following the discussion in section \ref{twod} for two dimensions, at time $\tau=0$, we have a standard thermofield double state of the two CFTs on hyperbolic spatial geometries, including the conformal defects at $\chi=0$. If we restrict the observations to either the left or right side, the reduced state is a thermal one and in particular, the bath CFT is in thermal equilibrium with the corresponding conformal defect, with temperature $T=1/(2\pi R)$.

Using the brane perspective and an appropriate choice of parameters,\footnote{Recall that we obtain a good aproximation to (semiclassical) Einstein gravity on the brane if we choose $\frac{L}{\ell_B} \ll 1$ and $\lambda_\mt{b}$ not too close to $-1$ -- see footnote \ref{bigtoe}.} we can describe the conformal defects are replaced by (two copies of) the boundary CFT coupled to Einstein gravity on an AdS$_d$ region. For the configuration described above, this yields a topological black hole solution shown in eq.~\reef{eq:metric_rindler_brane}. We emphasize that the latter  really describes an AdS$_d$ geometry in AdS-Rindler coordinates, and hence the thermal equilibrium between this `black hole' and the finite temperature CFT on the asymptotic boundary can be understood as arising because the two systems are coupled along an accelerated trajectory in the region of dynamical gravity. While the black hole is in equilibrium with the bath CFT,  under time evolution, the two systems are constantly exchanging thermal quanta. The immediate effect of this process after $\tau=0$ is to increase the entanglement between one side of the black hole, \ie one of the AdS-Rindler wedges on the brane, and its respective bath CFT. 

A standard measure for the entanglement between both AdS-Rindler wedges and their respective baths is given by the entanglement entropy of the complement of two belt subregions centered around the conformal defects in the boundary as discussed above. In section \ref{sec:time_evolution_non_extremal}, we saw that by a judicious change of coordinates (and cutoff), the calculation of the entanglement entropy of these regions can be mapped at late times (\ie $\tau_\Sigma\ge \chi_\Sigma$) to the case of a zero-width belt.
Further, in section \ref{sec:no_island_phase_2_non_extremal}, we found that the entanglement entropy grows linearly in time, as shown in eq.~\reef{carrot2}.\footnote{Implicitly, to apply eq.~\reef{carrot2}, we must also show $\partial_{\tau_\Sigma'}\simeq\partial_{\tau_\Sigma}$. The latter follows at late times from eq.~\reef{eq:new_time}, which yields
\beq
\frac{\partial\ }{\partial\tau_\Sigma'}=\(1-\frac{\sinh^2\chi_\Sigma}{\sinh^2\tau_\Sigma}\)^{1/2}\frac{\partial\ }{\partial\tau_\Sigma}\,.
\eeq
Alternatively, the same result  also follows by simply observing that eq.~\reef{eq:new_time} implies that at late times: $\tau' = \tau -  \log\left(\cosh \chi_\Sigma \right) + O(e^{-2 \tau})$. Let us add that this linear growth is analogous to that found for planar black holes in \cite{Hartman:2013qma}.}

As in the two-dimensional case \cite{Almheiri:2019yqk, Mathur:2014dia}, this linear growth of entropy would lead to an information paradox for our eternal black holes, if it was valid for all times.
The reason is that the entanglement entropy must be bound from above by the defect entropy, since the defects need to purify the bath system. In the case of interest, the theory is well approximated by weakly coupled Einstein gravity. This allows us to view the quantum fields on the gravitational background as giving a small correction to the entropy and thus, the defect entropy is well-approximated by two times the black hole entropy.\footnote{The black hole entropy is proportional to the horizon area of the black hole, which in our case is infinite. Hence to be precise, we must consider an IR regulated entropy, as discussed with eq.~\eqref{eq:cutoff_IR}.}

The appearance of an island in the effective gravity theory from the brane perspectice is simply related to a phase transition of the RT surfaces in the bulk description of our system. The RT surface changes from the no-island phase, in which it connects both CFTs through the horizon, to the island phase, in which it connects both sides of the defect in a single Rindler wedge. The fact that there will always be an extremal surface crossing the brane is easy to see: Before we invoke the extremization condition at the brane, there is an infinite family of candidate RT surfaces, which start in the bath and meet at the brane. To get the correct RT surface, we need only extremize the area by varying the position of the surface where  they meet the brane. Subregion duality and the homology constraint guarantee that there will be one extremal surface for every belt configuration (although the boundary of the island might sit at the horizon or at the CFT defect).

In order to establish unitarity of the Page curve, we still need to argue that the island appears before the black hole fails to purify the bath region $\bdyReg$ under consideration. In the case of interest here, we have that $\frac{\ell_B}{\LAdS}\gg1$. In this approximation, it follows from eq.~\eqref{eq:condition_1} that
\begin{align}
\label{eq:small_angle_zeta}
\zeta_{\QES} = \zeta_\text{hor} \left( 1 + \frac {\zeta_*^2(1+ \zeta_*^2)^{d-2}}{2 \zeta_\text{hor}^{(2d-4)} (1 + \lambda_\mt{b})^2} + \dots\right).
\end{align}
In deriving this equation, we have used that the location of the horizon on the brane is at $\zeta_\text{hor} \sim \frac{\ell_\mt{B}}{L} \gg 1$ and that $\zeta_*$ cannot scale with $\frac {\ell_\mt{B}} L$ at leading order. The reason is that $\zeta_*$ is bounded from above by a function of the belt width. We can see that the location of the new quantum extremal surface will always be close to the horizon -- see also the next section for numerical plots. The leading order contribution to the generalized entropy is given by the area of the horizon which gives the black hole entropy. While a more involved analysis is needed to demonstrate that the appearance of the island saves unitarity, this shows that the island mechanism has the right qualitative behaviour to unitarize the Page curve.


%% file: images/xizeta_updateSecondPart.pdf_tex
\begingroup%
  \makeatletter%
  \providecommand\color[2][]{%
    \errmessage{(Inkscape) Color is used for the text in Inkscape, but the package 'color.sty' is not loaded}%
    \renewcommand\color[2][]{}%
  }%
  \providecommand\transparent[1]{%
    \errmessage{(Inkscape) Transparency is used (non-zero) for the text in Inkscape, but the package 'transparent.sty' is not loaded}%
    \renewcommand\transparent[1]{}%
  }%
  \providecommand\rotatebox[2]{#2}%
  \newcommand*\fsize{\dimexpr\f@size pt\relax}%
  \newcommand*\lineheight[1]{\fontsize{\fsize}{#1\fsize}\selectfont}%
  \ifx\svgwidth\undefined%
    \setlength{\unitlength}{473.38582677bp}%
    \ifx\svgscale\undefined%
      \relax%
    \else%
      \setlength{\unitlength}{\unitlength * \real{\svgscale}}%
    \fi%
  \else%
    \setlength{\unitlength}{\svgwidth}%
  \fi%
  \global\let\svgwidth\undefined%
  \global\let\svgscale\undefined%
  \makeatother%
  \begin{picture}(1,0.37125749)%
    \lineheight{1}%
    \setlength\tabcolsep{0pt}%
    \put(0,0){\includegraphics[width=\unitlength,page=1]{images/xizeta_updateSecondPart.pdf}}%
    \put(0.25456912,0.29947594){\color[rgb]{0,0,0}\makebox(0,0)[lt]{\lineheight{1.25}\smash{\begin{tabular}[t]{l}$\RT$\end{tabular}}}}%
    \put(0.31947953,0.16189959){\color[rgb]{0,0,0}\makebox(0,0)[lt]{\lineheight{1.25}\smash{\begin{tabular}[t]{l}$\zeta_*$\end{tabular}}}}%
    \put(0,0){\includegraphics[width=\unitlength,page=2]{images/xizeta_updateSecondPart.pdf}}%
    \put(0.42306336,0.10301017){\color[rgb]{0,0,0}\makebox(0,0)[lt]{\lineheight{1.25}\smash{\begin{tabular}[t]{l}($\zeta_\mt{QES}, \,\xi_\mt{QES}$)\end{tabular}}}}%
    \put(0,0){\includegraphics[width=\unitlength,page=3]{images/xizeta_updateSecondPart.pdf}}%
    \put(0.72477142,0.2243614){\color[rgb]{0,0,0}\makebox(0,0)[lt]{\lineheight{1.25}\smash{\begin{tabular}[t]{l}horizon\end{tabular}}}}%
    \put(0,0){\includegraphics[width=\unitlength,page=4]{images/xizeta_updateSecondPart.pdf}}%
    \put(0.0916574,0.01245099){\color[rgb]{0,0,0}\makebox(0,0)[lt]{\lineheight{1.25}\smash{\begin{tabular}[t]{l}($\infty, \,\xi_\Sigma$)\end{tabular}}}}%
    \put(0.85218382,0.03382645){\color[rgb]{0,0,0}\makebox(0,0)[lt]{\lineheight{1.25}\smash{\begin{tabular}[t]{l}CFT$_\mt{L}$\end{tabular}}}}%
    \put(0.85218226,0.33801964){\color[rgb]{0,0,0}\makebox(0,0)[lt]{\lineheight{1.25}\smash{\begin{tabular}[t]{l}CFT$_\mt{R}$\end{tabular}}}}%
    \put(0,0){\includegraphics[width=\unitlength,page=5]{images/xizeta_updateSecondPart.pdf}}%
    \put(0.26996581,0.14820944){\color[rgb]{0,0,0}\makebox(0,0)[lt]{\lineheight{1.25}\smash{\begin{tabular}[t]{l}$\xi$\end{tabular}}}}%
    \put(0,0){\includegraphics[width=\unitlength,page=6]{images/xizeta_updateSecondPart.pdf}}%
    \put(0.07528671,0.18320013){\color[rgb]{0,0,0}\makebox(0,0)[lt]{\lineheight{1.25}\smash{\begin{tabular}[t]{l}$\bdyReg$\end{tabular}}}}%
    \put(0.89391185,0.18319775){\color[rgb]{0,0,0}\makebox(0,0)[lt]{\lineheight{1.25}\smash{\begin{tabular}[t]{l}$\bdyReg$\end{tabular}}}}%
    \put(0.43185032,0.34984749){\color[rgb]{0,0,0}\makebox(0,0)[lt]{\lineheight{1.25}\smash{\begin{tabular}[t]{l}conformal\\\end{tabular}}}}%
    \put(0.45688497,0.31613287){\color[rgb]{0,0,0}\makebox(0,0)[lt]{\lineheight{1.25}\smash{\begin{tabular}[t]{l}defect\end{tabular}}}}%
    \put(0,0){\includegraphics[width=\unitlength,page=7]{images/xizeta_updateSecondPart.pdf}}%
    \put(0.43185032,0.02664276){\color[rgb]{0,0,0}\makebox(0,0)[lt]{\lineheight{1.25}\smash{\begin{tabular}[t]{l}conformal\\\end{tabular}}}}%
    \put(0.45688497,-0.00707277){\color[rgb]{0,0,0}\makebox(0,0)[lt]{\lineheight{1.25}\smash{\begin{tabular}[t]{l}defect\end{tabular}}}}%
    \put(0,0){\includegraphics[width=\unitlength,page=8]{images/xizeta_updateSecondPart.pdf}}%
  \end{picture}%
\endgroup%

%% file: images/xizeta_update.pdf_tex
\begingroup%
  \makeatletter%
  \providecommand\color[2][]{%
    \errmessage{(Inkscape) Color is used for the text in Inkscape, but the package 'color.sty' is not loaded}%
    \renewcommand\color[2][]{}%
  }%
  \providecommand\transparent[1]{%
    \errmessage{(Inkscape) Transparency is used (non-zero) for the text in Inkscape, but the package 'transparent.sty' is not loaded}%
    \renewcommand\transparent[1]{}%
  }%
  \providecommand\rotatebox[2]{#2}%
  \newcommand*\fsize{\dimexpr\f@size pt\relax}%
  \newcommand*\lineheight[1]{\fontsize{\fsize}{#1\fsize}\selectfont}%
  \ifx\svgwidth\undefined%
    \setlength{\unitlength}{473.38582677bp}%
    \ifx\svgscale\undefined%
      \relax%
    \else%
      \setlength{\unitlength}{\unitlength * \real{\svgscale}}%
    \fi%
  \else%
    \setlength{\unitlength}{\svgwidth}%
  \fi%
  \global\let\svgwidth\undefined%
  \global\let\svgscale\undefined%
  \makeatother%
  \begin{picture}(1,0.37125749)%
    \lineheight{1}%
    \setlength\tabcolsep{0pt}%
    \put(0,0){\includegraphics[width=\unitlength,page=1]{images/xizeta_update.pdf}}%
    \put(0.31984708,0.15195835){\color[rgb]{0,0,0}\makebox(0,0)[lt]{\lineheight{1.25}\smash{\begin{tabular}[t]{l}$\zeta_*$\end{tabular}}}}%
    \put(0,0){\includegraphics[width=\unitlength,page=2]{images/xizeta_update.pdf}}%
    \put(0.06794846,0.18683263){\color[rgb]{0,0,0}\makebox(0,0)[lt]{\lineheight{1.25}\smash{\begin{tabular}[t]{l}$\bdyReg$\end{tabular}}}}%
    \put(0.91509069,0.18683026){\color[rgb]{0,0,0}\makebox(0,0)[lt]{\lineheight{1.25}\smash{\begin{tabular}[t]{l}$\bdyReg$\end{tabular}}}}%
    \put(0.88132874,0.0332923){\color[rgb]{0,0,0}\makebox(0,0)[lt]{\lineheight{1.25}\smash{\begin{tabular}[t]{l}CFT$_\mt{L}$\end{tabular}}}}%
    \put(0.88132719,0.33748887){\color[rgb]{0,0,0}\makebox(0,0)[lt]{\lineheight{1.25}\smash{\begin{tabular}[t]{l}CFT$_\mt{R}$\end{tabular}}}}%
    \put(0,0){\includegraphics[width=\unitlength,page=3]{images/xizeta_update.pdf}}%
  \end{picture}%
\endgroup%

%% file: images/xizetaRTPhase.pdf_tex
\begingroup%
  \makeatletter%
  \providecommand\color[2][]{%
    \errmessage{(Inkscape) Color is used for the text in Inkscape, but the package 'color.sty' is not loaded}%
    \renewcommand\color[2][]{}%
  }%
  \providecommand\transparent[1]{%
    \errmessage{(Inkscape) Transparency is used (non-zero) for the text in Inkscape, but the package 'transparent.sty' is not loaded}%
    \renewcommand\transparent[1]{}%
  }%
  \providecommand\rotatebox[2]{#2}%
  \newcommand*\fsize{\dimexpr\f@size pt\relax}%
  \newcommand*\lineheight[1]{\fontsize{\fsize}{#1\fsize}\selectfont}%
  \ifx\svgwidth\undefined%
    \setlength{\unitlength}{430.86614173bp}%
    \ifx\svgscale\undefined%
      \relax%
    \else%
      \setlength{\unitlength}{\unitlength * \real{\svgscale}}%
    \fi%
  \else%
    \setlength{\unitlength}{\svgwidth}%
  \fi%
  \global\let\svgwidth\undefined%
  \global\let\svgscale\undefined%
  \makeatother%
  \begin{picture}(1,0.40789474)%
    \lineheight{1}%
    \setlength\tabcolsep{0pt}%
    \put(0,0){\includegraphics[width=\unitlength,page=1]{images/xizetaRTPhase.pdf}}%
    \put(0.30839438,0.15163422){\color[rgb]{0,0,0}\makebox(0,0)[lt]{\lineheight{1.25}\smash{\begin{tabular}[t]{l}$\RT$\end{tabular}}}}%
    \put(0.20454863,0.31508506){\color[rgb]{0,0,0}\makebox(0,0)[lt]{\lineheight{1.25}\smash{\begin{tabular}[t]{l}$\lambda_\mt{UV}$\end{tabular}}}}%
    \put(0.24455377,0.25428652){\color[rgb]{0,0,0}\makebox(0,0)[lt]{\lineheight{1.25}\smash{\begin{tabular}[t]{l}$\lambda_*$\end{tabular}}}}%
    \put(0,0){\includegraphics[width=\unitlength,page=2]{images/xizetaRTPhase.pdf}}%
    \put(0.74209921,0.24053072){\color[rgb]{0,0,0}\makebox(0,0)[lt]{\lineheight{1.25}\smash{\begin{tabular}[t]{l}horizon\end{tabular}}}}%
    \put(0,0){\includegraphics[width=\unitlength,page=3]{images/xizetaRTPhase.pdf}}%
    \put(0.86214929,0.01587918){\color[rgb]{0,0,0}\makebox(0,0)[lt]{\lineheight{1.25}\smash{\begin{tabular}[t]{l}CFT$_\mt{L}$\end{tabular}}}}%
    \put(0.86214738,0.36973261){\color[rgb]{0,0,0}\makebox(0,0)[lt]{\lineheight{1.25}\smash{\begin{tabular}[t]{l}CFT$_\mt{R}$\end{tabular}}}}%
    \put(0.05384249,0.19044193){\color[rgb]{0,0,0}\makebox(0,0)[lt]{\lineheight{1.25}\smash{\begin{tabular}[t]{l}$\bdyReg$\end{tabular}}}}%
    \put(0.91147563,0.19043932){\color[rgb]{0,0,0}\makebox(0,0)[lt]{\lineheight{1.25}\smash{\begin{tabular}[t]{l}$\bdyReg$\end{tabular}}}}%
    \put(0,0){\includegraphics[width=\unitlength,page=4]{images/xizetaRTPhase.pdf}}%
  \end{picture}%
\endgroup%

%% file: images/sphere_draft.pdf_tex
\begingroup%
  \makeatletter%
  \providecommand\color[2][]{%
    \errmessage{(Inkscape) Color is used for the text in Inkscape, but the package 'color.sty' is not loaded}%
    \renewcommand\color[2][]{}%
  }%
  \providecommand\transparent[1]{%
    \errmessage{(Inkscape) Transparency is used (non-zero) for the text in Inkscape, but the package 'transparent.sty' is not loaded}%
    \renewcommand\transparent[1]{}%
  }%
  \providecommand\rotatebox[2]{#2}%
  \newcommand*\fsize{\dimexpr\f@size pt\relax}%
  \newcommand*\lineheight[1]{\fontsize{\fsize}{#1\fsize}\selectfont}%
  \ifx\svgwidth\undefined%
    \setlength{\unitlength}{595.27559055bp}%
    \ifx\svgscale\undefined%
      \relax%
    \else%
      \setlength{\unitlength}{\unitlength * \real{\svgscale}}%
    \fi%
  \else%
    \setlength{\unitlength}{\svgwidth}%
  \fi%
  \global\let\svgwidth\undefined%
  \global\let\svgscale\undefined%
  \makeatother%
  \begin{picture}(1,0.58095238)%
    \lineheight{1}%
    \setlength\tabcolsep{0pt}%
    \put(0,0){\includegraphics[width=\unitlength,page=1]{images/sphere_draft.pdf}}%
    \put(0.01831908,0.32947584){\color[rgb]{0,0,0}\makebox(0,0)[lt]{\lineheight{1.25}\smash{\begin{tabular}[t]{l}boundary \\of $H_{d-1}$\end{tabular}}}}%
    \put(0,0){\includegraphics[width=\unitlength,page=2]{images/sphere_draft.pdf}}%
    \put(0.78415091,0.07793746){\color[rgb]{0,0,0}\makebox(0,0)[lt]{\lineheight{1.25}\smash{\begin{tabular}[t]{l}defect\end{tabular}}}}%
    \put(0,0){\includegraphics[width=\unitlength,page=3]{images/sphere_draft.pdf}}%
    \put(0.74879942,0.51866908){\color[rgb]{0,0,0}\makebox(0,0)[lt]{\lineheight{1.25}\smash{\begin{tabular}[t]{l}origin in Poincar\'e\\     coordinares\end{tabular}}}}%
    \put(-0.10402243,0.11618918){\color[rgb]{0,0,0}\makebox(0,0)[lt]{\lineheight{1.25}\smash{\begin{tabular}[t]{l}point at infinity in\\ Poincar\'e coordinates\end{tabular}}}}%
    \put(0.69023554,0.26182211){\color[rgb]{0,0,0}\makebox(0,0)[lt]{\lineheight{1.25}\smash{\begin{tabular}[t]{l}$\chi_\Sigma$\end{tabular}}}}%
  \end{picture}%
\endgroup%

%% file: images/boost_drafts.pdf_tex
\begingroup%
  \makeatletter%
  \providecommand\color[2][]{%
    \errmessage{(Inkscape) Color is used for the text in Inkscape, but the package 'color.sty' is not loaded}%
    \renewcommand\color[2][]{}%
  }%
  \providecommand\transparent[1]{%
    \errmessage{(Inkscape) Transparency is used (non-zero) for the text in Inkscape, but the package 'transparent.sty' is not loaded}%
    \renewcommand\transparent[1]{}%
  }%
  \providecommand\rotatebox[2]{#2}%
  \newcommand*\fsize{\dimexpr\f@size pt\relax}%
  \newcommand*\lineheight[1]{\fontsize{\fsize}{#1\fsize}\selectfont}%
  \ifx\svgwidth\undefined%
    \setlength{\unitlength}{771.02362205bp}%
    \ifx\svgscale\undefined%
      \relax%
    \else%
      \setlength{\unitlength}{\unitlength * \real{\svgscale}}%
    \fi%
  \else%
    \setlength{\unitlength}{\svgwidth}%
  \fi%
  \global\let\svgwidth\undefined%
  \global\let\svgscale\undefined%
  \makeatother%
  \begin{picture}(1,0.21323529)%
    \lineheight{1}%
    \setlength\tabcolsep{0pt}%
    \put(0,0){\includegraphics[width=\unitlength,page=1]{images/boost_drafts.pdf}}%
    \put(-0.00992665,0.18296696){\color[rgb]{0,0,0}\makebox(0,0)[lt]{\lineheight{1.25}\smash{\begin{tabular}[t]{l}$\tilde{t}$\end{tabular}}}}%
    \put(0.06193397,0.18523932){\color[rgb]{0,0,0}\makebox(0,0)[lt]{\lineheight{1.25}\smash{\begin{tabular}[t]{l}$\tilde{x}_2$\end{tabular}}}}%
    \put(0.06772054,0.14051552){\color[rgb]{0,0,0}\makebox(0,0)[lt]{\lineheight{1.25}\smash{\begin{tabular}[t]{l}$\tilde{x}_1$\end{tabular}}}}%
    \put(0.51876388,0.01003243){\color[rgb]{0,0,0}\makebox(0,0)[lt]{\lineheight{1.25}\smash{\begin{tabular}[t]{l}$\tau_\Sigma=0$\end{tabular}}}}%
    \put(0.06121027,0.0100312){\color[rgb]{0,0,0}\makebox(0,0)[lt]{\lineheight{1.25}\smash{\begin{tabular}[t]{l}$\tau_\Sigma\neq0$\end{tabular}}}}%
    \put(0,0){\includegraphics[width=\unitlength,page=2]{images/boost_drafts.pdf}}%
    \put(0.44783929,0.08425941){\color[rgb]{0,0,0}\makebox(0,0)[lt]{\lineheight{1.25}\smash{\begin{tabular}[t]{l}boost\end{tabular}}}}%
  \end{picture}%
\endgroup%

%% file: images/boost_drafts2.pdf_tex
\begingroup%
  \makeatletter%
  \providecommand\color[2][]{%
    \errmessage{(Inkscape) Color is used for the text in Inkscape, but the package 'color.sty' is not loaded}%
    \renewcommand\color[2][]{}%
  }%
  \providecommand\transparent[1]{%
    \errmessage{(Inkscape) Transparency is used (non-zero) for the text in Inkscape, but the package 'transparent.sty' is not loaded}%
    \renewcommand\transparent[1]{}%
  }%
  \providecommand\rotatebox[2]{#2}%
  \newcommand*\fsize{\dimexpr\f@size pt\relax}%
  \newcommand*\lineheight[1]{\fontsize{\fsize}{#1\fsize}\selectfont}%
  \ifx\svgwidth\undefined%
    \setlength{\unitlength}{805.03937008bp}%
    \ifx\svgscale\undefined%
      \relax%
    \else%
      \setlength{\unitlength}{\unitlength * \real{\svgscale}}%
    \fi%
  \else%
    \setlength{\unitlength}{\svgwidth}%
  \fi%
  \global\let\svgwidth\undefined%
  \global\let\svgscale\undefined%
  \makeatother%
  \begin{picture}(1,0.34507042)%
    \lineheight{1}%
    \setlength\tabcolsep{0pt}%
    \put(0,0){\includegraphics[width=\unitlength,page=1]{images/boost_drafts2.pdf}}%
    \put(0.01638381,0.22048514){\color[rgb]{0,0,0}\makebox(0,0)[lt]{\lineheight{1.25}\smash{\begin{tabular}[t]{l}$\tilde{t}$\end{tabular}}}}%
    \put(0.0852081,0.22783568){\color[rgb]{0,0,0}\makebox(0,0)[lt]{\lineheight{1.25}\smash{\begin{tabular}[t]{l}$\tilde{x}_2$\end{tabular}}}}%
    \put(0.09075014,0.1933087){\color[rgb]{0,0,0}\makebox(0,0)[lt]{\lineheight{1.25}\smash{\begin{tabular}[t]{l}$\tilde{x}_1$\end{tabular}}}}%
    \put(0.82620271,0.00221924){\color[rgb]{0,0,0}\makebox(0,0)[lt]{\lineheight{1.25}\smash{\begin{tabular}[t]{l}$\tau_\Sigma'\neq0$, $\chi_\Sigma'=0$\end{tabular}}}}%
    \put(0.0294155,0.05939301){\color[rgb]{0,0,0}\makebox(0,0)[lt]{\lineheight{1.25}\smash{\begin{tabular}[t]{l}$\tau_\Sigma\neq0$, $\chi_\Sigma\neq0$\end{tabular}}}}%
    \put(0.82620588,0.22086264){\color[rgb]{0,0,0}\makebox(0,0)[lt]{\lineheight{1.25}\smash{\begin{tabular}[t]{l}$\tau_\Sigma'=0$, $\chi_\Sigma'\neq0$\end{tabular}}}}%
    \put(0,0){\includegraphics[width=\unitlength,page=2]{images/boost_drafts2.pdf}}%
    \put(0.51766323,0.15838048){\color[rgb]{0,0,0}\makebox(0,0)[lt]{\lineheight{1.25}\smash{\begin{tabular}[t]{l}$\tau_\Sigma<\chi_\Sigma$\end{tabular}}}}%
    \put(0.39422473,0.07892251){\color[rgb]{0,0,0}\makebox(0,0)[lt]{\lineheight{1.25}\smash{\begin{tabular}[t]{l}$\tau_\Sigma>\chi_\Sigma$\end{tabular}}}}%
    \put(0,0){\includegraphics[width=\unitlength,page=3]{images/boost_drafts2.pdf}}%
  \end{picture}%
\endgroup%

%% file: sections/04_numerics.tex
In the previous section, we found a phase transition between the no-island and island phases that has the right qualitative properties to yield a Page curve consistent with unitarity. The calculations involved differential equations  which have no known closed form solution. However, the reader might have realized that all of these equations were ordinary differential equations and are thus easily solved numerically. In this section, we will first present numerical solutions to the equations for the RT surface in the island phase, and then use the arguments of the previous section to obtain the Page curve for massless, topological black holes in equilibrium with a bath.

\subsection{General behavior of the islands}
As discussed previously, by choosing entangling surfaces with the hyperbolic symmetry of $H_{d-2}$, the problem of finding the corresponding RT surfaces reduces to a two-dimensional problem. Choosing the convenient coordinates in eq.~\eqref{eq:new_coords_non_extremal}, we can express the profile of the RT surface as $\zeta(\xi)$. We start here by discussing examples of extremal surfaces in the island phase for different choices of parameters. Instead of working with $\zeta$ as a radial coordinate, we conformally compactify the geometry and use the coordinate 
\begin{align}
\label{eq:conf_radius}
\varrho = \arctan(\zeta)\,,
\end{align}
which maps timeslices of AdS to a finite region.
In order to calculate the profile of the RT surface, we fix the location of the entangling surface $\chi_\Sigma$ at the boundary. Applying the large $r$ limit of eq.~\eqref{eq:new_coords_non_extremal}, we relate this to $\xi_\Sigma$, the location of the entangling surface in $\zeta,\xi$ coordinates. We can then use eqs.~\eqref{eq:condition_1} and \eqref{eq:condition_2} to determine $\zeta_*$ and $\zeta_\QES$ numerically as a function of $\xi_\Sigma$. The shape of the RT surface is obtained by integrating eq.~\eqref{eq:first_order_RT_surface} from the boundary. 

\begin{figure}[hp]
\centering
\begin{subfigure}[t]{\textwidth}
\centering
\includegraphics[width=0.9\textwidth]{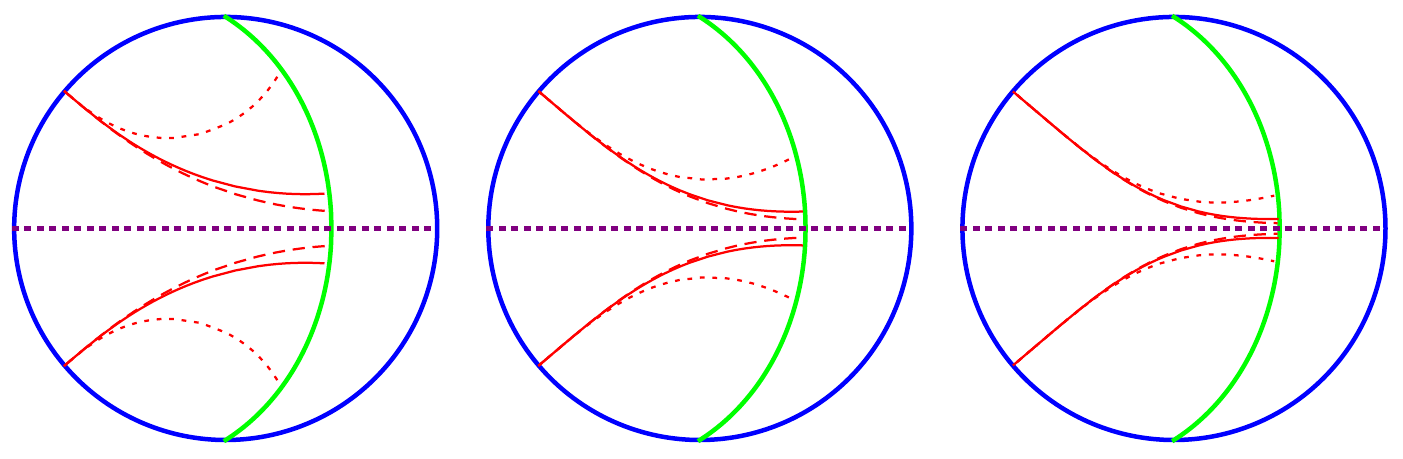}
\caption{RT surfaces for the island phase in (left to right) $d = 3,4,5$. The DGP coupling $\DGPRatio$ is chosen to be $1$/$0$/$-0.9$ for the dashed/solid/dotted curves. The brane angle is $\braneAngle = \frac \pi 4$ and the location of the entangling surface is $\chi_\Sigma = 1$.}
\label{fig:islands_varying_d}
\end{subfigure}
\par\bigskip
\begin{subfigure}[t]{\textwidth}
\centering
\includegraphics[width=0.9\textwidth]{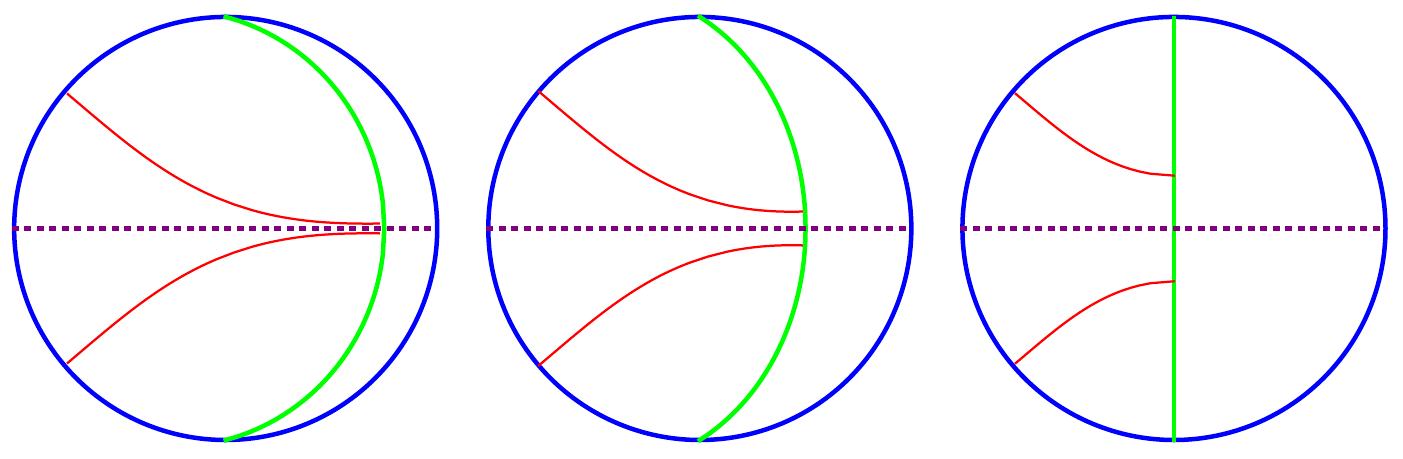}
\caption{RT surfaces in $d = 4$ with $\chi_\Sigma = 1$ and brane angle of (left to right) $\braneAngle = \frac 1 8 \pi, \frac 1 4 \pi, \frac 1 2 \pi$. The DGP coupling is set to zero.}
\label{fig:islands_varying_tension}
\end{subfigure}
\par\bigskip
\begin{subfigure}[t]{\textwidth}
\centering
\includegraphics[width=0.9\textwidth]{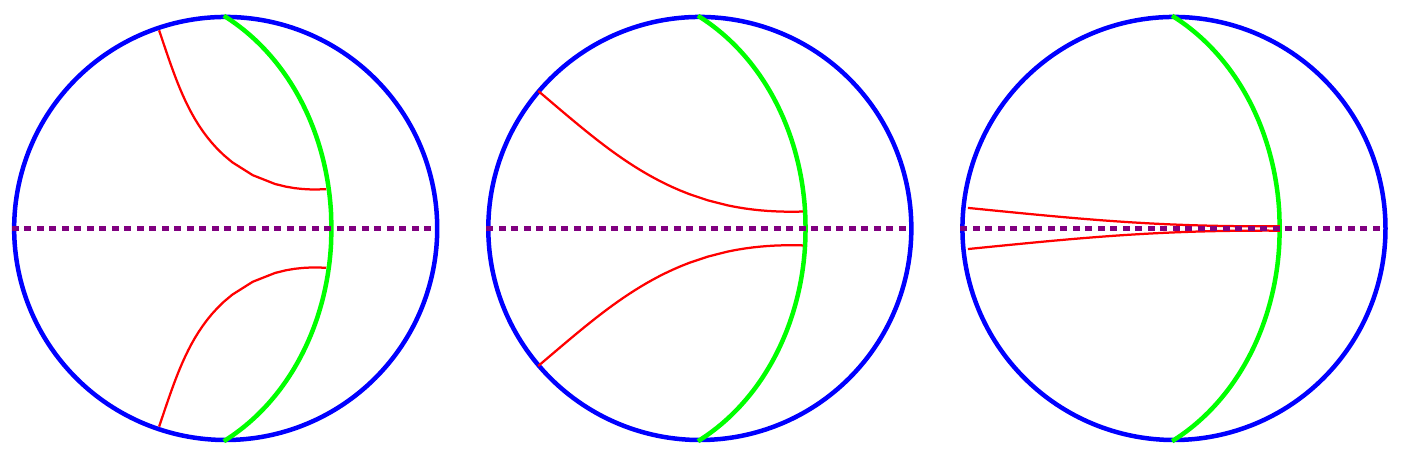}
\caption{RT surfaces in $d = 4$ with brane angle $\braneAngle = \frac \pi 4$ and (left to right) $\chi_\Sigma = \frac 1 3, 1, 3$. The DGP coupling is set to zero.}
\label{fig:islands_varying_belt_width}
\end{subfigure}
\caption{RT surfaces in the island phase in higher dimensions. We only show one side of the brane. The asymptotic boundary of the spacetime is shown in blue, the Planck brane in green and the RT surfaces in red. The radial coordinate is $\varrho$ defined in eq.~\eqref{eq:conf_radius}. On each side of the horizon (dashed purple line) the angular coordinate $\xi$ runs between $-\frac \pi 2$ and $\frac \pi 2$.}
\label{fig:islands}
\end{figure}

Figure \ref{fig:islands} shows a few examples of RT surfaces in the connected phase for $d = 3$, 4 and 5, \ie in four, five and six bulk dimensions, respectively. Here, we only show the geometry on one side of the brane. The other side is determined by a reflection across the brane. Since the RT surfaces do not cross the horizon, the configuration is independent of the choice of Rindler time $\tau$. 

Figure \ref{fig:islands_varying_d} shows RT surfaces with fixed $\chi_\Sigma$ for different values of the dimension and selected values of the DGP coupling $\DGPRatio$. We can see that positive DGP coupling pushes the point of intersection between brane and RT surface towards the horizon, \ie it reduces the area of the island's boundary. Similarly, negative DGP coupling causes the island to become bigger. This behaviour is readily explained through eq.~\eqref{Newton3} which shows that by increasing (decreasing) the value of $\DGPRatio$, the gravitational coupling in the brane theory, \ie the effective Newton's constant, becomes smaller (bigger). In turn, the coefficient of the Bekenstein-Hawking contribution is bigger (smaller) in the island rule \reef{eq:islandformula} and therefore creating an island of fixed size becomes harder (easier). 

Figure \ref{fig:islands_varying_tension} shows how the RT surface in the island phase behaves as we vary the brane angle given by $\sin \braneAngle = \LAdS/\ell_\mt{B}$ (or equivalently the brane tension -- see eq.~\reef{curve2}). Recall that Einstein gravity is a good approximation when $\braneAngle$ is small. As we depart from the limit of small brane angle, the island grows. 

Finally, figure \ref{fig:islands_varying_belt_width} shows that the size of the island varies with $\chi_\Sigma$, the location of the entangling surface in the bath. Moreover, as we will discuss momentarily, we see that an island phase for the RT surface seems to exist for all values of the belt width, although of course it will generally not dominate at early times. 
\begin{figure}[h!]
\centering
\vspace{2cm}
\begin{subfigure}[t]{\textwidth}
\includegraphics[width=\textwidth]{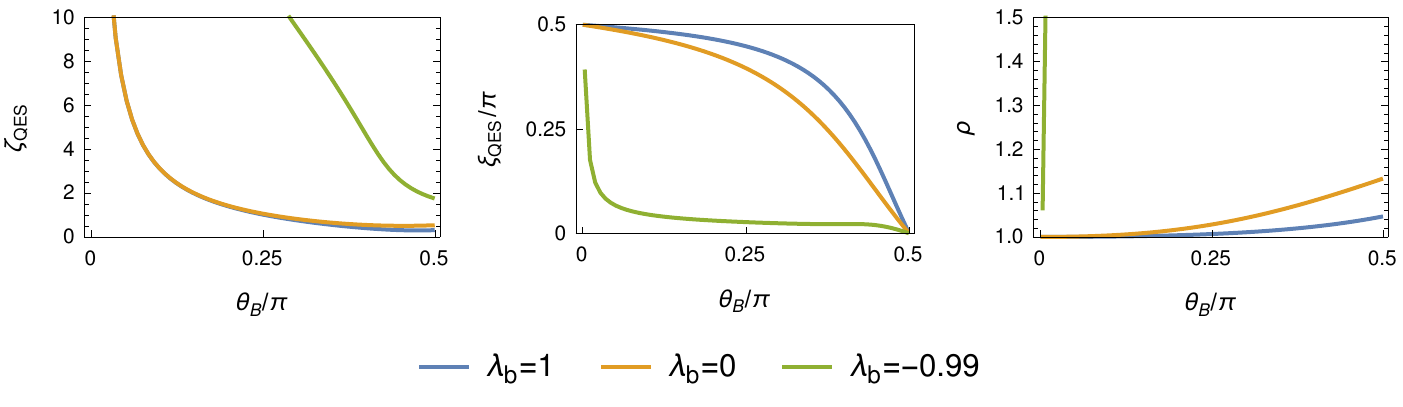}
\caption{The dependence of RT surface parameters on the brane angle $\braneAngle$ for $d=3$.}
\label{fig:detailed_islands_3d}
\end{subfigure}
\par\bigskip\bigskip
\begin{subfigure}[t]{\textwidth}
\includegraphics[width=\textwidth]{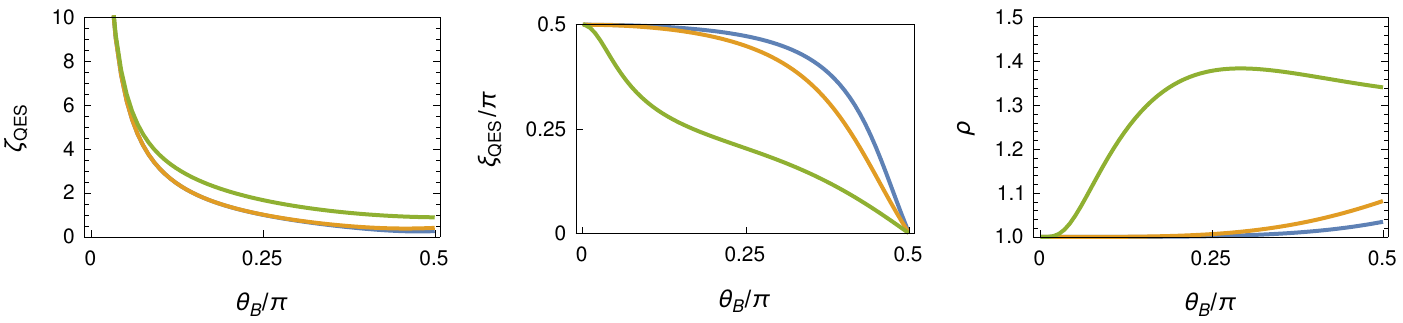}
\caption{The dependence of RT surface parameters on the brane angle $\braneAngle$ for $d=4$.}
\end{subfigure}
\par\bigskip\bigskip
\begin{subfigure}[t]{\textwidth}
\includegraphics[width=\textwidth]{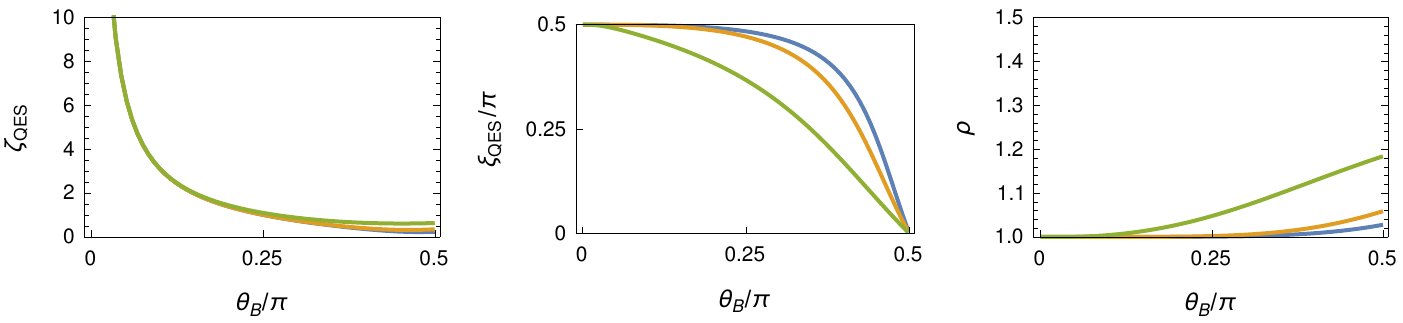}
\caption{The dependence of RT surface parameters on the brane angle $\braneAngle$ for $d=5$.}
\end{subfigure}
\par\bigskip
\caption{The dependece of the RT surface and the quantum extremal surface on the brane angle $\braneAngle$ for $d=3$, $4$ and $5$. The location of the entangling surface is chosen to be $\chi_\Sigma = 1$.}
\label{fig:detailed_islands}
\end{figure}

We can get an even better idea of the qualitative features of the islands in higher dimensions by plotting the turning point $\zeta_*$ and the QES position $(\zeta_\QES,\xi_\QES)$ as a function of the brane angle $\braneAngle$ for different dimensions -- see figure \ref{fig:detailed_islands}. A general feature is that in the $\braneAngle \to 0$ limit, the QES always approaches the horizon on the brane at $\xi=\pi/2$, as discussed around eq.~\eqref{eq:small_angle_zeta}. In terms of $\xi_\QES$ and the distance from the horizon on the brane, $\rho_\QES$, we have
\begin{align}
	\xi_\QES =& \frac{\pi}{2} - \frac{\zeta_* (1+\zeta_*^2)^{\frac{d-2}{2}}}{1+\DGPRatio} \braneAngle^{d-2} + O(\braneAngle^d) \label{eq:xi_small_angle}\\
	\rho_\QES =& 1 + \frac{\zeta_*^2 (1+\zeta_*^2)^{d-2}}{2 (1+\DGPRatio)^2} \braneAngle^{2(d-2)} + O(\braneAngle^{2(d-1)}) \label{eq:varrho_small_angle},
\end{align}
where the first terms on the RHSs give the location of the horizon. Granted $\zeta_*$ tends towards a finite value as $\braneAngle \to 0$, the above formulas tell us that the QES tends towards the horizon on the brane. Applying eq.~\eqref{eq:xi_small_angle} to eq.~\eqref{eq:condition_2} and noting from eq.~\eqref{eq:first_order_RT_surface} that
\beq
	\frac{d\zeta}{d\xi} \sim \mp \frac{\zeta^{d+1}}{\zeta_*(1+\zeta_*^2)^{\frac{d-2}{2}}} \qquad\text{for }\zeta\gg 1\,,
\eeq
we find that $\zeta_*$ at small $\braneAngle$ is determined by the equation
\begin{align}
	\frac{\pi}{2}
	- \frac{
	\zeta_* (1+\zeta_*^2)^{\frac{d-2}{2}}
	}{
	1+\DGPRatio
	}
	\braneAngle^{d-2}
  + O(\braneAngle^d)
  =& \xi_\QES
	= \xi_{\Sigma} + 2\int_{\zeta_*}^\infty d\zeta\; \left|\frac{d\zeta}{d\xi}\right|^{-1}
	+ O(\braneAngle^d),
\end{align}
with $d\zeta/d\xi$ given by eq.~\eqref{eq:first_order_RT_surface}. At leading order in $\braneAngle$, the second term on the LHS can be ignored and the above equation is just the statement that the RT surface should stretch from the belt boundary to approximately the bifurcation surface on the brane.


\subsection{The Page curve in $d > 2$}
\label{sec:page_curve_in_d_gtr_two}

As discussed in section \ref{sec:nonextremal}, the benefit of our model is that
calculating the entropy of (the complement of) the belt-shaped subregions
centered on the conformal defects reduces to calculating areas in an effectively
two-dimensional geometry. Further we produced explicit formulas for the areas of
a number of special RT surfaces, which -- as shown in section
\ref{sec:time_evolution_non_extremal} -- are sufficient to calculate the full
time evolution of the RT surfaces and thus of the entanglement entropy.

\begin{figure}
  \centering
  \begin{subfigure}[t]{\textwidth}
    \includegraphics[width=\textwidth]{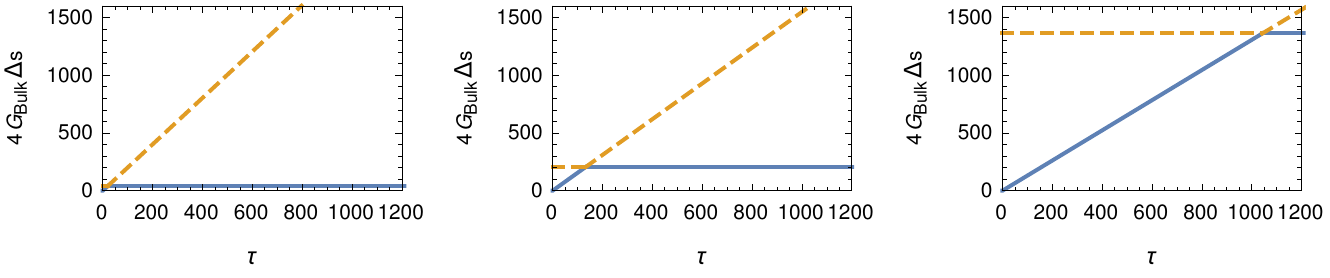}
    \caption{The Page curve for dimensions $d = 3,4,5$ (left to right). The
      entangling surface is located at $\chi_\Sigma = 1$ and the DGP coupling is
      set to zero. The brane angle is chosen as $\braneAngle = 0.1$.}
    \label{fig:page_curve_varying_d}
  \end{subfigure}
  \par\bigskip
  \begin{subfigure}[t]{\textwidth}
    \includegraphics[width=\textwidth]{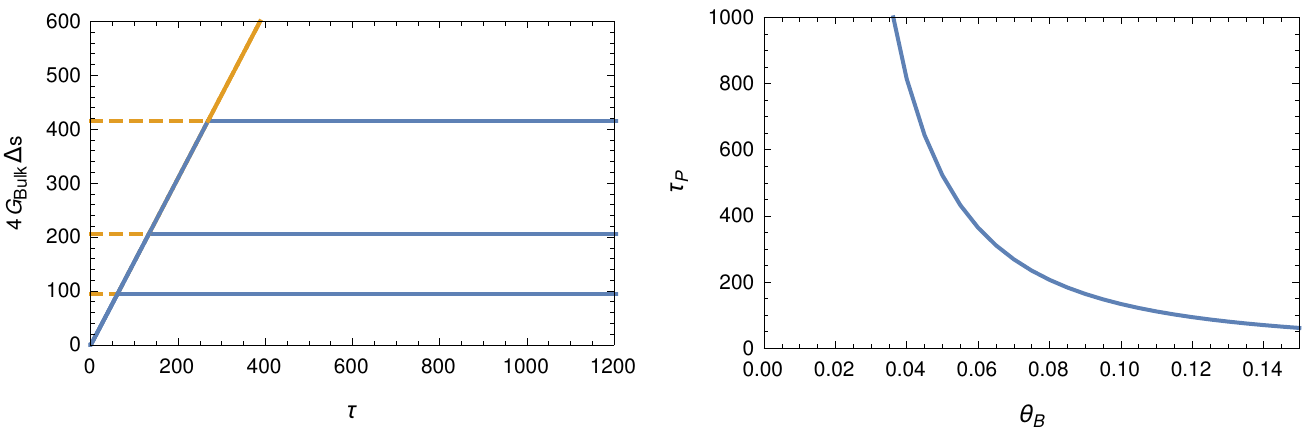}
    \caption{Left: The Page curve for selected brane angles $\braneAngle= 0.07,
      0.10, 0.15$ (top to bottom). Right: The Page time $\tau_P$ as a function
      of the brane angle $\braneAngle$. The constant parameters are set to
      $\DGPRatio = 0$, $\chi_\Sigma = 1$, and $d=4$.}
    \label{fig:page_curve_varying_angle}
  \end{subfigure}
  \par\bigskip
  \begin{subfigure}[t]{\textwidth}
    \includegraphics[width=\textwidth]{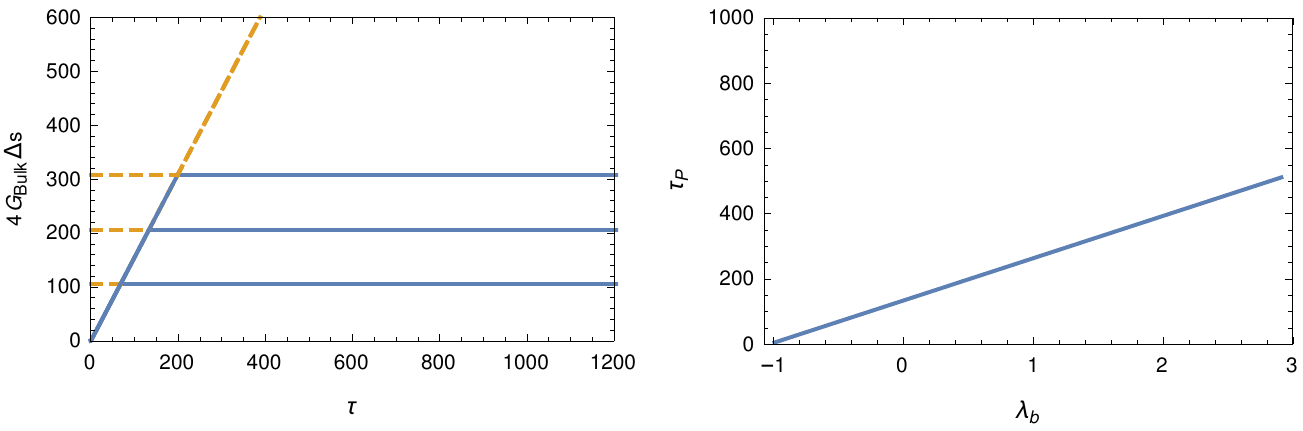}
    \caption{Left: The Page curve for selected values of the DGP coupling
      $\DGPRatio = 0.5, 0, -0.5$ (top to bottom). Right: The Page time $\tau_P$
      as a function of the DGP coupling $\DGPRatio$. The constant parameters are
      set to $\braneAngle = 0.1$, $\chi_\Sigma = 1$, and $d=4$.}
    \label{fig:page_time_varying_dgp}
  \end{subfigure}
  \caption{The Page curve in various dimensions. The solid blue line indicates
    the physical Page curve. The dashed orange lines correspond to entropies
    calculated by non-minimal extremal surfaces. At early times, the RT surface
    in the no-island phase is the minimal surface. After some time, the minimal
    surface transitions to the RT surface in the island phase. }
  \label{fig:page_curves}
\end{figure}

Figure \ref{fig:page_curve_varying_d} shows the Page curves for $d$-dimensional
topological black holes, coupled to a bath on a hyperbolic background, for the
cases $d = 3,4,5$. More precisely, we consider the entropy of the region defined
by $\chi_\Sigma = 1$, which is given by
\begin{align}
	\label{eq:page_curve_entropy}
	4 G_\text{bulk} S (\tau) = \min \left( \left[ \area(\RT) + \frac{2 \LAdS \DGPRatio}{(d-2)} \area(\RTbrn) \right]_\island\!\!, \, \left[\area(\RT)\right]_{\noisland} \right)	\,.
\end{align}
Here $\area(\Sigma)$ are the regulated areas of the RT surfaces, and the
subscript indicates whether we consider the extremal surface in the island or
no-island phase. Since eq.~\eqref{eq:page_curve_entropy} is a cutoff dependent
quantity, it is convenient to subtract off $\left[\area(\RT)\right]_{\noisland,
  \tau=0}$. That is, we subtract off the value of the entropy at $\tau = 0$, at
which point the minimal RT surfaces in the no-island phase, to define
\begin{align}
  \Delta S(\tau) = S(\tau) - S(\tau = 0).
\end{align}
Even though the UV divergences have been removed,
eq.~\eqref{eq:page_curve_entropy} would still be infinite, as a result of the
infinite extend of the entangling surface. Hence the plots in figure
\ref{fig:page_curves} show the change in the entropy density,
\begin{align}
	\Delta s = \frac {\Delta S}{\vol_{H_{d-2}} L^{d-2}}\,,
\end{align}
with respect to the entropy at $\tau = 0$.\footnote{Note that we are actually
  plotting $4 G_\text{bulk}\,\Delta s$, which is a dimensionless quantity. For
  the horizontal axes, also recall that the AdS-Rindler time $\tau$ is also
  dimensionless -- see further comments below.} The kinks in the plots of figure
\ref{fig:page_curves} indicate the time at which the island phase of the RT
surface begins to dominate. The corresponding time is, of course, the natural
analog of the Page time for eternal black holes coupled to a bath at finite
temperature. The slope of the (linearly) rising portion of the Page curve has
been determined in section \ref{sec:no_island_phase_2_non_extremal} and is given
by
\begin{align}
  \label{eq:late_time_slope}
  4 G_N \Delta s / \tau \sim 4 \frac{(d-2)^{(d-2)/2}}{(d-1)^{(d-1)/2}}\,.
\end{align}
Moreover, recall that $\tau$ is a dimensionless time such that the temperature
of the hyperbolic black hole is $\frac 1 {2\pi}$ (cf. the discussion in section
\ref{sec:geometry_on_brane_non_extremal}). The dimensionful time $t$ is related
to $\tau$ by
\begin{align}
  t = \tau R = \frac{\tau}{2 \pi T} ,
\end{align} 
where $R$ is the curvature scale for the spatial sections in the bath CFT, as
defined in eq.~\eqref{bmetric}, and the bath CFT is taken at temperature $T =
\frac{1}{2\pi R}$.

The calculation of the RT surfaces is performed as follows: the area in the
island phase is computed by substituting eqs.~\eqref{eq:condition_1} and
\eqref{eq:brane_trajectory_2} into eq.~\eqref{eq:condition_2} and numerically
solving for $\zeta_\QES$. The result is then used together with
eq.~\eqref{eq:condition_1} to numerically integrate the area in
eq.~\eqref{eq:RT_area}. There are three different regimes for the calculation of
the are in the \emph{no-island phase}. At early times, $\tau_\Sigma \leq
\chi_\Sigma$, the calculation of the entropy of the subregion with boundaries at
$\pm \chi_\Sigma$ can be translated to the calculation of the entropy of a belt
with boundary $\chi_\Sigma' = \pm \text{arccosh}\left(\frac{\cosh
    \chi_\Sigma}{\cosh \tau_\Sigma}\right)$ in the $\tau = 0$ time-slice, as
explained in section \ref{sec:time_evolution_non_extremal}. As also explained in
the same section, we need choose a different cutoff on $r$ in this case.
However, working in $\zeta, \xi$ coordinates, it turns out that the cutoff on
$\zeta$ does not change. At intermediate times, $\tau_\Sigma \gtrsim
\chi_\Sigma$, the entropy can be computed by calculating the area of an RT
surface for a zero-belt-width entangling surface at a time given in
eq.~\eqref{eq:new_time}. Accidentally, the relation between $r$ and $\zeta$
works out in such a way that the cutoff of $r$ agrees with the cutoff on $\zeta$
in the previous calculation. As $\tau_\Sigma$ becomes larger, the numerics
become less reliable. However, for moderately sized belt widths we are already
well into the regime in which the area of the RT surface grows linearly in time.
Therefore, we use a linear fit to extrapolate the last few numeric data points
to late times, $\tau_\Sigma \gg \chi_\Sigma$. We verified that the resulting
slope agrees with the analytic result given in eq.~\eqref{eq:late_time_slope}.

In figure \ref{fig:page_curve_varying_angle}, we show how the Page curve and
Page time change as we vary the brane angle. As we see, increasing $\braneAngle$
decreases the Page time, or in other words decreases the number of microstates
available to the black hole on the brane. This can also be understood from the
CFT point of view where the defect entropy is given in terms of an RT surface in
the island phase \cite{Tak11,Fujita:2011fp}. As the brane angle approaches zero,
the Page time diverges. The reason is that in this limit the area of the island
diverges. The absence of islands in this limit was already noted in
\cite{Geng:2020qvw}. The divergence as $\braneAngle \to 0$ goes like
$\braneAngle^{2-d}$, and in the small-angle approximation we find that
\begin{align}
  \tau_P \sim \frac{(d-1)^{\frac {d-1} 2}}{(d-2)^{\frac {d} 2}} \frac{1}{\braneAngle^{d-2}}\,.
\end{align}
For example, the numerical coefficient which multiplies $\braneAngle^{2-d}$ can
be estimated from the above formula to be $1.30$ for $d=4$. A fit to the
numerical data plotted in figure \ref{fig:page_curve_varying_angle} agrees with
this value.

Figure \ref{fig:page_time_varying_dgp} shows the dependence of the Page curve
and Page time on the DGP coupling. As we decrease the DGP coupling (\ie increase
$G_\mt{eff}$) the Page time goes to zero. The linearity can be easily explained
be recalling that in the small $\braneAngle$ regime we are interested in the
island sits close to the horizon and thus has a fixed location for varying
values of $\lambda_b$. The Page transition occurs whenever the area of the RT
surface in the no-island phase exceeds the area of the RT surface in the island
phase. Since the area in the no-island phase approximately grows linearly with
time and the area in the island phase depends approximately linearly on
$\DGPRatio$, c.f. eq.~\eqref{eq:page_curve_entropy}, we obtain a linear
relationship between the Page time $\tau_P$ and $\DGPRatio$. Based on this
argument, we can estimate the slope of the graph to be
\begin{align}
  \tau_P / \DGPRatio \sim \frac{(d-1)^{\frac {d-1} 2}}{(d-2)^{\frac {d} 2}} \frac{1}{\braneAngle^{d-2}} \,,
\end{align}
which for the parameters in \ref{fig:page_time_varying_dgp} (\ie $\braneAngle =
0.1$ and $d=4$) evaluates to $\tau_P \sim 130 \, \DGPRatio$ and agrees with the
fitted value of the slope.

\begin{figure}
  \centering
  \begin{subfigure}[t]{0.45\textwidth}
    \includegraphics[width=\textwidth]{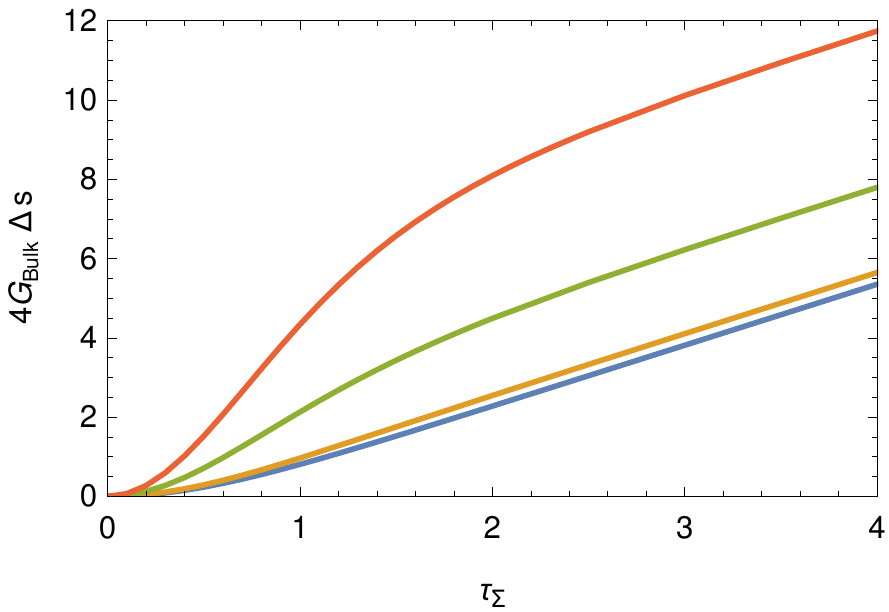}
    \caption{This figure shows the onset of the Page curve for different values
      of the location of the entangling surface $\chi_\Sigma = 0.1, 1, 2, 2.5$
      (bottom to top) in four dimensions.}
    \label{fig:page_curve_vs_belt_width}
  \end{subfigure}
  \hspace{0.05\textwidth}
  \begin{subfigure}[t]{0.45\textwidth}
    \includegraphics[width=\textwidth]{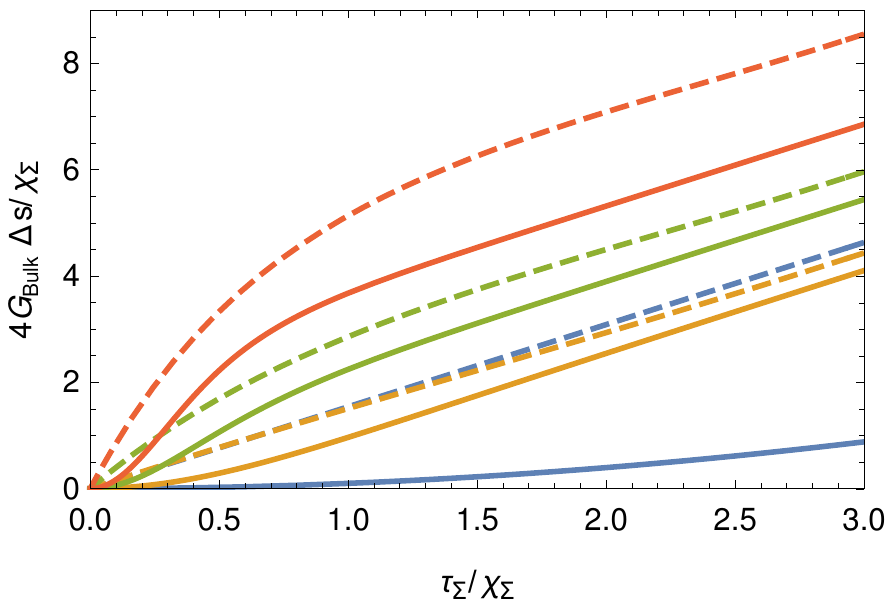}
    \caption{The same plot, but with axes rescaled by $\chi_\Sigma$. The solid
      lines are numerical results, while the dashed lines are the bounds explained
      in the main text.}
    \label{fig:page_curve_bounds}
  \end{subfigure}
  \caption{The initial behaviour of the Page curve in four dimensions (left) and
    a rescaled version of the same plot with bounds (dashed) on the onset
    (right).}
\end{figure}
The Page curve and Page time only depends very weakly on the belt size. In fact,
the only significant effect can be seen at very early times of the evaporation.
Figure \ref{fig:page_curve_vs_belt_width} shows that for wide belts, the
entanglement between the belts and baths starts growing convexly (\ie
$\partial^2\Delta s/\partial\tau_\Sigma^2>0$), then enters a period of concave
growth (\ie $\partial^2\Delta s/\partial\tau_\Sigma^2<0$) before entering the
linear regime.

Generally, we can separate the time-dependence of the
Page curve into four different regimes. At times of the order of the thermal
scale $\beta$ ($\sim 0.16$ in figure \ref{fig:page_curve_vs_belt_width}) the
entanglement growth increases until it enters a phase of fast growth between
$\tau_\Sigma \sim O(\beta)$ and $\tau_\Sigma \sim O(\chi_\Sigma)$. This fast
growth depends on the belt size. At time $\tau_\Sigma \sim O(\chi_\Sigma)$ a
universal, linear behavior takes over, which is independent of the belt width.
The entanglement keeps growing until at the Page time $\tau_P$ it saturates and
stays constant.

In the following we will explain the region of fast growth and its transition
into the region of universal linear growth. To understand the behaviour of the
Page curve, first consider a few characteristics of our belt geometries. As can
be seen from the metric in eq.~\eqref{eq:metric_bdry_slicing}, points on any of
our entangling surfaces are a fixed distance $\chi_\Sigma$ from the surface at
$\chi = 0$, where the defect is located, \ie{}where the bath is coupled to the
black hole.\footnote{The proper distance would be $R\chi_\Sigma$ in the boundary
  metric \reef{bmetric}.} However, the extrinsic curvature of the entangling
surfaces which we consider depends on this distance. Similarly, the entangling
surfaces with larger $\chi_\Sigma$ have a larger regulated volume.

In \cite{Mezei:2016wfz}, it was proposed that the growth of entropy $S[\Sigma]$
for an arbitrary entangling surface $\Sigma$ is bound by
\begin{align}
  \label{eq:entanglement_spreading}
  \frac 1 R \frac{dS[\Sigma]}{d\tau}
  =
  \frac{dS[\Sigma]}{dt}
  \le& s_{\therm}\, v_{\ent}\, \area(\Sigma) \,,
\end{align}
where $\area(\Sigma)$ is the area of the entangling surface $\Sigma$, as measured by the
boundary metric in eq.~\eqref{eq:metric_bdry_slicing}. The thermal entropy density $s_{\therm}$ and the
entanglement velocity $v_{\ent}$ are region independent constants. The entropy
density is given by the black hole entropy (\ie{}$\frac{1}{4 G_\bulk}$ times
horizon area) divided by the CFT volume of the spatial slices (again, measured
by the metric \eqref{eq:metric_bdry_slicing}):
\begin{align}
   s_{\therm} = \frac{1}{4 G_\bulk}\,\frac{L^{d-1}}{R^{d-1}}.
\end{align} 

In
\cite{Mezei:2016wfz} which primarily considers flat
space, $v_{\ent}$ is defined such that
eq.~\eqref{eq:entanglement_spreading} is saturated at times just above the
thermal scale for sufficiently straight entangling surfaces -- this definition is well-defined in the sense that $v_{\ent}$ turns out to be
independent of the shape of the entangling surface, provided it is sufficiently
straight \cite{Liu:2013iza,Liu:2013qca}. In hyperbolic space, $v_{\ent}$
can be similarly defined by demanding that the straight
surface $\chi=0$ saturates eq.~\eqref{eq:entanglement_spreading} -- we shall
justify this choice further below -- specifically,
\begin{align}
 v_{\ent} =& \frac{(d-2)^{\frac{d-2}{2}}}{(d-1)^{\frac{d-1}{2}}}\,,
             \label{eq:andDesertYou}
\end{align}
obtained by comparison of eq.~\eqref{eq:entanglement_spreading} with the
zero-width belt result in eq.~\eqref{carrot2}.

It is clear that \eqref{eq:entanglement_spreading} cannot be tight at late times for belts of finite width. The reason is that the area factor on the right hand side $\area[\chi>0]$ is
exponentially large compared to $\area[\chi=0]$, while, as can be seen from figure \ref{fig:page_curve_vs_belt_width}, all belts share the
same rate of entanglement growth at late times. To more tightly bound the late time behavior of finite width belts, we will therefore need to combine eq.~\eqref{eq:entanglement_spreading} with the monotonicity of mutual information. It will turn out that the optimal bound obtained
in this way for finite-width belts uses eq.~\eqref{eq:entanglement_spreading}, but always evaluated for the
$\chi=0$ surface $\Sigma$ at late times; thus we will find that the
$\chi=0$ surface acts as a bottleneck for entanglement growth even for finite
width belts.

To see why the surface at $\chi_\Sigma = 0$ acts as a bottleneck, let us formulate the more refined combined bound
now, following closely \cite{Mezei:2016wfz}. To this end, it will be less helpful to
consider the entanglement entropy of the bath intervals $\bdyReg$;
instead we will consider their complement $\bar \bdyReg$, \ie{}belts surrounding
the defects, whose entropy is the
same as $\bdyReg$ since the state of both Rindler patches is pure. Considering $\bar \bdyReg$
instead of $\bdyReg$ is equivalent to looking at the Page curve of the black
hole instead of that of the radiation. It is useful to rewrite the entropy
displayed in the Page curve as
\begin{align}
  \Delta S =  I[\bar \bdyReg_L: \bar \bdyReg_R](0) - I[\bar \bdyReg_L: \bar \bdyReg_R](\tau),
\end{align}
where $I[\bar \bdyReg_L:\bar \bdyReg_R](\tau) = S(\bar \bdyReg_L) + S(\bar
\bdyReg_R) - S(\bar \bdyReg_L \cup \bar \bdyReg_R)$ is the mutual information
between the regions $\bar \bdyReg$ in the left ($L$) and right ($R$) CFT at time
$\tau$.

Similar to \cite{Mezei:2016wfz}, we now assume that information is only
transported with the butterfly velocity $v_\but$ or less.\footnote{The butterfly velocity
  is defined as the spread of the region in which the commutator of an operator
  $O_1(t)$ with $O_2(t)$ is bigger than $1$ \cite{RobSta15}.} For the hyperbolic
geometries considered here and the temperature $T=\frac{1}{2\pi R}$, this
velocity is given by \cite{Perlmutter:2016pkf,Ahn:2019rnq}
\begin{align}
	\label{eq:butterfly_velocity}
  v_{\but} = \frac{1}{d-1}\,.
\end{align}
This implies that a belt region $\bar{\bdyReg}'$ at time $\tau'$
can be considered a subsystem of the original belt $\bar \bdyReg$ at $\tau$
if $\frac{\chi_\Sigma-\chi_\Sigma'}{v_\but} \geq |\tau-\tau'|$. We can then
use monotonicity of mutual information
\begin{align}
  I[\bar \bdyReg_L:\bar \bdyReg_R](\tau) \geq I[\bar \bdyReg'_L:\bar \bdyReg'_R](\tau') = S[\bar \bdyReg'_L](\tau') + S[\bar \bdyReg'_R](\tau') - S[\bar \bdyReg'_L \cup \bar \bdyReg'_R](\tau').
\end{align}
In our setup, we have that the one-sided entropies are time-independent, $S[\bar
\bdyReg'_{R/L}](\tau') = S[\bar \bdyReg'_{R/L}](0)$. Using eq.
\eqref{eq:entanglement_spreading} we can then bound $S[\bar \bdyReg'_L \cup \bar
\bdyReg'_R](\tau')$ from above
\begin{align}
  S[\bar \bdyReg'_L \cup \bar \bdyReg'_R](\tau') \leq R s_{\therm}\, v_{\ent}\, \area(\partial \bar \bdyReg')\, \tau' + S[\bar \bdyReg'_L \cup \bar \bdyReg'_R](0).
  \label{eq:meow}
\end{align} 
Collecting everything, we find a bound on the Page curve of the black hole,
\begin{align}
  \label{eq:final_inequality}
  \Delta S[\bar \bdyReg_L \cup \bar \bdyReg_R] \leq R s_{\therm}\, v_{\ent}\, \area(\partial \bar \bdyReg')\, \tau'  + I[\bar \bdyReg_L: \bar \bdyReg_R](0) - I[\bar \bdyReg'_L: \bar \bdyReg'_R](0)\,.
\end{align}
To find a tightest bound this has to be minimized over all choices of $\chi'_\Sigma$, see below. For any fixed $\chi'_\Sigma$ it is sufficient to focus on the case where $\tau' < \tau$, which will always give the smaller bound.
The mutual information appearing on the right hand side are
evaluated on the initial time slice and can be obtained numerically by
using the results of section \ref{sec:no_island_phase_1_non_extremal}.

From eq.~\eqref{eq:final_inequality}, it is now easy to see why the entanglement
growth becomes universal at late times. Note that eq.~\eqref{eq:final_inequality} is in fact a
family of inequalities, parametrized by a choice of regions $\bar \bdyReg'$. The
time $\tau'$ is chosen such that $\bar \bdyReg'$ at $\tau'$ is just
barely a subsystem of $\bar \bdyReg$ at $\tau$, in the sense described below eq.~\eqref{eq:butterfly_velocity}. For times before $\tau'$ we assume
that the mutual information of subregions $\bar \bdyReg'$ is allowed to decrease
as fast as possible, while still compatible with
eq.~\eqref{eq:entanglement_spreading}. Since the regions $\bar \bdyReg'$ at time
$\tau'$ are subregions of $\bar \bdyReg$ at time $\tau$, their mutual
information bounds the mutual information of regions $\bar \bdyReg$. We can find
a tight bound on the Page curve by minimizing over all choices of $\bar
\bdyReg'$, or in other words, by minimizing over all $\chi'_\Sigma$ with $\tau'
= \tau - \frac{\chi_\Sigma - \chi_\Sigma'} {v_\but}$. It turns out that, for
sufficiently large $\tau$, the tightest bound is obtained for $\chi_\Sigma' = 0$,
yielding the prescription stated below eq.~\eqref{eq:andDesertYou}. We thus see
from the first term on the right hand side of eq.~\eqref{eq:final_inequality}
that this surface acts as a bottle neck for information transfer and thus
controls the late time growth of entropy.  Matching this behaviour to the late time rate of growth of the exact Page curve provides further justification,
\textit{a posteriori}, for the choice of the entanglement velocity stated in eq.~\eqref{eq:andDesertYou}.

The bounds found in this way are
presented in figure \ref{fig:page_curve_bounds}. We see that a fast growth at early times is allowed by the bounds, before the linear growth
phase is entered. Further, as can be seen from the figure, these bounds are
fairly loose. It would be interesting to understand how to make them
tighter.\label{par:plots} Note that the blue curve in figure
\ref{fig:page_curve_bounds} behaves qualitatively different than the other
curves. The reason is that the early convex onset of the curve is controlled by
the thermal scale and thus lasts for roughly $\Delta \tau \sim O(\beta)$,
independent of the belt width. The rescaling in figure
\ref{fig:page_curve_bounds} magnifies the early time behavior of belts with
$\chi_\Sigma < 1$ while it reduces the early time behavior of belts of width
$\chi_\Sigma < 1$. Thus, while all other curves show the linear entanglement
spreading for time scales $\tau \sim O(\chi_\Sigma) > O(\beta)$, the behavior of
the blue curve is dominated by entanglement spreading through thermalization,
since the belt width is of order of the thermalization scale. The
quadratic\footnote{Note that time-reflection symmetry demands that the Page
  curve have an early time expansion containing only even powers of $\tau$. For
  the zero-width, it is easily verified, at least numerically, from
  eq.~\eqref{vbound3} that $\sqrt{1-r_*} \sim \tau$ so that the growth is indeed
  quadratic by eq.~\eqref{simple0}. For finite-width belts,
  plugging
  eqs.~\eqref{eq:b_prime} and \eqref{eq:cutoff_case_1} into
  eq.~\eqref{eq:new_coords_non_extremal} shows that early time evolution is
  equivalent to holding the cutoff at fixed $\zeta$ and shifting the $\xi$ of the
  entangling surface by $\sim \tau^2$, again leading to quadratic entanglement growth.} growth at times below the thermal scale is reminiscent of the
`pre-local-equilibration growth' described in \cite{Liu:2013iza,Liu:2013qca}. 

Let us end with a few observations regarding the structure of entanglement
spreading in our system. First, we note that the entanglement velocity
\eqref{eq:andDesertYou} for Rindler spacetime with hyperbolic spatial slices
differs from the analogous velocity
$\sqrt{d}(d-2)^{\frac{1}{2}-\frac{1}{d}}/[2(d-1)]^{1-\frac{1}{d}}$ in flat space
\cite{Hartman:2013qma} dual to AdS planar black holes. Furthermore, for $d>3$,
the entangling velocity for a CFT on hyperbolic space exceeds the butterfly
velocity, eq. \eqref{eq:butterfly_velocity}. Typically, whenever $v_{\ent}>
v_{\but}$, one might worry about contradictions to entanglement monotonicity
laws \cite{Hartman:2015apr,Mezei:2016wfz} which apply above the thermal scale.
However, no immediate contradictions appear in the present case, as we now
explain.

For concreteness, let us interpret eq.~\eqref{carrot2} as describing the
entanglement growth in hyperbolic space without defects, specifically, computing
the entropy for a region consisting of half-spaces $\chi>0$ on either side of
the TFD.\footnote{To be precise, we should multiply eq.~\eqref{carrot2} by
  $\frac{1}{2}\cdot\frac{1}{4G_\bulk}$ with the factor of $1/2$ due to our focus
  on just two copies of the entangling surface $\chi=0$ --- one on either side
  of the TFD.} This growth saturates
eq.~\eqref{eq:entanglement_spreading} with $v_{\ent}>v_{\but}$ in $d>3$ so one
might worry that entanglement is spreading faster than the speed $v_{\but}$
permitted by operator commutator growth \cite{RobSta15}. Specifically, by
applying an analysis similar to the one reviewed around
eq.~\eqref{eq:butterfly_velocity} to thermal relative entropies,
\cite{Hartman:2015apr,Mezei:2016wfz} argue that, for regions and times above the
thermal scale, entanglement growth must be bounded by the thermal entropy
density $s_{\therm}$ times the volume between the entangling surface and a
tsunami wavefront propagating with speed $v_{\but}$ away from the entangling
surface (in either direction). Said differently, the rate $dS/dt$ of
entanglement growth is bounded by $s_{\therm}v_{\but}$ times the area of the
tsunami wavefront --- this is essentially eq.~\eqref{eq:entanglement_spreading}
with $v_{\but}$ replacing $v_{\ent}$ and the tsunami wavefront replacing the
entangling surface. In flat space, the tsunami wavefront can be typically chosen
to propagate in a direction away from the entangling surface such that it
shrinks or does not grow in time (\eg{}propagating inward from a spherical
entangling surface). Thus, for the flat space equivalent of
eq.~\eqref{eq:entanglement_spreading} to be saturated, one must require
$v_{\ent}<v_{\but}$. In hyperbolic space however, it is possible for the tsunami
wavefront to grow in both directions away from the entangling surface. Indeed,
this is precisely what happens for the hyperbolic half-space which has an
entangling surface $\chi=0$ of minimal area; within a few thermal times, the
tsunami wavefront propagating in either direction grows to an area exponentially
large compared to the entangling surface. We thus see that, though the
hyperbolic half-space saturates eq.~\eqref{eq:entanglement_spreading} with
$v_{\ent}>v_{\but}$, this does not contradict the bound on entanglement
spreading due to the butterfly velocity.

%% file: sections/05_extremal_bh.tex

\begin{figure}[h]
	\def\svgwidth{0.4\linewidth} \centering{ 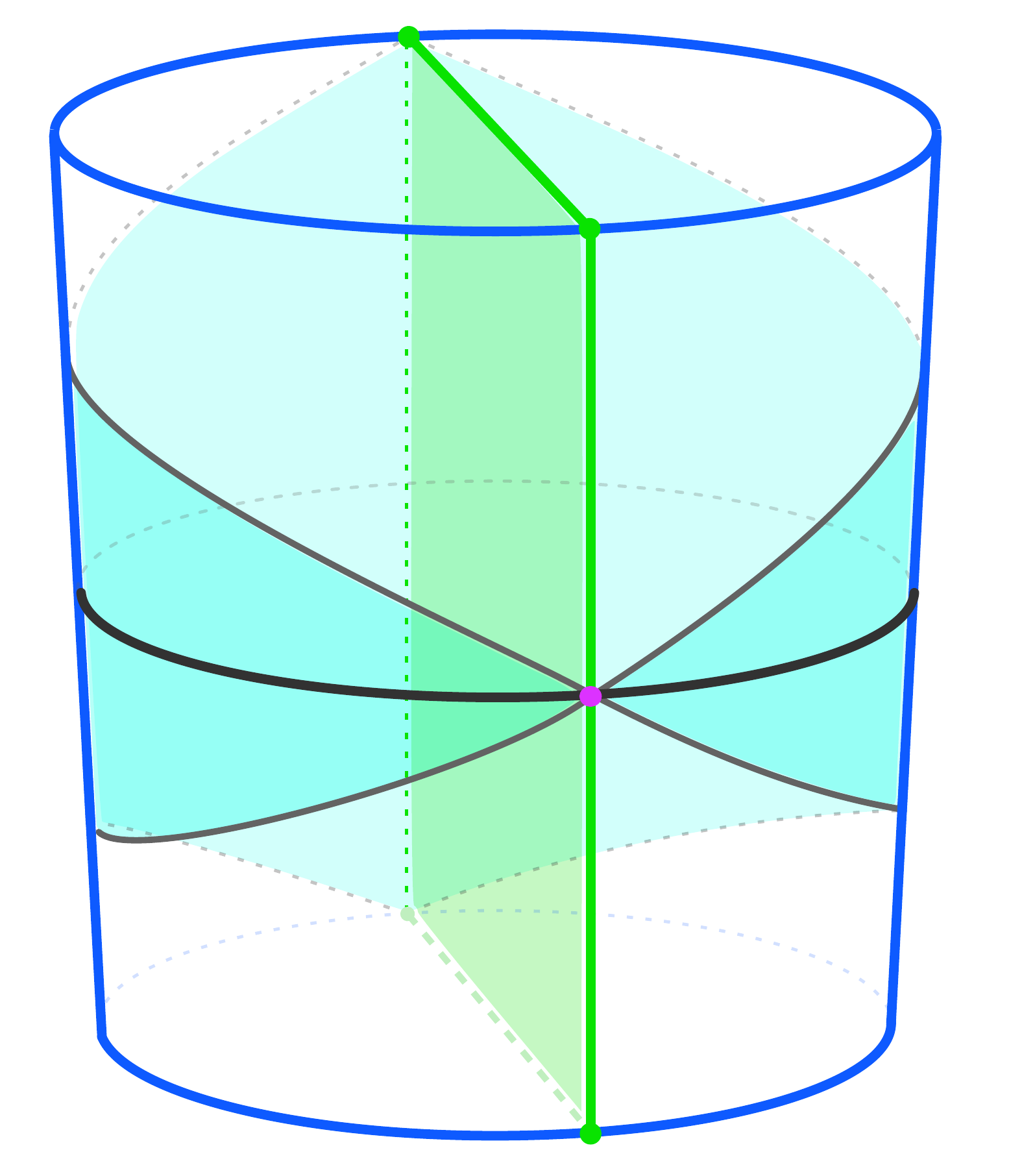
		\caption{The Poincare patch models a zero temperature extremal black hole.
      The brane intersects the CFT Poincar\'e patch at the origin and infinity.
    }
		\label{fig:PoincarePatches}
	}
\end{figure}

Here we turn our attention to extremal black holes. In particular, we consider the same bulk geometry described in section \ref{sec:RS}, \ie a backreacting codimension-one brane extending across the spacetime which locally has the geometry of AdS${}_{d+1}$. However, we replace the AdS-Rindler coordinates introduced in eq.~\reef{eq:metric_rindler_bulk} with Poincar\'e
coordinates, 
\beq 
ds^2=\frac{L^2}{z^2}\(dz^2-dt^2+dx_1^2+ \cdots+dx_{d-1}^2\)\,.
\label{metric0}
\eeq 
Of course, the coordinate singularity at $z\to\infty$ corresponds to an extremal $T=0$ horizon. Figure
\ref{fig:PoincarePatches} illustrates the Poincar\'e patch in our bulk geometry. 

\begin{figure}[t]
	\def\svgwidth{0.8\linewidth} \centering{ 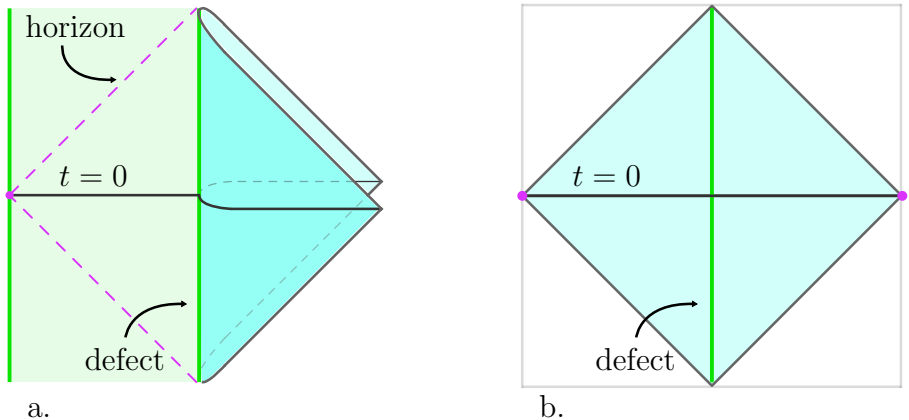
		\caption{The brane and boundary perspectives of the extremal black hole setup.}
		\label{fig:PoincarePatches2}
	}
\end{figure}

For the most part, we will be interested in limit of large tension (\ie $\ell_\mt{B}\gg L$), for which the brane theory can be described as Einstein gravity coupled to two copies of the boundary CFT. As we describe in a moment, the brane geometry naturally inherits a Poincar\'e metric from the bulk geometry. Hence the brane supports an extremal black hole which is equilibrium with the $T=0$ bath CFT on the asymptotic AdS boundary. We note that with Poincar\'e coordinates, we are examining the system in a new conformal frame where the bath CFT is living on flat $d$-dimensional Minkowski space,
\beq
\label{flat}
ds_\mt{CFT}^2=-dt^2+dx_1^2+ \cdots+dx_{d-1}^2\,.
\eeq
This brane perspective is illustrated in figure \ref{fig:PoincarePatches2}a.

Of course, we may also have the boundary perspective where the $d$-dimensional CFT in Minkowski space is coupled to a codimension-one conformal defect. For simplicity, we insert the latter at $x_1=0$ for the metric in eq.~\reef{flat} and so the induced geometry on the defect is also flat, \ie ($d$--1)-dimensional Minkowski space. The Penrose diagram for this perspective is shown in figure \ref{fig:PoincarePatches2}b. Note that in contrast to the finite temperature TFD state (entangling two copies of the bath CFT) in section \ref{sec:nonextremal}, here for the $T=0$ scenario, we only have a single copy of the bath CFT, \eg compare the above to figures \ref{fig:Eternal} and \ref{fig:bndrypenrose}. Of course, at $T=0$, we are simply studying the vacuum state of the defect CFT in flat space (analogous to what was done in \cite{Chen:2020uac} but in a different conformal frame).\footnote{Of course, this is a pure state, as is manifest in bulk since the Poincar\'e time slices constitute complete Cauchy slices.}

We may recall from \cite{Almheiri:2019yqk} that for the extremal case in $d=2$, one always finds islands for the analogous belt regions. This result is a consequence of two features which hold for $d=2$: firstly, there always exists a bulk RT surface intersecting the brane to
produce an island; secondly, the alternative no-island RT candidate surface has an additional IR divergence\footnote{Coming from integrating the length of the surface down to the extremal horizon.} and this surface is therefore subdominant. However, neither of
these statements hold in $d\ge 3$. Indeed, we will find in higher dimensions that quantum extremal islands do not appear in the large tension limit. Nonetheless,
no information paradox arises since extremal black holes do not radiate, \ie the black hole and the bath are not exchanging radiation. This contrasts with the non-extremal case in
section \ref{sec:nonextremal}, where the information paradox for the eternal black hole in the effective $d$-dimensional gravity theory arises because of the continuous exchange of quanta between the black hole and the bath. Of course, the paradox is avoided by the appearance of quantum extremal islands.

The remainder of this section is organized as follows. We shall begin by first
explicitly constructing the bulk and brane metrics to be used in the extremal
case and by introducing the entanglement entropy calculation which we wish to
consider. Then, in subsections \ref{sec:lonely} and \ref{sec:suicide}, we carry
out this calculation using RT surfaces corresponding to island and no-island
phases, respectively. Finally, we collect these results in subsection \ref{sec:hopeless} to
determine when each phase dominates.\\

The Poincar\'e coordinates \reef{metric0} cover a wedge of the AdS$_{d+1}$ vacuum geometry. However, in the present geometry with a backreacting brane, a portion of two such wedges would appear on either side of the brane -- see figure \ref{fig:PoincarePatches}. If we consider the coordinate transformation 
\beq z = y \sin\theta,\qquad x_1 = y\cos\theta\,,
\label{house}
\eeq 
the metric \reef{metric0} is transformed to the form given in eq.~\eqref{metric33}, where the $\AdS_d$ slices each inherit a Poincar\'e metric. As described in section \ref{sec:RS}, the brane spans one such slice at a
fixed $\theta=\braneAngle$ determined by the brane tension $T_o$ according to
eq.~\eqref{curve2}, \ie
\beq \sin^2 \braneAngle = 2\,\veps\(1-\veps/2\)\,.
\label{house22}
\eeq 
The induced metric on the brane then becomes
\beq 
ds_{\AdS_d}^2=\frac{L^2}{y^2
  \sin^2\!\theta_\mt{B}}\(dy^2-dt^2+dx_2^2+ \cdots+dx_{d-1}^2\)\,,
\label{metric1}
\eeq 
and we may then read off the curvature scale of the brane as $\ell_{\mt{B}} = L/\sin \braneAngle$, as expected from eq.~\reef{curve2}. Here, $y$ is
interpreted as the radial Poincar\'e coordinate running along the brane, and the
Poincar\'e horizon on the brane, located at $y \to \infty$, is inherited from
the bulk. As usual, we wish to work in the regime $L^2/\ell_{\mt{B}}^2\ll1$, or alternatively $\braneAngle\ll1$.
\begin{figure}[t]
  \centering \includegraphics[width=\textwidth]{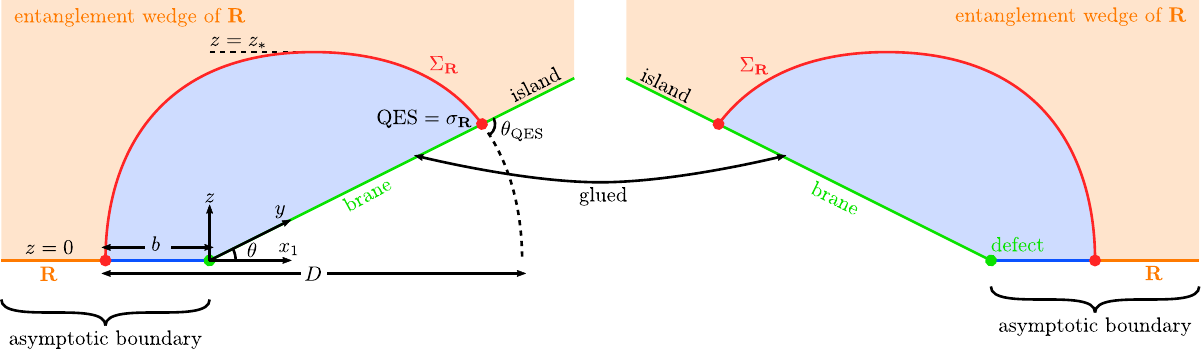}
  \caption{The bulk dual to a $d$-dimensional Minkowski CFT with a defect (green
    dot) along a line $x_1=0$. The CFT lives on the asymptotic boundary of a
    Poincar\'e $\AdS_{d+1}$ spacetime with a brane (green line) running through
    it. We consider the entanglement entropy of the complement
    $\bdyReg=(-\infty,-b]\cup[b,\infty)$ of a belt geometry in the CFT. As
    considered in section \ref{sec:lonely}, one candidate RT surface $\RT$,
    shown in red, intersects the brane at a QES $\RTbrn$, forming an island on
    the brane belonging to the entanglement wedge of $\bdyReg$. Various
    quantities defined in section \ref{sec:lonely} are marked in this figure.
  }
  \label{fig:braneflat}
\end{figure}

Following the brane perspective described above (and in section \ref{sec:RS}), eq.~\reef{metric1} is interpreted as an extremal black hole solution of the gravity theory  induced  brane at $\theta=\braneAngle$ and the CFT of the flat asymptotic boundary at $z=0$ becomes the zero temperature bath. This then provides a direct extension of the extremal scenario  in \cite{Almheiri:2019yqk} to $d$ dimensions. The question which interests us here is then whether the entanglement wedge of certain subregions in the bath includes 
islands residing on the brane. 

Specifically, we consider the entanglement entropy calculation for a
boundary region $\bdyReg$ that is the complement of a ``belt" geometry centered on
the defect at $x_1=0$, \ie the boundary
subregion $\bdyReg=(-\infty,-b]\cup[b,\infty)$. According to the RT formula we should consider
codimension-two surfaces $\blkSurf$ sharing the same boundary $\partial\blkSurf
=\partial\bdyReg\equiv \RTbdy$. To
determine RT surface candidates among these surfaces, we must search for
surfaces which extremize their area. As we discussed in the introduction, there are generally
two sets of surfaces which achieves this extremization; the RT prescription then
instructs us to choose the one with the smallest area. The first class of
surfaces are those which intersect the brane, forming a quantum extremal island
on the brane which belongs to the entanglement wedge of $\bdyReg$ -- see figure \ref{fig:braneflat}. We will say that this RT surface is in the \emph{island phase}. The second set of surfaces fall
trivially into the bulk and do not produce islands on the brane, \ie these surfaces are in
the \emph{no-island phase}.

\subsection{Island phase}
\label{sec:lonely}
As a starting point, let us review the calculation for RT surfaces of belt geometries in pure AdS \cite{Ryu:2006ef}. That is, we are considering the complement of $\bdyReg$, but the RT calculations for this region and for its complement, $\overline\bdyReg=[-b,b]$,  are equivalent. Integrating out the
$x_2,\ldots,x_{d-1}$ directions in which the brane is constant, the area
functional of a codimension-2 surface $\blkSurf$ becomes
\begin{align}
  \area(\blkSurf)
  = & \LAdS^{d-1} \volPerp \int_\blkSurf dx_1
      \frac{\sqrt{1+\left(\frac{dz}{dx_1}\right)^2}}{z^{d-1}},
      \label{eq:banana}
\end{align}
where $\volPerp$ is the volume of transverse directions 
$\{x_2,\ldots,x_{d-1}\}$.\footnote{Note that in contrast to $\vol_{H_{d-2}}$ introduced in section \ref{sec:nonextremal}, $\volPerp$ has the dimensions of {\it length}$^{d-2}$ and so is essentially given by $\cutoffIR^{d-2}$ where $\cutoffIR$ is an IR cutoff in the $x_2,\ldots,x_{d-1}$ directions.}

\begin{figure}
  \centering
  \begin{subfigure}[b]{0.49\textwidth}
    \centering \includegraphics[width=0.7\textwidth]{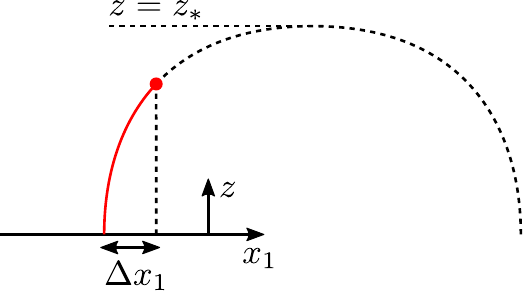}
    \caption{Heading into the bulk $\pm=+$.}
  \end{subfigure}
  \begin{subfigure}[b]{0.5\textwidth}
    \centering \includegraphics[width=0.7\textwidth]{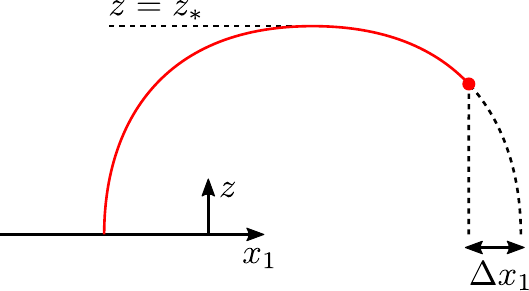}
    \caption{Heading out of the bulk $\pm=-$.}
  \end{subfigure}
  \caption{Definitions for the choice of $\pm$ in eq.~\reef{eq:chips} and for the corresponding $\Delta x_1\ (>0)$ from eq.~\reef{eq:crisps} on the two branches of the RT
    surface. }
  \label{fig:andHurtYou}
\end{figure}

The RT surface $\RT$ is obtained by extremizing the area functional \reef{eq:banana} with respect to the profile $z(x_1)$. This functional, viewed as a
Lagrangian, contains no explicit dependence on $x_1$ and hence the corresponding Hamiltonian is
a constant along $\RT$, allowing us to deduce
\begin{align}
  \frac{dz}{dx_1}
  =& \pm \frac{\sqrt{z_*^{2(d-1)}-z^{2(d-1)}}}{z^{d-1}}
     \label{eq:chips}
\end{align}
for some constant $z_*$. Further, the sign $\pm$ above is determined by whether we are on
the portion of the RT surface heading into the bulk ($+$) or heading out of the
bulk ($-$) with increasing $x_1$ -- see figures \ref{fig:braneflat} and \ref{fig:andHurtYou}.\footnote{As noted previously, if we restrict our attention to positive tension $T_o$, we will have $0<\braneAngle<\pi/2$. In this case, the RT surface must be increasing in $x_1$ as one heads away from the boundary $(z,x_1)=(0,-b)$, in order for the RT surface to meet the  brane.}
From eq.~\reef{eq:chips}, we see that $dz/dx_1=0$ at $z=z_*$ and therefore $z_*$ is the maximal $z$-value attained by $\RT$. We can integrate eq.~\eqref{eq:chips} to obtain the trajectory of the RT
surface:
\begin{align}
  \Delta x_1
  ={} & \frac{z^d}{d\;z_*^{d-1}}\  \hyperF\!\left[
        \frac{1}{2}, \frac{d}{2(d-1)}; \frac{d}{2(d-1)} + 1;
        \left(\frac{z}{z_*}\right)^{2(d-1)}
        \right] \label{eq:crisps}
\end{align}
Here $\Delta x_1>0$ is the absolute $x_1$-separation between a point on the RT
trajectory and the initial (final) endpoint on the asymptotic boundary, on the
portion of the RT surface heading into (out of) the bulk --- see figure \ref{fig:andHurtYou}. If we evaluate this expression at $z=z_*$, we obtain half of the width of the boundary strip (in the $x_1$ direction)
defined by the RT surface. Denoting this width as $\RTDiam$, which we emphasize is in the empty AdS vacuum (see figure \ref{fig:braneflat}), we have 
\begin{align}
  \frac{\RTDiam}{2}
  = & \ 
   \frac{\sqrt{\pi}\,\Gamma\!\left[\frac{d}{2(d-1)}\right]}
      {\Gamma\!\left[\frac{1}{2(d-1)}\right]}\ 
      z_*.
      \label{eq:chocolate}
\end{align}

Now returning to the geometry with the backreacting brane, each half of the RT surface $\RT$ on
either side of the brane will follow the trajectory given in
eq.~\eqref{eq:crisps} for pure AdS prior to meeting the
brane.
We have placed the defect at $x_1=0$ and the RT surface begins on the asymptotic boundary at $x_1=-b$. Further, if we were to extend the RT surface past the brane, it would hit the asymptotic boundary again at  $x_1=-b+D$.
In terms of eq.~\eqref{eq:crisps}, $x_1$ along the trajectory is then given by
\begin{align}
  x_1
  =& -\beltHalfWidth+\begin{cases}
    \Delta x_1
    & \text{when heading into bulk (towards $z=z_*$)}
    \\
    D-
    \Delta x_1
    & \text{when heading out of bulk (away from $z=z_*$)}
  \end{cases}.
      \label{eq:tortilla}
\end{align}

In general, as illustrated in figure
\ref{fig:braneflat}, $\frac{\RTDiam}{2}\ne \beltHalfWidth$, rather, the relation
between $\RTDiam$ (or $z_*$) and $\beltHalfWidth$ must be determined by demanding that the choice of the intersection $\RTbrn$ of the RT surface with the brane should extremize the RT surface's area (plus the area of the QES, when brane action includes an extra DGP term). As described in \cite{Chen:2020uac} and
reviewed around eq.~\eqref{eq:variation_RT_surface_intersection}, this extremization leads to a boundary condition restricting the angle at which the RT surface meets the
brane. Again, we may reduce this to a two-dimensional problem where we view the RT surface as a geodesic in an effective two-dimensional geometry
\begin{align}
      ds_{2D}^2
      = & \LAdS^{2(d-1)}\, (\volPerp)^2\ 
          \frac{dz^2 + dx_1^2}{z^{2(d-1)}}\,,
          \label{eq:neverGonnaMakeYouCry}
\end{align}
and the area becomes the length of the geodesic in this geometry. 

\begin{figure}
  \centering
  \begin{subfigure}[t]{0.48\textwidth}
    \includegraphics[scale=1]{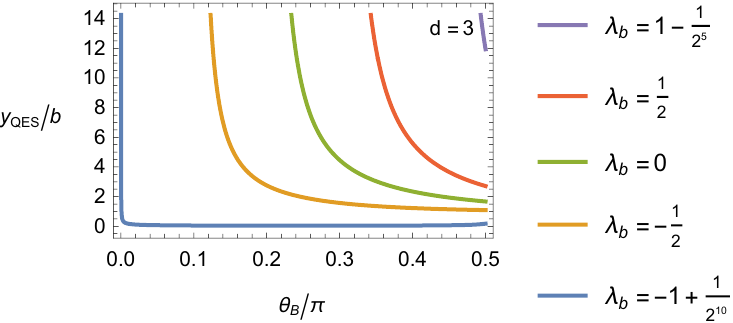}
    \caption{Position of $\RTbrn$ as a function of brane angle
    }
    \label{fig:a}
  \end{subfigure}
  \hfill
  \begin{subfigure}[t]{0.48\textwidth}
    \includegraphics[scale=1]{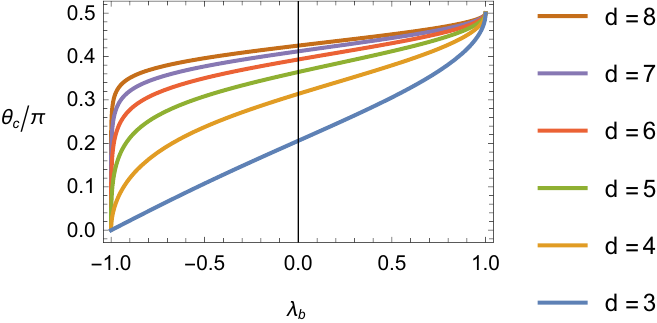}
    \caption{Critical brane angle as a function of the ratio of $G_\bulk$ to
      $G_\brane$.}
    \label{fig:aa}
  \end{subfigure}
  \caption{Plots of the position of $\RTbrn$, the intersection of the RT surface with
    the brane, and the critical brane angle at which this surface runs off
    to $y_{\QES}\to +\infty$.}
  \label{fig:aan}
\end{figure}

As before, we may use
eq.~\eqref{eq:variation_RT_surface_intersection} to determine the variation of the
RT surface area under perturbations of $\RTbrn$, the QES on the brane. Here, $h_{ij}$ is given by eq.~\eqref{eq:neverGonnaMakeYouCry}, the deviation vector $X^j$ is chosen to be $\partial_y$, and the tangent $T^i$ determined from eq.~\eqref{eq:chips}, with both $X^j$ and $T^i$ normalized with respect to
$h_{ij}$. Hence, upon perturbing the
intersection of the RT surface with the brane, the RT area varies as
\begin{align}
  \frac{\partial \area(\RT)}{\partial y_{\QES}}
  = & \frac{2 L^{d-1} \volPerp}{z_{\QES}^{d-1}} \cos\theta_{\QES}
      = 2 \LAdS^{d-1} \volPerp
      \left(
      \frac{\cos\braneAngle}{z_*^{d-1}}
      \pm\sqrt{\frac{1}{z_{\QES}^{2(d-1)}} - \frac{1}{z_*^{2(d-1)}}}
      \sin\braneAngle
      \right),
      \label{eq:chickPea}
\end{align}
where $\theta_{\QES}$ is the angle between the RT surface and the brane,
$y_{\QES}$ is the $y$ coordinate of $\RTbrn$ -- see figure \ref{fig:braneflat}
--- and the $\pm$ sign is the same one as introduced in eq.~\eqref{eq:chips} and
illustrated in figure \ref{fig:andHurtYou}.
An extra factor of $2$ is included to account for the two components of the RT surface on either side of the brane.
From eq.~\eqref{metric1}, we read off the area of $\RTbrn$:
\begin{align}
  \area(\RTbrn)
  = & \volPerp \left(\frac{\LAdS}{y_{\QES}\sin\braneAngle}\right)^{d-2},
  &
    \frac{\partial \area(\RTbrn)}{\partial y_{\QES}}
    = & - \frac{(d-2) \volPerp \LAdS^{d-2} \sin\braneAngle}{z_{\QES}^{d-1}}.
        \label{eq:microwave}
\end{align}
The extremality condition
\begin{align}
	0 =&  \frac{\partial }{\partial y_{\QES}}\left(
       \frac{\area(\RT)}{4 G_\bulk}
       + \frac{\area(\RTbrn)}{4 G_\brane}
       \right)
       \label{eq:abr}
\end{align}
is satisfied if
\begin{align}
  \cos\theta_{\QES}
  =&
     \DGPRatio \sin\braneAngle
  &
  &\iff
  &
    z_{\QES}
    =& z_* \left[
       \sin\braneAngle\,\left(
       \DGPRatio\, \cos\braneAngle
       + \sqrt{1-\DGPRatio^2\, \sin^2\!\braneAngle}
       \right)
       \right]^{\frac{1}{d-1}}\,,
       \label{eq:vanilla}
\end{align}
where $\DGPRatio$ is defined in eq.~\reef{Newton3}. 
The relationship between $z_*$ and $\beltHalfWidth$ may then be determined by substituting $(x_1,z) =(z_{\QES}\cot\braneAngle,z_{\QES})$ into
eq.~\eqref{eq:tortilla}, and using eqs.~\eqref{eq:chocolate}, \eqref{eq:crisps} and \eqref{eq:vanilla} to find
\begin{align}
         \begin{split}
           \beltHalfWidth =& \pm\Delta x_1 + \frac{1\mp 1}{2} \RTDiam -
           z_{\QES}\, \cot \braneAngle = \bByzs(d,\DGPRatio,\braneAngle)\, z_*
         \end{split}
                             \label{eq:dieAlone}
  \\
  \begin{split}
    \bByzs(d,\DGPRatio,\braneAngle) \equiv& \pm \frac{z_{\QES}^d}{d\;z_*^d}\;
    \hyperF\left[ \frac{1}{2}, \frac{d}{2(d-1)}; \frac{d}{2(d-1)} + 1;
      \left(\frac{z_{\QES}}{z_*}\right)^{2(d-1)} \right]
    \\
    & +(1\mp1)\frac{\sqrt{\pi}\Gamma\left[\frac{d}{2(d-1)}\right]}
    {\Gamma\left[\frac{1}{2(d-1)}\right]} -\frac{z_{\QES}}{z_*} \cot \braneAngle
  \end{split}
      \label{eq:oven}
\end{align}
where the top (bottom) signs chosen above if the RT surface intersects the brane to the left (right) of the extremal point $z=z_*$. We have noted in the second equality of
eq.~\eqref{eq:dieAlone} that all terms of the previous expression are linear in
$z_*$; in particular, note in eq.~\eqref{eq:oven} that the ratio $z_{\QES}/z_*$
is determined by eq.~\eqref{eq:vanilla}. In figure \ref{fig:a}, we have plotted
the position of the intersection $\RTbrn$ between the RT surface and the brane
as a function of the brane angle $\braneAngle$ for various $\DGPRatio$ and
$d=3$. In section \ref{sec:hopeless}, we shall discuss the fact that, for
$\braneAngle$ below some critical angle $\theta_c$, the extremal surfaces discussed here fail to exist. That is, $y_{\QES}$, the position of the QES on the brane, runs off to infinity as $\braneAngle\to\theta_c$ from above.

Having determined the profile of the RT surfaces, we may proceed to evaluate their corresponding entropies using the RT formula \reef{eq:island2} -- keeping in mind that we have not shown that these surfaces minimize the entropy functional yet. Inserting eqs.~\eqref{eq:chips} and \eqref{eq:microwave} into the generalized entropy functional, we find that the entropy of the
belt geometry $\overline\bdyReg$ and hence of the complementary bath region $\bdyReg$ is given by
\begin{align}
  \begin{split}
    \MoveEqLeft[0.8] \frac{\area(\RT)}{4 G_\bulk} + \frac{\area(\RTbrn)}{4
      G_\brane}
    \\
    ={}&\frac{\LAdS^{d-2}}{4G_\brane}\,\frac{\volPerp}{z_{\QES}^{d-2}} + \frac{\LAdS^{d-1}}{4 G_\bulk}\Bigg\{ \frac{ (1\mp 1)
      \sqrt{\pi}\Gamma\left[\frac{2-d}{2(d-1)}\right] }{
      (d-1)\Gamma\left[\frac{1}{2(d-1)}\right] } \,
    \frac{\volPerp}{z_*^{d-2}}
    \\
    &+\frac{2}{d-2} \left[ \frac{\volPerp}{\delta^{d-2}}
      \mp\frac{\volPerp}{z_{\QES}^{d-2}}\ \hyperF\!\left( \frac{1}{2},
        \frac{d}{2(d-1)} - 1; \frac{d}{2(d-1)};
        \left(\frac{z_*}{z_{\QES}}\right)^{2(d-1)} \right) \right] \Bigg\}
  \end{split}
      \label{eq:butter}
\end{align}
where $z=\delta$ defines the UV cutoff surface near the asymptotic AdS boundary, and $z_{\QES}$ and $z_*$
are linearly related to $\beltHalfWidth$ by eqs.~\eqref{eq:vanilla} and
\eqref{eq:oven}. 
For $z_\QES\ll z_*$, the hypergeometric function becomes $1+O[(z_{\QES}/z_*)^{2(d-1)}]$, giving
\begin{align}
  \begin{split}
    \frac{\area(\RT)}{4 G_\bulk} + \frac{\area(\RTbrn)}{4 G_\brane} =&
    \frac{\LAdS^{d-1}}{4 G_\bulk}\Bigg\{ \frac{2}{d-2} \frac{\volPerp}{\delta^{d-2}}
    -\frac{2}{d-2}\( \frac{ 2
    \sqrt{\pi}\,\Gamma\!\left[\frac{d}{2(d-1)}\right] }{
    (d-2)\,\Gamma\!\left[\frac{1}{2(d-1)}\right] }\)^{d-1} \frac{\volPerp}{D^{d-2}}
    \Bigg\}
    \\
    &+ \volPerp\left(\frac{\LAdS}{z_{\QES}}\right)^{d-2}\left\{
      \frac{1}{4G_\eff} +
      O\left[\frac{1}{G_\bulk}\left(\frac{z_{\QES}}{z_*}\right)^{2(d-1)}\right]
    \right\}\,,
  \end{split}
      \label{eq:korma}
\end{align}
where we have used eq.~\reef{eq:chocolate} to replace $z_*$ with $D$ in the first line.
Note from eq.~\eqref{eq:vanilla} that $z_\QES/z_* \sim [(\DGPRatio+1)\braneAngle]^{1/(d-1)}$ so the correction is
indeed smaller than the other terms shown here in high tension limit.

Using the brane perspective, let us examine the various contribution to the generalized entropy on the right-hand side of eq.~\eqref{eq:korma}. Beginning with the leading term of the second line in
eq.~\eqref{eq:korma}, we find that it corresponds to  Bekenstein-Hawking of the QES, \ie
$\frac{1}{4G_{\eff}}$ times the area of $\RTbrn$. It is interesting to note that
that there are no higher curvature corrections to the generalized entropy of the
QES as might have been expected from the Wald-Dong entropy formula.\footnote{One
  can argue that all of the higher curvature corrections to the Wald-Dong
  entropy must cancel against one another as follows: In the present case, these
  terms would arise from integrating out the boundary CFT on the gravitating
  brane and so should be conformally invariant, \eg see
  \cite{Solodukhin:2008dh}. However, by a simply Weyl transformation, the brane
  metric becomes flat and further  both the intrinsic and the extrinsic
  curvatures of $\RTbrn$ vanish. Hence in this flat conformal frame, the higher
  curvature corrections to the Wald-Dong entropy  individually vanish. Hence
  while these curvatures do not vanish in the original conformal frame, the
  higher curvature entropy corrections must all cancel against one another.}
Turning to the first term in the first line of eq.~\eqref{eq:korma}, we have the
area law divergence associated with the two components of the entangling surface $\RTbdy$ at $x_1=\pm b$. 
This leaves us with the second term in the first line. Upon closer examination can be recognized as the finite 
contribution to the entanglement entropy for a belt of width $D$, up to an additional factor of 2, \eg see \cite{Ryu:2006bv,Ryu:2006ef}. 
Further, we note that both contributions on the first line of eq.~\eqref{eq:butter} contain a prefactor proportional to $\LAdS^{d-1}/G_{\bulk}\sim c_T$, which measures the number of degrees of freedom in the boundary CFT, \eg \cite{Buchel:2009sk}.

We can see that these results correspond
approximately to the expected entropy from the brane perspective as follows: We begin by considering the contribution from the CFT to one side of the conformal defect, say $x_1<0$. Imagine we begin with a single copy of the CFT in flat space \reef{flat}, and evaluate 
the entropy of a belt of width $D$ with entangling surfaces at $x_1=-b$ and $x_1=D-b$. For this geometry, the holographic entanglement entropy becomes \cite{Ryu:2006bv,Ryu:2006ef}
\beq
S_\mt{EE}=
    \frac{\LAdS^{d-1}}{4 G_\bulk}\Bigg\{ \frac{1}{d-2} \frac{\volPerp}{\delta^{d-2}} + \frac{1}{d-2} \frac{\volPerp}{\delta^{d-2}}
    -\frac{1}{d-2}\( \frac{ 2
    \sqrt{\pi}\,\Gamma\!\left[\frac{d}{2(d-1)}\right] }{
    (d-2)\,\Gamma\!\left[\frac{1}{2(d-1)}\right] }\)^{d-1} \frac{\volPerp}{D^{d-2}}\Bigg\}\,,
\label{flatter1}
\eeq
where we have separated the area law contributions of the two components of the entangling surface.
Now from the brane perspective in our system, the bath CFT reside in flat space for $x_1<0$ but the corresponding copy of the CFT resides onto the AdS$_d$ geometry of the brane for $x_1>0$. However the latter can be produced by making a local Weyl transformation in the positive $x_1$ domain:
\beq
\label{notflat}
ds^2=\frac{\delta^2}{x_1^2\sin^2\theta_\mt{B}}\,ds_\mt{CFT}^2=\frac{\delta^2}{x_1^2\sin^2\theta_\mt{B}}\(-dt^2+dx_1^2+ \cdots+dx_{d-1}^2\)\,.
\eeq
Note that this is geometry is not the induced metric \reef{metric1} but rather we are considering the standard conformal frame where one strips off the factor of $(L/\delta)^2$ from the bulk metric.\footnote{Further, we are only performing the Weyl transformation \reef{notflat} for $x_1>\delta/\tan\theta_B$, which corresponds to the intersection of the brane with the UV cutoff surface $z=\delta$.} Now the net effect of this Weyl transformation on the entanglement entropy \reef{flatter1} is to modify the cutoff appearing in the area law contribution for the surface at $x_1=D-b$, \ie
$\delta\to (D-b)\sin\theta_\mt{B}\simeq z_\QES$, where the latter assumes that $\theta_\mt{B}\ll 1$. Hence the entropy \reef{flatter1} becomes
\beqa
S'_\mt{EE}&\simeq&
    \frac{\LAdS^{d-1}}{4 G_\bulk}\Bigg\{ \frac{1}{d-2} \frac{\volPerp}{\delta^{d-2}} 
    -\frac{1}{d-2}\( \frac{ 2
    \sqrt{\pi}\,\Gamma\!\left[\frac{d}{2(d-1)}\right] }{
    (d-2)\,\Gamma\!\left[\frac{1}{2(d-1)}\right] }\)^{d-1} \frac{\volPerp}{D^{d-2}}\Bigg\}
    \nonumber\\
    &&\qquad\qquad + \frac14\,\frac{L}{(d-2)G_\bulk}\(\frac{L}{z_\QES}\)^{d-2}\frac{\volPerp}{\delta^{d-2}}\,.
\label{flatter2}
\eeqa
Now using eq.~\reef{Newton3}, the term on the second line can be recognized as the contribution of one of the boundary CFTs to the Bekenstein-Hawking entropy of the quantum extremal surface on the brane. Hence combining the above contribution \reef{flatter2} with that from the other copy of the boundary CFT (which extends to the bath for $x_1>0$) and the DGP contribution to the Bekenstein-Hawking entropy, we precisely recover the leading contributions in eq.~\reef{eq:korma}. Hence this simple CFT argument allows us to match the leading contributions in the holographic result with the expected
entanglement entropy.

\subsection{No-island phase}
\label{sec:suicide}
Above, we studied the set of candidate RT surfaces which intersect the brane. In fact (for $\braneAngle<\pi/2$), there exists another set of simple extremal surfaces which must also be considered under the RT prescription \reef{eq:island2}. These surfaces are
constant $x_1$ planes anchored on the entangling surface $\RTbdy$ on the
asymptotic boundary and fall straight into the bulk. By reflection symmetry about $x_1=\pm \beltHalfWidth$, these planes trivially extremize the area functional, which becomes
\beq
  \area(\blkSurf)   =  2\LAdS^{d-1} \volPerp \int_\blkSurf 
      \frac{dz}{z^{d-1}}\,.
      \label{pineappleX}
\eeq
A factor of $2$ has been included above to account for the two planes at $x_1=\pm\beltHalfWidth$.\footnote{Further, let us note that for the special case $d=2$, the integral produces an IR divergence at $z\to\infty$. However, there is no such IR divergence for $d\ge3$.} Unlike the surfaces considered in
section \ref{sec:lonely}, these planes do not intersect the brane and thus no islands are formed on the brane.
The entropy in this no-island phase is easily obtained from evaluating the area functional \reef{pineappleX}, which then yields
\begin{align}
  \frac{\area(\RT)}{4 G_\bulk}
  =& \frac{\LAdS^{d-1} }{2(d-2) G_\bulk }\,
  \frac{\volPerp}{\delta^{d-2}}\,,
     \label{eq:depression}
\end{align}
where $\delta$ is again the UV cutoff in the boundary CFT.

\subsection{Islands at $T=0$ for $d>2$}
\label{sec:hopeless}

Altogether, we have two candidate RT surfaces: the extremal surfaces described
in section \ref{sec:lonely} which intersect the brane to form a quantum extremal
island, and the extremal planes described in section \ref{sec:suicide}
corresponding to the no-island phase. To determine which is the correct RT
surface, we must first study the parameter space for which each surface exists.
When both types surfaces exist simultaneously, the correct RT surface is given
by the one which has the smallest generalized entropy, as in
eq.~\reef{eq:island2}. Below, we first observe that on a brane at angle
$\braneAngle< \pi/2$, an island can only possibly exist when $-1<\DGPRatio < 1$;
more specifically, for this range of the DGP parameter $\DGPRatio$, there is a
critical angle $\theta_c<\pi/2$ which gives the minimum $\braneAngle$ that
supports the island phase -- recall that this critical angle was plotted in
figure \ref{fig:aa}. For $\braneAngle>\theta_c$, the island phase exists and is
dominant. At $\braneAngle=\theta_c$ the entropies computed by the island and
no-island RT surfaces equalize, leading to a transition to the no-island phase
below $\theta_c$. As we shall find that $\theta_c$ scales as
$(1+\lambda_b)^{\frac{1}{d-2}}$ at its smallest, this precludes the possibility
of islands in the regime where the brane is well-described by QFT on
semiclassical gravity -- see footnote \ref{bigtoe}. This differs from the $d=2$
case, where the island phase always exists; furthermore, while the no-island RT
surface in $d>3$ has an IR-finite area, the analogous surfaces in $d=2$ produce
an IR divergence and thus are never dominant.

Let us begin our analysis by constraining the parameter space in which each type
of RT candidate surface exists. It is easy to see that the extremal planes of
the no-island phase exist if and only if $\braneAngle \le \pi/2$.\footnote{Of
  course, this was our regime of interest, as this was the regime where a
  (nearly) massless graviton is induced on the brane.} It is slightly more
involved to determine when the extremal surfaces in the island phase exists. For
a start, the first equality of eq.~\eqref{eq:vanilla} indicates that for
$\braneAngle<\pi/2$, sensible extremal surfaces intersecting the brane can only
possibly exist when
$-1<\DGPRatio <1$.\footnote{Specifically, this can be seen as follows: Let us take the extreme case of $\DGPRatio=1$ ($\DGPRatio=-1$). Then eq.~\eqref{eq:vanilla} indicates that $\theta_\QES=\braneAngle-\pi/2$  ($\theta_\QES=\braneAngle+\pi/2$). For $\DGPRatio=1$, this implies that when $\braneAngle > \pi/2$, the RT surface falls straight into the bulk until it hits the brane, \ie{} $z_*=\infty$ -- see figure \ref{fig:braneflat}. Now as $\braneAngle\to \pi/2$ from above, the QES runs off towards the horizon and consequently no QES exists for $\braneAngle<\pi/2$. For $\DGPRatio=-1$, one can argue that for $\braneAngle<\pi/2$, the QES is stuck to the defect, \ie $z_\QES=\infty$. As increasing (decreasing) $\DGPRatio$ beyond $1$ ($-1$) means the DGP entropy contribution exerts a greater force pushing the QES towards the horizon (the defect), it follows that no QES exists for $\braneAngle<\pi/2$ when $\DGPRatio>1$
  ($\DGPRatio<-1$). In these parameter ranges, the naive `solutions' obtained from eq.~\eqref{eq:vanilla} are unphysical, \ie have the
  RT surface anchored in the unphysical region behind the brane.} From figure \ref{fig:a}, we see that this is the range of $\DGPRatio$ for which
there exists some $\braneAngle<\pi/2$ such that the DGP gradient has not
overpowered the bulk term of eq.~\eqref{eq:abr} to push the QES to the
asymptotic boundary $y=0$ or to the horizon $y=\infty$.

To be more precise, we must consider properties of the $\bByzs$ function
introduced in eq.~\eqref{eq:oven}. For $-1<\DGPRatio < 1$, some (numerically deduced) facts about $\bByzs(d,\DGPRatio,\braneAngle)$ are
that it is decreasing in $\DGPRatio$ and increasing in $\braneAngle$. Moreover,
\begin{align}
	\bByzs(d,\DGPRatio,\text{$\braneAngle$ close to $0$})
	={}& -(1+\DGPRatio)^{\frac{1}{d-1}}
       \braneAngle^{-\frac{d-2}{d-1}} [1+O(\braneAngle^2)] + \frac{\RTDiam}{z_*}
       \label{eq:alcoholic}
	\\
	\bByzs(d,\DGPRatio,\text{$\braneAngle$ close to $\pi$})
	={}& (1-\DGPRatio)^{\frac{1}{d-1}}
       (\pi-\braneAngle)^{-\frac{d-2}{d-1}} \{1+O[(\pi-\braneAngle)^2]\}.
\end{align}
Since the former diverges negatively while the latter diverges positively, it
follows that there exists a critical angle $\theta_c$ for which
$\bByzs(d,\DGPRatio,\theta_c)=0$. For $-1<\DGPRatio<1$, we have
$0<\theta_c<\pi/2$ with $\theta_c\to 0,\pi/2$ as $\DGPRatio\to -1,1$,
respectively.\footnote{In particular then, no islands form with $\DGPRatio>1$ in
  the regime of interest with $\theta_\mt{B}\le\pi/2$.} The physical
significance of $\theta_c$ can be seen from the second equality of
eq.~\eqref{eq:dieAlone}: for $\braneAngle$ above $\theta_c$, there exist
extremal surfaces which intersect the brane; as $\braneAngle\to \theta_c$ from
above, $z_*,z_{\QES}, y_{\QES}$ run off to $+\infty$ as $\sim
(\braneAngle-\theta_c)^{-1}$; finally, for $\braneAngle < \theta_c$, no extremal
surfaces exist which intersect the brane. In Figure \ref{fig:aa}, we plot the
critical angle $\theta_c$ as a function of $\DGPRatio$ for various $d$.

Before continuing, let us briefly note a number of peculiarities which arise
when $|\DGPRatio|>1$. First, for $\DGPRatio>1$, there exists a range of
$\braneAngle\gtrsim\pi/2$ for which no extremal surfaces of any kind exists, \ie
the RT prescription fails completely. This may indicate that there is no CFT
plus defect theory which can be dual to a bulk with this range of parameters --
of course, the brane has a negative tension in this regime and so there is no
effective gravitational theory on the brane. Second, recall that as
$\DGPRatio\to -1$ from above, the coefficient of the Einstein-Hilbert action
vanishes, leading to a breakdown of semiclassical Einstein gravity, as mentioned
in footnote \ref{bigtoe}. Further taking $\DGPRatio <-1$ then corresponds leads
to an unphysical ghost-like gravity action in the brane theory. At any rate,
from here on, we shall restrict our focus to $-1< \DGPRatio <1$.

Now we have two competing possible RT surfaces: for $\braneAngle \le \pi/2$,
extremal planes anchored on the entangling surfaces to either side of the brane,
which correspond to a no-island phase; and, for $\braneAngle > \theta_c$,
extremal surfaces which intersect the brane, corresponding to an island phase.
As both types of surfaces exist for $\theta_c < \braneAngle < \pi/2$, the RT
formula instructs us to choose the surface with the smallest area in this
parameter space. Thus, we consider the area difference:
\begin{align}
  \left[
  \frac{\area(\RT)}{4 G_\bulk}
  + \frac{\area(\RTbrn)}{4 G_\brane}
  \right]_{\island}
  -\left[
  \frac{\area(\RT)}{4 G_\bulk}
  \right]_{\noisland}
  =& -\frac{\LAdS^{d-1} \volPerp}{2(d-2) G_\bulk z_*^{d-2}} \bByzs(d,\DGPRatio,\braneAngle)
     \label{eq:loneliness}
\end{align}
where we have used eqs.~\eqref{eq:oven}, \eqref{eq:butter},
\eqref{eq:depression}, and the hypergeometric function identity\footnote{This
  can be proven using eq. (15.1.8) and (15.2.25) of \cite{abramowitz}.}
\begin{align}
  \begin{split}
    \label{eq:hypergeometric_identity}
    \MoveEqLeft[3] \hyperF\left[ \frac{1}{2}, \frac{d}{2(d-1)}-1;
      \frac{d}{2(d-1)}; w \right]
    \\
    =& \sqrt{1-w} + \left(\frac{w}{d}\right) \hyperF\left[ \frac{1}{2},
      \frac{d}{2(d-1)}; \frac{d}{2(d-1)} + 1;
      w \right].
  \end{split}
\end{align}
From eq.~\eqref{eq:loneliness}, we see that whenever the island- and
no-island-type surfaces coexist, the island-type surface always gives a lower
area and is thus the surface picked out by the RT formula. Moreover, we see that
entropy transitions continuously between the island and no-island phases at the
critical angle $\theta_c$ where $\bByzs(d,\DGPRatio,\theta_c)=0$. Altogether, we
find that, for $\braneAngle<\theta_c$, we are in the no-island phase where the
RT surface is given by planes falling straight into the bulk, and, for
$\braneAngle >\theta_c$, we transition to an island phase where the RT surface
is given by extremal surfaces which intersect the brane and form an island.

To gain intuition for the critical angle $\theta_c$ from the brane perspective,
we note from eq.~\eqref{eq:korma} that eq.~\eqref{eq:loneliness} can may be
approximated as
\begin{align}
  \begin{split}
    \MoveEqLeft[3] \left[ \frac{\area(\RT)}{4 G_\bulk} + \frac{\area(\RTbrn)}{4
        G_\brane} \right]_{\island} -\left[ \frac{\area(\RT)}{4 G_\bulk}
    \right]_{\noisland}
    \\
    =&
    -\frac{\LAdS^{d-1}}{4 G_\bulk} \frac{ 4
    \sqrt{\pi}\,\Gamma\!\left[\frac{d}{2(d-1)}\right] }{
    (d-2)\,\Gamma\!\left[\frac{1}{2(d-1)}\right] } \frac{\volPerp}{z_*^{d-2}}
    \\
    &+ \volPerp\left(\frac{\LAdS}{z_{\QES}}\right)^{d-2}\left\{
      \frac{1}{4G_\eff} +
      O\left[\frac{1}{G_\bulk}\left(\frac{z_{\QES}}{z_*}\right)^{2(d-1)}\right]
    \right\}
  \end{split}
      \label{eq:citalopram}
\end{align}
in the small
$\braneAngle$ limit. Building upon the discussion given below
eq.~\eqref{eq:korma}, we interpret the RHS as giving a change in generalized
entropy due to the introduction of the island in the effective theory of the
asymptotic boundary and brane. Namely, comparing with the island rule
\reef{eq:islandformula}, the first term on the RHS of eq.~\eqref{eq:citalopram} gives the change in
$S_\mt{QFT}$ due to the introduction of the island,
and the second term gives Bekenstein-Hawking entropy of the QES. Hence, for
$\braneAngle>\theta_c$, the island phase is favoured as the introduction of the
island reduces generalized entropy. For $\braneAngle < \theta_c$, the QES ceases
to exist and only the no-island phase is possible.

We briefly comment that, unlike for the CFT region considered in
\cite{Chen:2020uac}, the addition of topological terms to the bulk gravity
theory does not change the favourability between the island and no-island phases
of the belt geometry. This is because such a modification can only effect a
topological contribution to the Wald-Dong entropy formula and, for the belt
geometry, the RT surfaces in both phases have vanishing Euler characteristic.
Namely, the RT surface of the island phase has the topology of an infinite strip
while the RT surface of the no-island phase consists of two half-planes. Thus,
the topological contribution would not favour one phase over the other.

In closing, we note that, unlike the $d=2$ case \cite{Almheiri:2019yqk}, we have
found that in the small $\braneAngle$ limit, where an effective theory of
gravity plus quantum matter emerges on the brane, islands typically do not exist
for extremal black holes in $d\ge 3$. To be more precise,
eq.~\eqref{eq:alcoholic} and figure \ref{fig:aa} suggest that
$\theta_c^{d-2}\sim 1+\DGPRatio$. It is still possible to stay in the island
phase by tuning $1+\DGPRatio$ to scale as $\sim \braneAngle^{d-2}$.
However, from eq.~\reef{Newton3}, we see that this limit $\DGPRatio\to -1^+$
corresponds to $G_\eff\to +\infty$, leading to a breakdown of the semiclassical
description of the effective brane theory \cite{Chen:2020uac} (as mentioned in
footnote \ref{bigtoe}.). We remark that, unlike for non-extremal black holes to
be discussed in section \ref{sec:nonextremal}, there is no immediate information
paradox that arises as a result of the lack of islands in the extremal case
here.

%% file: images/ExtremalBlackHole.pdf_tex
\begingroup%
  \makeatletter%
  \providecommand\color[2][]{%
    \errmessage{(Inkscape) Color is used for the text in Inkscape, but the package 'color.sty' is not loaded}%
    \renewcommand\color[2][]{}%
  }%
  \providecommand\transparent[1]{%
    \errmessage{(Inkscape) Transparency is used (non-zero) for the text in Inkscape, but the package 'transparent.sty' is not loaded}%
    \renewcommand\transparent[1]{}%
  }%
  \providecommand\rotatebox[2]{#2}%
  \newcommand*\fsize{\dimexpr\f@size pt\relax}%
  \newcommand*\lineheight[1]{\fontsize{\fsize}{#1\fsize}\selectfont}%
  \ifx\svgwidth\undefined%
    \setlength{\unitlength}{453.54330709bp}%
    \ifx\svgscale\undefined%
      \relax%
    \else%
      \setlength{\unitlength}{\unitlength * \real{\svgscale}}%
    \fi%
  \else%
    \setlength{\unitlength}{\svgwidth}%
  \fi%
  \global\let\svgwidth\undefined%
  \global\let\svgscale\undefined%
  \makeatother%
  \begin{picture}(1,1.15)%
    \lineheight{1}%
    \setlength\tabcolsep{0pt}%
    \put(0,0){\includegraphics[width=\unitlength,page=1]{images/ExtremalBlackHole.pdf}}%
  \end{picture}%
\endgroup%

%% file: images/Penrose_braneperspectiveExtremal.pdf_tex
\begingroup%
  \makeatletter%
  \providecommand\color[2][]{%
    \errmessage{(Inkscape) Color is used for the text in Inkscape, but the package 'color.sty' is not loaded}%
    \renewcommand\color[2][]{}%
  }%
  \providecommand\transparent[1]{%
    \errmessage{(Inkscape) Transparency is used (non-zero) for the text in Inkscape, but the package 'transparent.sty' is not loaded}%
    \renewcommand\transparent[1]{}%
  }%
  \providecommand\rotatebox[2]{#2}%
  \newcommand*\fsize{\dimexpr\f@size pt\relax}%
  \newcommand*\lineheight[1]{\fontsize{\fsize}{#1\fsize}\selectfont}%
  \ifx\svgwidth\undefined%
    \setlength{\unitlength}{1026.14173228bp}%
    \ifx\svgscale\undefined%
      \relax%
    \else%
      \setlength{\unitlength}{\unitlength * \real{\svgscale}}%
    \fi%
  \else%
    \setlength{\unitlength}{\svgwidth}%
  \fi%
  \global\let\svgwidth\undefined%
  \global\let\svgscale\undefined%
  \makeatother%
  \begin{picture}(1,0.46132597)%
    \lineheight{1}%
    \setlength\tabcolsep{0pt}%
    \put(0,0){\includegraphics[width=\unitlength,page=1]{images/Penrose_braneperspectiveExtremal.pdf}}%
    \put(0.07787226,0.26200548){\color[rgb]{0,0,0}\makebox(0,0)[lt]{\lineheight{1.25}\smash{\begin{tabular}[t]{l}$t=0$\end{tabular}}}}%
    \put(0.04188418,0.42191076){\color[rgb]{0,0,0}\makebox(0,0)[lt]{\lineheight{1.25}\smash{\begin{tabular}[t]{l}horizon\end{tabular}}}}%
    \put(0,0){\includegraphics[width=\unitlength,page=2]{images/Penrose_braneperspectiveExtremal.pdf}}%
    \put(0.10521995,0.06596231){\color[rgb]{0,0,0}\makebox(0,0)[lt]{\lineheight{1.25}\smash{\begin{tabular}[t]{l}defect\end{tabular}}}}%
    \put(0,0){\includegraphics[width=\unitlength,page=3]{images/Penrose_braneperspectiveExtremal.pdf}}%
    \put(0.62675438,0.26200502){\color[rgb]{0,0,0}\makebox(0,0)[lt]{\lineheight{1.25}\smash{\begin{tabular}[t]{l}$t=0$\end{tabular}}}}%
    \put(0.65122944,0.06596396){\color[rgb]{0,0,0}\makebox(0,0)[lt]{\lineheight{1.25}\smash{\begin{tabular}[t]{l}defect\end{tabular}}}}%
    \put(0,0){\includegraphics[width=\unitlength,page=4]{images/Penrose_braneperspectiveExtremal.pdf}}%
    \put(0.04359212,0.01533437){\color[rgb]{0,0,0}\makebox(0,0)[lt]{\lineheight{1.25}\smash{\begin{tabular}[t]{l}a.\end{tabular}}}}%
    \put(0.59102744,0.01533491){\color[rgb]{0,0,0}\makebox(0,0)[lt]{\lineheight{1.25}\smash{\begin{tabular}[t]{l}b.\end{tabular}}}}%
  \end{picture}%
\endgroup%

%% file: sections/06_JT.tex

In this section, we specialize to the case of $d=2$ which, as mentioned in the
main text, requires a slightly different treatment. We begin with a discussion
of the induced action on the brane, supplemented with JT gravity. Next, we review the bulk
AdS$_3$ and brane AdS$_2$ geometries. Finally, we study extremal
surfaces serving as candidate RT surfaces to determine the entropy in the two phases, with and without an island,
leading to the Page curve. At leading order in an expansion in terms of small brane angles, \ie $\theta_\mt{B}\to 0$, our results precisely agree to those of \cite{Almheiri:2019yqk}. However, we can also retain the subleading terms, which produce
corrections due to the finite UV cutoff on the brane. 

\subsection{Brane action}
\label{sec:insideWeBothKnow}
We begin by briefly reviewing the modifications for the induced brane action in
two dimensions -- a more complete discussion can be found in
\cite{Chen:2020uac}.

Let us start in the absence of JT gravity, considering only the brane action
$I_\induced$ induced by the bulk Einstein-Hilbert action (with cosmological
constant) given in eq.~\eqref{act2}, its corresponding Gibbons-Hawking action on
the brane, and the brane tension term
\begin{align}
  I_{\brane}
  =& - T_o \int d^2x\; \sqrt{-\tilde{g}}.
     \label{eq:weVeKnownEachOther}
\end{align} 
As we saw in section \ref{sec:RS}, the induced action evaluated for higher
dimensions contains coefficients with factors of $(d-2)$ (see eq.~\reef{act3}),
which prevent a naive substitution $d\to 2$. Instead, redoing the calculation
specifically in two dimensions, the induced brane action is found to
be \begin{equation}
  \label{induct}
  I_{\mt{induced}} = \frac{1}{16 \pi G_\mt{eff}}\int d^2x\sqrt{-\tilde{g}}
  \Big[\frac{2}{\ell_\mt{eff}^2} - \tilde{R} \,\log\left(-\frac{L^2
    }{2}\tR\right)+\tR +\frac{L^2}{8}\,\tilde{R}^2 +\cdots\Big]\,. \end{equation}
where the two effective scales are
\begin{align}\label{2deff}
  \left( \frac{\LAdS}{\ell_\mt{eff}} \right)^2
  =& 2\left( 1-4\pi \Gbk \LAdS T_o \right)\;,
  &
    G_\mt{eff}=&\Gbk/\LAdS\,.
\end{align}
Notice that while the first equality follows the same definition as in higher
dimensions, the second one must be redefined for $d=2$  (c.f.~eq.~\eqref{Newton3}). The unusual logarithmic term above arises from the nonlocal Polyakov action \cite{Skenderis:1999nb}, which appears from integrating out the two-dimensional CFT on the brane -- see the discussion in \cite{Chen:2020uac}. In the absence of any DGP terms in the brane,
extremization of $I_{\induced}$ leads to an $\AdS_2$ brane with radius of
curvature $\ell_{\mt{B}}$ related to $\ell_{\mt{eff}}$ in the same way as in
higher dimensions (\ie{} through eqs.~\eqref{curve2} and
\eqref{Newton3}): \begin{equation}
  \label{curve33}
  \frac{L^2}{{\ell}_\mt{eff}^2}= f\!\(\frac{L^2}{\ell_\mt{B}^2}\)\equiv
  2\(1-\sqrt{1-\frac{L^2}{\ell_\mt{B}^2}}\,\) \,. \end{equation} Thus, as in the higher
dimensional case, the large tension limit leads to $\Lbrn\gg \LAdS$ and a small
brane angle $\braneAngle$ in eq.~\eqref{curve2}. In this limit, the brane moves
towards the would-be $\AdS_3$ boundary at $\theta=0$, giving rise to a
logarithmic UV divergence in eq.~\eqref{induct} as $\LAdS/\Lbrn \to 0$.

Throughout the main text, we considered supplementing the brane action with a
DGP term --- compare eqs.~\eqref{act1} and \reef{eq:weVeKnownEachOther}. In two
dimensions, an Einstein-Hilbert action is topological and so it is common to
instead consider Jackiw-Teitelboim (JT) gravity
\cite{Jackiw:1984je,Teitelboim:1983ux} in the brane theory (\eg see recent
discussions of quantum extremal islands in $d=2$, \eg \cite{Almheiri:2019psf,
  Almheiri:2019hni, Almheiri:2019yqk, Chen:2019uhq}). Following
\cite{Chen:2020uac}, we then choose the brane action as \begin{equation}
  \label{eq:ahhhMySoul}
  I_\mt{brane}= I_\mt{JT} -\frac{1}{4\pi \Gbk L}\int d^2x\sqrt{-\tilde{g}}\,, \end{equation}
with the JT action taking the usual form (again as in section \ref{sec:RS}, we
are omitting boundary terms), \begin{equation}\label{JTee} I_\mt{JT} =\frac{1}{16\pi
    G_\mt{brane}}\int d^2x\sqrt{-\tilde{g}}\left[\Phi_0\,\tilde{R}+
    \Phi\left(\tilde{R}+\frac{2}{\ell^2_\mt{JT}} \right)\right]\,. \end{equation} The
Einstein-Hilbert term, though topological, still contributes to the generalized
entropy with weight $\Phi_0$. With the addition of JT gravity on the brane in
eq.~\eqref{eq:ahhhMySoul}, we arrive at the following induced action on the brane,
\begin{align}
  &&I_\mt{induced}=\frac{1}{16 \pi G_\mt{eff}}\int d^2x\sqrt{-\tilde{g}} \Big[ -  \tilde{R} \,\log\left(-\frac{L^2 }{2}\tR\right) +\frac{L^2}{8}\,\tilde{R}^2 +\cdots\Big]
     \nonumber\\
  &&\qquad+\frac{1}{16\pi G_\mt{brane}}\int d^2x\sqrt{-\tilde{g}}\left[\tilde\Phi_0\,\tilde{R}+ \Phi\left(\tilde{R}+\frac{2}{\ell^2_\mt{JT}}
     \right)\right]\,,
     \label{fullindyact}
\end{align}
where we have redefined the topological part of the dilaton upon collecting the
coefficients multiplying an Einstein-Hilbert terms,
\ie \begin{equation}\label{shift0} \tilde \Phi_0 =\Phi_0 +
  G_\mt{brane}/G_\mt{eff}\,. \end{equation} Note that we have discarded the
usual tension coefficient $T_o$ in eq.~\eqref{eq:ahhhMySoul} and instead chosen the
tension such that no cosmological constant appears in the first line of eq.~\eqref{fullindyact} for simplicity. In eq.~\reef{fullindyact}, it is clear
that varying $\Phi$ yields an equation of motion simply setting the radius of
curvature on the brane to $\ell_{\mt{B}}=\ell_{\mt{JT}}$. The limit of small
brane angle $\braneAngle$, related to $\Lbrn$ still through the first equality
of eq.~\eqref{curve2}, is therefore obtained by taking $\LJT\ll \LAdS$. Note
that this leads to a logarithmic UV divergence in eq.~\eqref{fullindyact}
similar to the non-JT case, as mentioned below eq.~\eqref{curve33}. Similarly,
the source-free equations of motion for the dilaton can then be obtained by
varying the metric and further shifting the dilaton, as discussed in
\cite{Chen:2020uac}.

The above reviews our discussion of the induced action in \cite{Chen:2020uac}.
However, we would like to compare our results for the quantum extremal surfaces
and the Page curve to those derived in \cite{Almheiri:2019yqk}. To facilitate
this comparison, we make the following field
redefinitions
\begin{align}
  \phibare_0 =&\ \frac{\Phi_0}{4G_\brane}\,,
  &
    \phibare=&\ \phibare_0+\frac{\Phi}{4G_\brane}\,,
  \\
  \phiren_0
  =&\ \phibare_0 - \frac{1}{2G_\mt{eff}}
     \log\left(\frac{L}{\ell_{\mt{JT}}}\right)\,,
  &
    \phiren =&\ \phibare - \frac{1}{2G_\mt{eff}}
     \log\left(\frac{L}{\ell_{\mt{JT}}}\right),
           \label{newdil}
\end{align}
giving the bare and renormalized values of the dilaton --- we shall clarify the
meaning of this renormalization shortly. In terms of the latter, induced action \reef{fullindyact} now reads
\begin{align}
  \begin{split}
    I_{\induced} =& \frac{1}{16 \pi G_\mt{eff}} \int d^2x\; \sqrt{-\tilde{g}}
    \left[ 
    - \tilde{R} \,\log\left(-\frac{\ell_{\mt{JT}}^2
        }{2}\tR\right) + \tilde{R} + \frac{L^2}{8}\,\tilde{R}^2 +\cdots \right]
    \\
    &+\frac{1}{4\pi } \int d^2x\; \sqrt{-\tilde{g}} \left[ \phiren\, \tilde{R}
      +\frac{2}{\ell^2_\mt{JT}} (\phiren-\phiren_0) \right]\,.
  \end{split}
      \label{eq:youReTooShyToSayIt}
\end{align}
Here, the first line eq.~\eqref{eq:youReTooShyToSayIt} may be interpreted as the renormalized
effective action produced by integating out the brane CFT, and the second line contains the renormalized JT action, which can be compared to eq.~(2) in \cite{Almheiri:2019yqk}. Here, `renormalized' means that we have absorbed the logarithmic UV divergence that would otherwise appear in the induced action\footnote{Recall that we also removed the power law divergence corresponding to the induced cosmological constant term by introducing a counterterm in eq.~\reef{eq:ahhhMySoul}.}  as
$\LAdS/\Lbrn \to 0$ into the JT action, which was achieved by the renormalization of $\phibare_0\to\phiren_0$ in eq.~\eqref{newdil}.

As before, the dilaton $\phiren$ acts as a Lagrange multiplier which fixes the
brane geometry to be locally $\AdS_2$ with radius of curvature $\Lbrn=\LJT$. The equation
of motion for the induced metric $\tilde{g}_{ij}$, on the other hand, yields the
dilaton equation of motion
\beq
  \label{fulleom}
  -\nabla_{i}\nabla_{j}\phiren+\tilde{g}_{ij}\(\nabla^2\phiren-\frac{\phiren-\phiren_0}{\ell^2_\mt{JT}}\)
  =2\pi \, \widetilde{T}^\mt{CFT}_{ij}
  =-\frac{\tilde{g}_{ij}}{4\LAdS^2 \Geff}\,
     f\!\left( \frac{\LAdS^2}{\LJT^2} \right)
     \,.
\eeq
In the final expression, we evaluated the renormalized CFT stress tensor $\widetilde{T}^\mt{CFT}_{ij}$ using the function $f$ defined in eq.~\reef{curve33}.\footnote{As noted in \cite{Chen:2020uac}, $f(\LAdS^2/\LJT^2) = \LAdS^2/\LJT^2 + O(\LAdS^4/\LJT^4)$ and hence this expression yields the expected trace anomaly $\langle (\widetilde{T}^\mt{CFT})^i{}_i \rangle = 2\times\frac{c}{24\pi}\,\tilde R$ to leading order in $\LAdS/\LJT$. But the latter also  receives additional corrections due to
the finite UV cutoff on the brane -- see eq.~(2.45) in \cite{Chen:2020uac}. Recall that the central charge of the boundary CFT is given by $c={3\LAdS}/{2\Gbk}$ and the extra factor of two in the trace anomaly arises because the brane supports two copies of this CFT.}
The standard discussions of JT gravity (\eg \cite{Maldacena:2016upp,Almheiri:2019psf}) refer to the source-free dilaton equation, \ie the RHS vanishes, but this is easily accommodated by a further shift\footnote{Note that implementing this shift in the action \reef{eq:youReTooShyToSayIt} introduces a new cosmological constant term. Hence an alternative approach would be to introduce a general brane tension $T_o$  in eq.~\reef{eq:ahhhMySoul} and then tune the latter to absorb both the corresponding (power law) UV divergence in the induced action and the RHS of the dilaton equation \reef{fulleom}.}
\beq
\phinew_0=\phiren_0+\frac{\ell^2_\mt{JT}}{4\LAdS^2 \Geff}\, f\!\left( \frac{\LAdS^2}{\LJT^2} \right)\,.
\label{goner}
\eeq

\subsection{Bulk and brane geometries}
\label{sec:whatSBeenGoingOn}

Let us now review the geometry for our current setup. Due to the simplicity of
AdS$_3$, we will find it convenient to describe RT surfaces using global
coordinates, even though we will be considering Rindler time evolution, as in
the main text. In global coordinates, we may write the bulk $\AdS_3$ metric as
\begin{align}\label{metric-global}
  ds^2 = \frac{\LAdS^2}{\cos^2 \tilde{r}}
  \left[ -d\tilde\tau^2 + d \rg^2 + \sin^2\rg\, d\varphi^2 \right]
\end{align}
where $\tilde\tau\in\mathbb R$, $\rg\in[0,\pi/2]$ and $\varphi\in [-\pi,\pi]$.

In the AdS-Rindler coordinates, the AdS$_3$ geometry becomes
\begin{align}\label{metric-rindler}
  ds^2 
        &=L^2\left( -(\rr^2-1)\,d\tau^2+\frac{d\rr^2}{r^2-1}+\rr^2 d\chi^2 \right)\, ,
\end{align}
which is just the special case of eq.~\eqref{eq:metric_rindler_bulk} for $d$=2.
Here, $\tau,\chi \in (-\infty,\infty)$ and one exterior region is given by
$r>1$. As described in section \ref{sec:RS}, the AdS-Rindler coordinates are
useful for the description of vacuum AdS as a topological black hole, such that
the boundary CFT is in a thermofield double state.
The inverse temperature with respect to time $\tau$ is $2\pi$, giving the
periodicity of $i\tau$ necessary for a smooth Euclidean continuation --- we
shall also define a dimensionful time and temperature shortly. 
Indeed, these coordinates describe a horizon at
$\rr=1$. Note that in $d=2$, the boundary geometry is flat, \ie it is simply two
copies of ${\mathbb R}^2$. The AdS-Rindler coordinates $(\tau,r,\chi)$ are
related to the global coordinates $(\tilde{\tau},\tilde{r},\varphi)$ in
eq.~\reef{metric-global} by
\begin{align}
  \tanh \tau
  =&  \frac{\sin\tilde{\tau}}{\cos\varphi\,\sin\tilde{r}}\,,
  &
    \tanh\chi
    =&\frac{\sin\varphi\,\sin\tilde{r}}{\cos\tilde{\tau}}\,,
  &
    r^2
    =&  \frac{\cos^2\tilde{\tau}-\sin^2\varphi\,\sin^2\tilde r}{\cos^2\tilde{r}}\,.
       \label{onemore}
\end{align}

As described above in section \ref{sec:insideWeBothKnow}, extremizing the brane
action in eq.~\reef{fullindyact} with respect to $\Phi$ (or
eq.~\reef{eq:youReTooShyToSayIt} with respect to $\phiren$) fixes the intrinsic
brane geometry to be $\AdS_2$ with radius of curvature
$\ell_{\mt{B}}=\ell_{\mt{JT}}$. This becomes the $\theta=\theta_\mt{B}$ slice of
the AdS$_3$ metric written as in eq.~\reef{metric33}, where $\theta_\mt{B}$ is
determined by \begin{equation} \sin\theta_\mt{B}=\frac{L}{\LJT}\,,
  \label{curve99}
\end{equation} as in eq.~\reef{eq:brane_curvature_scale}. We write the induced metric on
the brane as
\begin{equation} \LJT^2 ds_{\AdS_2}^2 = \LJT^2 \left(-(\rho^2-1)\, d\tau^2 +
    \frac{d\rho^2}{\rho^2-1}\right) = -\frac{4\pi^2\,\LJT^2}{\beta^2 }\, \frac{
    d\yAlm^+ d\yAlm^- }{ \sinh^2\!\( \frac{\pi(\yAlm^+ - \yAlm^-)}{\beta}
  \right) }\,.
\label{eq:andWeReGonnaPlayIt}
\end{equation}
The first line element with $(\tau,\rho)$ is simply the special case of
AdS-Rindler coordinates given in eq.~\reef{eq:metric_rindler_brane} with $d=2$.
The light-cone coordinates $(\yAlm^+,\yAlm^-)$ in the second line element are
those used by \cite{Almheiri:2019yqk}, whose results we wish to compare against.
The relationship between $(\tau,\rho)$ and $(\yAlm^+,\yAlm^-)$ is given by
\begin{align}
  \tau
  =& \frac{\pi(\yAlm^+ + \yAlm^-)}{\beta}
     = \frac{2\pi t}{\beta}\, ,
  &
    \rho
    =& \coth\left[ \frac{\pi(\yAlm^+ - \yAlm^-)}{\beta} \right].
       \label{eq:twoHouseholds}
\end{align}
Given that the TFD has temperature $\frac{1}{2\pi}$ with respect to
dimensionless time $\tau$, we have introduced the dimensionful time
$t=\frac{\yAlm^+ + \yAlm^-}{2}$ where the temperature becomes
$T=1/\beta$.\footnote{This is the same time coordinate introduced below
  eq.~\eqref{eq:metric_rindler_bulk}, though the relation $\beta=2\pi\,R$ loses
  its meaning as there is no spatial curvature in $d=2$.}

On the brane, eq.~\eqref{fulleom} is easily solved for the dilaton profile in
terms of $\rho$ or $\yAlm^\pm$:
\begin{align}
  \phiren &=\phinew_0+ \frac{2\pi\phi_r}{\beta} \, \rho
            = \phinew_0
            + \frac{2\pi\phi_r}{\beta}\, \coth\!\left[
            \frac{\pi(\yAlm^+ - \yAlm^-)}{\beta}
            \right] \,,
            \label{eq:dilatonprof}
\end{align}
where $\phi_r$ is a constant introduced in \cite{Almheiri:2019yqk} (see eq.~(18)
and discussion below (2) there).

In the AdS-Rindler metric given in eq.~\reef{metric-rindler}, we introduce a
surface of large constant $\rr=\rr_\UV$ which will serve as the UV cutoff
surface. Then following \cite{Almheiri:2019yqk}, we take the induced metric on
this surface as the background metric for the bath CFT, \ie \begin{equation}
  ds^2_\mt{CFT}=\LAdS^2 r_\UV^2(-d\tau^2 + d\chi^2)\,,
  \label{bored22}
\end{equation} with the conformal defect at $\chi=0$. Now the light-cone coordinates
$\yAlm^\pm$ can be extended to describe the geometry of $\AdS_3$ bulk, and in
particular the bath region on the asymptotic boundary near $\theta=\pi$ as well
as the brane geometry given in eq.~\reef{eq:andWeReGonnaPlayIt} at $\theta=\theta_\mt{B}$, by
taking an AdS$_3$ metric in the form eq.~\reef{metric33}. Indeed, on the
asymptotic boundary, with metric given in eq.~\reef{bored22}, $\yAlm^\pm$ are related to $(\tau,\chi)$
with\footnote{We should note that the geometry in \cite{Almheiri:2019yqk} can be
  seen as a ${\mathbb Z}_2$ orbifold of our setup (see section \ref{twod}).
  Hence they would only consider $\chi<0$ 
  of the flat boundary geometry in eq.~\reef{bored22}. Therefore, the extension
  of the null coordinates that we are discussing here has to be considered
  separately for each side of the conformal defect. As a technical point, let us
  add that in \cite{Almheiri:2019yqk}, the sign of the spatial coordinate on the
  brane is reversed so that $\yAlm^+-\yAlm^->0$ describes the asymptotic
  boundary while $\yAlm^+ - \yAlm^-<0$ describes the brane. Here, $y^+ - y^-$ is
  always positive and $\theta = \pi,\braneAngle$ correspond respectively to the
  bath and brane. 
  \label{foot:inFairVerona}}
\begin{align}
  \yAlm^\pm
  =& \frac{\beta(\tau\mp \chi)}{2\pi}\,,
  &
    ds^2_\mt{CFT}
    =& -\left( \frac{2\pi\LAdS r_\UV}{\beta} \right)^2  d\yAlm^+ d\yAlm^-\,.
       \label{eq:bothAlikeInDignity}
\end{align}

As in higher dimensions, we are interested in computing the entanglement entropy
of a boundary region $\bdyReg$ comprised of all of the points with $|\chi|\ge
\chi_\Sigma$ in the two baths (associated with the two copies of the CFT
entangled in the TFD state). That is, this region is the complement of two
intervals (`belts') centered on the conformal defects in the two boundaries
(which corresponds to the intersection of the brane with the asymptotic boundary
-- see figure \ref{fig:timeslice}). Focusing on a single Rindler wedge and on
one side of the brane, the entangling surface is located at a fixed
$\chi=-\chi_\Sigma<0$, which we define as \begin{equation}
  \frac{\yAlm^+-\yAlm^-}{2}=\bAlm>0 \qquad{\rm with}\quad
  \bAlm=\frac{\beta}{2\pi}\,\chi_\Sigma\,,
  \label{bored23}
\end{equation} for all Rindler times $\tau$. Similar assignments apply for the patches
covering the other portions of the boundary.

Finally, we note that going to the asymptotic boundary (with $\tilde{r}\to\pi/2$
and $\rr\to \infty$), eq.~\reef{onemore} yields the relation of the global and
Rindler coordinates on the boundary:
\begin{align}\label{bdycoords}
  \tan \varphi&=\frac{\sinh \chi}{\cosh \tau}\ \ \ ,\ \ \ \tan \tilde\tau=\frac{\sinh \tau}{\cosh \chi}\ ,
\end{align}
which allow us to simplify some calculations below. It will be useful to denote
the (time-dependent) global coordinate angle of the entangling surface at
$\chi=-\chi_\Sigma$ as $\varphi_\Sigma$.

\subsection{Entropies: Island and no-island phases}
\label{sec:weKnowTheGame}

Now we turn to the problem of computing entropies using the RT formula in the
background of the hyperbolic AdS$_3$ black hole coupled to the AdS$_2$ brane
with JT gravity. Specifically, we wish to compute the entropy of the region
$\bdyReg$ complementary to belts centered on the defects, as described at the
end of subsection \ref{sec:whatSBeenGoingOn}. In the island and no-island phases
the RT formula equates the entropy to:
\begin{align}\label{SgenJT}
  \left[
  \frac{\area(\RT)}{4 G_\bulk}
  + \phibare_\QES
  \right]_{\island}
  \ \ ,\ \ \left[
  \frac{\area(\RT)}{4 G_\bulk}
  \right]_{\noisland} .
\end{align}
The RT variational problem instructs us to consider extremal co-dimension two
surfaces $\RT$ in the bulk, which in AdS$_3$ are simply geodesics. Although we
are primarily concerned with evolution in Rindler time, the boundaries of the
entangling surface are simply four points; these can always be simultaneously
placed on a surface of constant global time. This property, not present in
higher dimensions, allows us to simplify the analysis by using global
coordinates\footnote{The fact that the endpoints reside at constant global time,
  together with the conservation of the charge associated with the global time
  Killing vector (obtained by dotting with the RT tangent) implies that the RT
  surfaces themselves must reside on constant global time slices.} as seen
below.

Now just as in higher dimensions, the minimization procedure yields two
competing phases. At early times, the minimal surfaces cross the Rindler horizon
avoiding the brane and the entropy is given purely by the bulk length of the RT
surface, as in the second of the expressions \eqref{SgenJT}. This length
stretches with Rindler time and leads to a growing entropy. At late times the RT
surfaces go across the brane instead, leading to an island where the
contribution of the dilaton becomes important, as shown in the first of the
expressions \eqref{SgenJT}. As in the rest of the paper, we restrict to the
regime of small brane angle $\braneAngle$.

\begin{figure}[t]
	\def\svgwidth{.7\linewidth} \centering{ 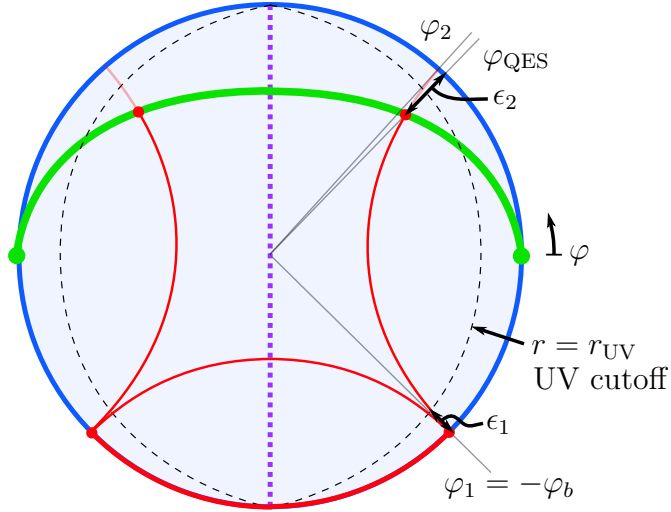 }
  \caption{A slice of constant global time in AdS$_3$, showing the two phases of
    the generalized entropy. The two cutoffs $\epsilon_{1,2}$ involved in the
    computation are associated to the UV cutoff at the asymptotic boundary and
    the brane, respectively. The global coordinate angles $\varphi_1,\varphi_2$
    relate to the RT surface opening angle, while $\varphi_\QES$ is the angle at
    which the RT surface intersects the brane and corresponds to the boundary of
    the island. Recall the geometry is cut at the brane and continued by gluing
    it to another copy.}
  \label{fig:timeslice}
\end{figure}

We begin by considering geodesics and their lengths in global coordinates. As is
well known, a convenient way to parametrize the RT surfaces on constant global
time $\tilde\tau$ is by using two anchoring points $\varphi_1,\varphi_2$, where
geodesics are given by
\begin{align}\label{RTsurface}
  \sin(\tilde r) \cos \left( \varphi-\frac{\varphi_1+\varphi_2}{2} \right)
  =&\cos \left( \frac{\varphi_2-\varphi_1}{2} \right)\,.
\end{align}
such that the curves hit the boundary $\tilde r\to \pi/2$ at $\varphi_1$ and
$\varphi_2$. The area (length in $d=2$) of an RT surface with this trajectory is
given by
\begin{align}\label{ART}
  \area
  =& \LAdS \sum_{i\in\{1,2\}}
     \tanh^{-1}\left[
     \csc\left( \frac{\Delta\varphi}{2} \right)
     \sqrt{
     -\cos\left( \frac{\Delta\varphi}{2} + \tilde{r}_i \right)
     \cos\left( \frac{\Delta\varphi}{2} - \tilde{r}_i \right)
     }
     \right]
  \\
  =& \LAdS \log \left[
     \frac{4\sin^2\left( \Delta\varphi/2 \right)}{\epsilon_1 \epsilon_2}
     \right]
     -\frac{\LAdS}{12}\left( 1+3\cot^2\frac{\Delta\varphi}{2} \right)
     (\epsilon_1^2+\epsilon_2^2)
     +O(\epsilon_1^4) + O(\epsilon_2^4)
     \,,
     \label{eq:youReTooBlindToSee}
\end{align}
where
\begin{align}
  \Delta \varphi =& |\varphi_1-\varphi_2|,
  & 
    \epsilon_i=\frac{\pi}{2}- \rg_i \qquad(i\in\{1,2\})
    \label{eq:epsilons}
\end{align}
are respectively the opening angle of the RT surface and the UV cutoffs (in the
global radial coordinate) at which the area integral is terminated, see figure
\ref{fig:timeslice}.

The leading order term in eq.~\eqref{eq:youReTooBlindToSee} corresponds to the
standard entanglement entropy formula of an interval on the circle
\cite{Calabrese:2004eu,Ryu:2006bv} (but allowing now for two different UV
cutoffs). We have also included the next-to-leading order terms as these will be
important for computing corrections to entropy formulas on the brane.

Now as usual, one must appropriately regularize the areas of the RT surfaces. As
explained above, we place the cutoff surface at a large holographic radius
$\rr=\rr_\UV$ in the Rindler radial coordinate. In terms of global coordinates,
this describes the surface
\begin{align}\label{cutoff}
  \sin^2(\tilde\tau)
  &= \left( \sin \rg \cos\varphi \right)^2-(\rr_\UV^2-1)\cos^2 (\rg)\ .
\end{align}
Expanding to leading order in $\rr_\UV$, one finds that the UV cutoff is
associated with a length in eq.~\eqref{eq:epsilons} given by

\begin{align}
  \epsilon_1
       = \frac{1}{\rr_\UV} \sqrt{
       \frac{
       2
       }{
       \cosh(2\tau) + \cosh(2\chi)
       }
       }
       + O(
       \rr_\UV^{-3}
       )\ .
       \label{eq:giveYouUp}
\end{align}
where we have used eq.~\eqref{bdycoords}. Here and below, we shall use
$\epsilon_1$ to denote the cutoff at the end-point of the RT surface at the
asymptotic boundary; $\epsilon_2$, on the other hand, will either be a cutoff at
the asymptotic boundary or due to the brane, depending on whether we are in the
no-island or island phase. Note that although the entropies diverge with the
regulator $\rr_\UV$, these contributions will cancel once we consider the
difference between the island and no island phases, as seen below.

Equipped with this, we can now compute the generalized entropy in the two phases
and reproduce the Page curve found in \cite{Almheiri:2019yqk}.

\paragraph{No-island phase.}We begin with the no-island phase. Here once again
due to the simplicity of AdS$_3$, the minimal surfaces lie on constant global
time slices. The RT surface consists of two pieces, one connected piece on
either side of the brane with trajectory given by eq.~\eqref{RTsurface} where
$\varphi_1=-\varphi_\Sigma$ and $\varphi_2=-\pi+\varphi_\Sigma$ (recall the
definition of $\varphi_\Sigma$ below eq.~\eqref{bdycoords}). The total RT length
is given by double eq.~\eqref{eq:youReTooBlindToSee} (due to the two pieces)
with both cutoffs $\epsilon_1,\epsilon_2$ given by eq.~\eqref{eq:giveYouUp}.
Substituting this into eq.~\eqref{eq:youReTooBlindToSee} with
$\Delta\varphi=\pi-2\varphi_\Sigma$ and using eq.~\eqref{bdycoords}, the associated
entanglement entropy in the no-island phase is

\begin{align}
  \left[
  \frac{\area(\RT)}{4 G_\bulk}
  \right]_{\noisland} &=\frac{L}{\Gbk}\log \left( 2 r_\UV \cosh\tau  \right)\nonumber\\
                      &=\frac{2c}{3}\log \left[ \frac{\beta }{\pi\cutoffbdy} \cosh\left( \frac{2\pi t}{\beta} \right)  \right],\label{eq:hawking2d}
\end{align}
where we used the Brown-Henneaux central charge
\begin{align}
  c=\frac{3\LAdS}{2\Gbk}.
  \label{eq:whereWeLayOurScene}
\end{align}
In the second line of eq.~\eqref{eq:hawking2d}, we have expressed the answer in
terms of the dimensionful time $t$, as in eq.~\eqref{eq:twoHouseholds} (see also
below eq.~\eqref{eq:metric_rindler_bulk}) and the short-distance cutoff in the
boundary CFT
\begin{align}
  \delta = \frac{\beta}{2\pi\, r_\UV}
  \label{eq:fromAncientGrudge}
\end{align}
in the $\yAlm^\pm$ coordinates on the boundary\footnote{ To be precise,
  eq.~\eqref{eq:hawking2d} computes the entropy of $\bdyReg$ in a CFT with
  metric $-d\yAlm^+ d\yAlm^-$ and short distance cutoff $\cutoffbdy$ --- here,
  $\cutoffbdy$ is both the proper distance cutoff and the cutoff in
  $\yAlm^\pm$.
  We may equivalently take the CFT metric to be the induced metric $-\left(
    \frac{\LAdS}{\cutoffbdy} \right)^2 d\yAlm^+ d\yAlm^-$, in
  eq.~\eqref{eq:bothAlikeInDignity}, with coordinate cutoff $\cutoffbdy$ in
  $\yAlm^\pm$, corresponding to a proper distance cutoff $\LAdS$ as measured by
  the induced metric.
}. 
Eq.~\eqref{eq:hawking2d} matches the entropy from eq.~(29) of
\cite{Almheiri:2019yqk}, accounting for the fact that here the central charge is
doubled since we include the regions on both sides of the brane\footnote{There
  is a typo in eq.~(29) of \cite{Almheiri:2019yqk}: inside the logarithm, it
  should be $\beta/\pi$ rather than $\pi/\beta$. The UV cutoff $\cutoffbdy$ is
  also hidden. The full answer is obtained by applying the conformal
  transformation
  \begin{align}
    w^\pm
    =&\tanh\left( \frac{\pi y_R^\pm}{\beta} \right)
       = -\coth\left( \frac{\pi y_L^\pm}{\beta} \right)
       \label{eq:makesCivilHandsUnclean}
  \end{align}
  (mapping the vacuum to a TFD) to the entropy formula
  \begin{align}
    S[-dw^+ dw^-,\cutoffbdy]
    =& \frac{c}{6}\log\left[
       -\frac{(w_R^+ - w_L^+)(w_R^- - w_L^-)}{\cutoffbdy^2}
       \right]
    \\
    \begin{split}
      \to S[-d\yAlm^+ d\yAlm^-,\cutoffbdy] =& S[-dw^+ dw^-,\cutoffbdy] -
      \frac{c}{12}\log( \partial_{y_R^+} w_R^+ \partial_{y_R^-} w_R^-
      \partial_{y_L^+} w_L^+ \partial_{y_L^-} w_L^- )
      ,
    \end{split}
           \label{eq:whereCivilBlood}
  \end{align}
  where $y_R^\pm=t\pm b$ and $y_L^\pm = t\mp b$ are the entangling surfaces on
  the $R$ and $L$ sides respectively. We have used the notation $S[ds^2,\delta]$
  to denote entropy in a CFT living in $ds^2$ with proper distance cutoff
  $\delta$ as measured by $ds^2$. Eq.~\eqref{eq:whereCivilBlood} gives the
  length of the piece of the RT surface to one side of the brane;
  eq.~\eqref{eq:hawking2d} is then exactly double
  eq.~\eqref{eq:whereCivilBlood}. \label{foot:ofTheseTwoFoes}}. For times much
larger than the thermal scale,
\begin{align}
  \left[
  \frac{\area(\RT)}{4 G_\bulk}
  \right]_{\noisland}
  =& \frac{2c}{3}\left[
     \log\left( \frac{\beta}{2\pi\cutoffbdy} \right)
     +\frac{2\pi t}{\beta}
     \right]
     + O\left(ce^{-4\pi t/\beta}\right)\ ,
     \label{eq:hawking2d2}
\end{align}
which corresponds to the linear growth predicted by Hawking.

\paragraph{Island phase.} Let us next consider the island phase. As explained in
section \ref{sec:RS}, since translations in Rindler time are an isometry, we can
use this symmetry to bring the problem to the $\tilde\tau=0=\tau$ slice. Notice
that this is also a symmetry of the dilaton profile as is clear from
eq.~\eqref{eq:dilatonprof}.

We will leave point 1 anchored on the cutoff surface near the asymptotic
boundary at global coordinate $\varphi_1=-\varphi_\Sigma$, as in the no-island
phase. But, the RT surface will now intersect the brane at its other endpoint.
Here it is important to distinguish between two different angles appearing in
the island calculation -- see figure \ref{fig:timeslice}. First, $\varphi_2$
(together with $\varphi_1$) characterize the trajectory of the RT surface, as in
eq.~\eqref{RTsurface}, such that the trajectory, when maximally extended (even
behind the brane), reaches the asymptotic boundary at $\varphi_1$ and
$\varphi_2$. The opening angle $\Delta\varphi$ appearing in
eq.~\eqref{eq:youReTooBlindToSee} is defined in terms of $\varphi_1$ and
$\varphi_2$ as per eq.~\eqref{eq:epsilons}. Second, there is the global angular
coordinate $\varphi=\varphi_\QES$ of the QES where the RT surface intersects the
brane. In the limit of vanishing brane angle $\braneAngle\to 0$,
$\varphi_\QES\to\varphi_2$ but, at finite $\braneAngle$, $\varphi_\QES \ne
\varphi_2$.

While $\epsilon_1$ is still given by eq.~\eqref{eq:giveYouUp}, the regulator
$\epsilon_2$ is now provided by the brane position and is given by
\begin{align}\label{epsilon2}
  \epsilon_2
  =&\tan^{-1}\left[ \tan(\braneAngle) \sin( \varphi_\QES ) \right]
  \\
  =& \braneAngle \sin(\varphi_\QES)
     + \frac{\braneAngle^3}{3} \sin(\varphi_\QES) \cos^2(\varphi_\QES)
     + O(\braneAngle^5)\,,
\end{align}
which we use below perturbatively in the regime of $\theta_\mt{B}\ll 1$.
From eq.~\eqref{eq:youReTooBlindToSee}, the area of the RT surface (including
the pieces to either side of the brane and to either side of the horizon) is
given in terms of $\varphi_\Sigma$ and $\varphi_\QES$ by
\begin{align}
  \begin{split}
    \left[ \frac{\area(\RT)}{4 G_\bulk} \right]_{\island} =& \frac{L}{G_\bulk}
    \log \left[ \frac{4}{\epsilon_1 \braneAngle} \frac{
        \sin^2\left(\frac{\varphi_\Sigma+\varphi_\QES}{2}\right) }{
        \sin(\varphi_\QES) } \right]
    \\
    &+ \frac{\LAdS \braneAngle^2}{G_\bulk}\left[ -\frac{1}{3} +\frac{
        \sin^2\varphi_\QES }{ 4\sin^2\left(
          \frac{\varphi_\Sigma+\varphi_\QES}{2} \right) } \right] +
    O\left(\frac{\LAdS\braneAngle^4}{G_\bulk}\right)\,.
  \end{split}
      \label{A_I}
\end{align}
We can also write this in terms of the $\yAlm^\pm$ coordinates of
\cite{Almheiri:2019yqk}, reviewed around eqs.~\eqref{eq:andWeReGonnaPlayIt} and
\eqref{eq:bothAlikeInDignity} (see also footnote \ref{foot:inFairVerona}).
Placing the belt boundary at $\theta=\pi$, $\frac{\yAlm^+-\yAlm^-}{2}=\bAlm$ and
the QES at $\theta=\braneAngle$, $\frac{\yAlm^+-\yAlm^-}{2}=\aAlm$ (matching the
$a$ and $b$ of \cite{Almheiri:2019yqk}), we find
\begin{align}
  \MoveEqLeft[2]\left[
  \frac{\area(\RT)}{4 G_\bulk}
  \right]_{\island}
  \nonumber\\
  =& \frac{\LAdS}{G_\bulk} \log\left\{ 
     \frac{4r_\UV}{\braneAngle}
     \frac{
     \sinh^2\left[\frac{\pi(\aAlm+\bAlm)}{\beta}\right]
     }{
     \sinh \frac{2\pi \aAlm}{\beta}
     }
     \right\}
     + \frac{\LAdS \braneAngle^2}{12 G_\bulk}\left\{
     \frac{
     3\sinh^2\left[ \frac{\pi(\aAlm - \bAlm)}{\beta} \right]
     }{
     \sinh^2\left[ \frac{\pi(\aAlm + \bAlm)}{\beta} \right]
     }
     - 1
     \right\}
     + O\left(\frac{\LAdS\braneAngle^4}{G_\bulk}\right)
  \\
  =& \frac{2c}{3} \log\left\{ 
     \frac{2\beta \LJT}{\pi \cutoffbdy \cutoffbrn}
     \frac{
     \sinh^2\left[\frac{\pi(\aAlm + \bAlm)}{\beta}\right]
     }{
     \sinh \frac{2\pi \aAlm}{\beta}
     }
     \right\}
     - \frac{c \cutoffbrn^2}{6 \LJT^2}
     \frac{
     \sinh\left( \frac{2\pi \aAlm}{\beta} \right)
     \sinh\left( \frac{2\pi \bAlm}{\beta} \right)
     }{
     \sinh^2\left[
     \frac{\pi(\aAlm + \bAlm)}{\beta}
     \right]
     }
     +O\left(\frac{c\cutoffbrn^4}{\LJT^4}\right),
     \label{eq:breakToNewMutiny}
\end{align} 
where, in the second line, we have written the answer in terms of the CFT
central charge $c$ and cutoff $\cutoffbdy$ (in $\yAlm^\pm$) in the bath, given
in eqs.~\eqref{eq:whereWeLayOurScene} and \eqref{eq:fromAncientGrudge}; we have
also used the proper distance UV cutoff $\cutoffbrn=\LAdS$ on the brane (hinted
at earlier below eq.~\eqref{fulleom}) with the induced metric given in eq.~\reef{eq:andWeReGonnaPlayIt} --- see discussion in
\cite{Chen:2020uac}. Using eq.~\reef{curve99} (as well as $\Lbrn=\ell_\mt{JT}$), we can write
\begin{equation}
  \theta_\mt{B}
  = \frac{\cutoffbrn}{\ell_\mt{JT}}\(1
  + \frac{\cutoffbrn^2}{6\ell^2_\mt{JT}}
  + O\left(\frac{\cutoffbrn^4}{\ell^4_\mt{JT}}\right)\)\,.
\label{bored24}
\end{equation}
Eq.~\eqref{eq:breakToNewMutiny} is to be interpreted as the von Neumann
entropy of the effective CFT spanning the asymptotic boundary and the brane. The
first term of eq.~\eqref{eq:breakToNewMutiny} precisely recovers the expected
CFT result\footnote{ To see this, we may apply the transformation between $w$
  and $y_R$ written in eq.~\eqref{eq:makesCivilHandsUnclean} to
  \begin{align}
    S\left[
    \left\{
    \begin{aligned}
      &\textstyle{-\frac{\LAdS^2 dw^+ dw^-}{\cutoffbdy^2}} & \bath
      \\
      &\textstyle{-\frac{4 \LJT^2 dw^+ dw^-}{(w^+ - w^-)^{2}}} & \brane
    \end{aligned}\right\},\LAdS\right]
                                                                  =&\frac{c}{6}\log\left[
                                                                     \frac{2\LJT}{\LAdS(w_\QES^+ - w_\QES^-)}
                                                                     \frac{-(w_\Sigma^+ - w_\QES^+)(w_\Sigma^- - w_\QES^-)}{\cutoffbdy}
                                                                     \right]
    \\
    \to
    S\left[
    \left\{ \begin{aligned}
        &\textstyle{-\frac{\LAdS^2 d\yAlm^+ d\yAlm^-}{\cutoffbdy^2}} & \bath
        \\
        &\textstyle{-\frac{4 \LJT^2 dw^+ dw^-}{(w^+ - w^-)^{2}}} & \brane
      \end{aligned}\right\},\LAdS \right]
                                                                    =& S\left[
                                                                       \left\{ \begin{aligned}
                                                                           &\textstyle{-\frac{\LAdS^2 dw^+ dw^-}{\cutoffbdy^2}} & \bath
                                                                           \\
                                                                           &\textstyle{-\frac{4 \LJT^2 dw^+ dw^-}{(w^+ - w^-)^{2}}} & \brane
                                                                         \end{aligned}\right\},\LAdS\right]
                                                                                                                                       - \frac{c}{12}\log(
                                                                                                                                       \partial_{y_\Sigma^+} w_\Sigma^+
                                                                                                                                       \partial_{y_\Sigma^-} w_\Sigma^-
                                                                                                                                       ),
                                                                                                                                       \label{eq:fromForthTheFatalLoins}
  \end{align}
  where we have used the notation $S[\bullet,\bullet]$ introduced in footnote
  \ref{foot:ofTheseTwoFoes}, and $y_\Sigma^\pm = t\pm b$ and $y_\QES^\pm = t\mp a$
  correspond to the entangling surface and the QES respectively. (In this
  footnote, we have swapped the sign of $\yAlm^+ - \yAlm^-$ on the $\AdS_2$
  brane relative to the main text, so that here $\yAlm^+ - \yAlm^->0$ and
  $\yAlm^+ - \yAlm^-<0$ correspond respectively to the bath and brane.) Then,
  the first term of eq.~\eqref{eq:breakToNewMutiny} is precisely four times
  eq.~\eqref{eq:fromForthTheFatalLoins}. }, while the higher orders in
$\cutoffbrn/\LJT$ may be interpreted as corrections due to the finite UV cutoff
on the brane. Curiously, the leading order correction in
eq.~\eqref{eq:breakToNewMutiny} vanishes for the case of a zero-width belt
$\bAlm=0$, \ie{}when $\bdyReg$ completely contains the baths. We may add eq.~\eqref{eq:breakToNewMutiny} to the bare dilaton
profile $\phi$,
given by eqs.~\eqref{newdil} and \eqref{eq:dilatonprof}, evaluated at the QES, to obtain the
generalized entropy
\begin{align}
  \begin{split}
    \left[
      \frac{\area(\RT)}{4 G_\bulk}
      + \phibare_\QES
    \right]_{\island}
    =&
    2\phinew_0
    + \frac{4\pi\phi_r}{\beta}\,
    \coth\left( \frac{2\pi \aAlm}{\beta} \right)
    +
    \frac{2c}{3} \log\left\{ \frac{2\beta}{\pi \cutoffbdy } \frac{
        \sinh^2\left[\frac{\pi(\aAlm + \bAlm)}{\beta}\right] }{ \sinh \frac{2\pi
          \aAlm}{\beta} } \right\}
    \\
    &- \frac{c \cutoffbrn^2}{6 \LJT^2} \frac{ \sinh\left( \frac{2\pi
          \aAlm}{\beta} \right) \sinh\left( \frac{2\pi \bAlm}{\beta} \right) }{
      \sinh^2\left[ \frac{\pi(\aAlm + \bAlm)}{\beta} \right] } +
    O\left(\frac{c\cutoffbrn^4}{\LJT^4}\right),
  \end{split}
      \label{eq:aPairOfStarCrossedLovers}
\end{align}
where we have included dilaton contributions from the QES points on both the
left and the right of the TFD. Recall that $\phinew_0$ conveniently absorbs the part of
eq.~\eqref{eq:breakToNewMutiny} which becomes logarithmically divergent on the
brane as we take the UV limit $\cutoffbrn/\LJT \to 0$
-- see eqs.~\eqref{newdil} and \reef{goner}. This is unsurprising given
that the renormalized entropy is derivable from the renormalized matter
effective action, and that the renormalization of $\phibare_0\to\phiren_0\sim\phinew_0$ is
precisely designed to eliminate the UV divergence of the matter effective action
on the brane. The first line of eq.~\eqref{eq:aPairOfStarCrossedLovers} matches
exactly\footnote{In fact, the match between the first line if
  eq.~\eqref{eq:aPairOfStarCrossedLovers} and (19) in \cite{Almheiri:2019yqk} is
  exact even after keeping all terms collected in their ``constant''. This can
  be checked by keeping all constant terms in the von Neumann entropy
  calculation, described in eq.~\eqref{eq:fromForthTheFatalLoins}, as well as
  the topological dilaton contribution $\phinew_0$.} eq.~(19) of
\cite{Almheiri:2019yqk}, accounting for the doubling and
quadrupling of the dilaton and von Neumann entropies here (since eq.~(19) of
\cite{Almheiri:2019yqk} considers only one side of the TFD and they work with an
end of the world brane with bulk spacetime only to one side). The terms of
higher order in $\cutoffbrn/\LJT$ are the corrections due to the UV cutoff,
inherited from the von Neumann entropy in eq.~\eqref{eq:breakToNewMutiny}.

To find the location $\frac{\yAlm^+- \yAlm^-}{2}=\aAlm$ of the QES, the RT
prescription instructs us to extremize the generalized entropy given in
eq.~\eqref{eq:aPairOfStarCrossedLovers}. Symmetry has already allowed us to
restrict the QES to the same slice of Rindler time $\tau\propto
t=\frac{\yAlm^++\yAlm^-}{2}$ as the anchoring point on the asymptotic boundary.
It thus remains only to extremize eq.~\eqref{eq:aPairOfStarCrossedLovers} in the
spacial direction. Setting the derivative of
eq.~\eqref{eq:aPairOfStarCrossedLovers} in $\frac{\yAlm^+ - \yAlm^-}{2}=\aAlm$
to zero, we obtain the extremization condition:
\begin{align}
  \frac{6 \pi \phi_r}{c \beta}
  =&
     \frac{
     \sinh\left( \frac{2\pi \aAlm}{\beta} \right)
     \sinh\left[ \frac{\pi(\aAlm - \bAlm)}{\beta} \right]
     }{
     \sinh\left[ \frac{\pi(\aAlm + \bAlm)}{\beta}  \right]
     }
     \left\{
     1+\frac{\cutoffbrn^2}{4\LJT^2}
     \frac{
     \sinh\left( \frac{2\pi \aAlm}{\beta} \right)
     \sinh\left( \frac{2\pi \bAlm}{\beta} \right)
     }{
     \sinh^2\left[ \frac{\pi(\aAlm + \bAlm)}{\beta} \right]
     }
     \right\}
     + O\left(\frac{\cutoffbrn^4}{\LJT^4}\right)\,.
\end{align}
At leading
order in $\cutoffbrn/\LJT$, this matches eq.~(20) in \cite{Almheiri:2019yqk}
accounting for
the fact that we have two
copies of the CFT versus a single copy of JT gravity.
This equation can be solved for the QES position $\aAlm$ in terms of
the belt width $\bAlm$ numerically or analytically with an additional expansion
in
$\frac{\phi_r}{c\beta} \gg 1$:
\begin{align}
  \aAlm
  =& \bAlm
     + \frac{\beta}{2\pi}\left[
     \log\left(\frac{12 \pi \phi_r}{c\beta}\right)
     - \frac{\cutoffbrn^2}{4\LJT^2}
     \left( 1 - e^{-\frac{4 \pi \bAlm}{\beta}} \right)
     + O\left(\frac{\cutoffbrn^4}{\LJT^4}\right)
     + O\left( \frac{c\beta}{\phi_r} \right)
     \right],
     \label{eq:buryTheirParentsStrife}
\end{align} 
matching eq.~(21) in \cite{Almheiri:2019yqk} at leading order in $\cutoffbrn/\LJT$,
again accounting for the doubling of the CFT. We see that the leading order
correction due to finite $\cutoffbrn/\LJT$ is to push the QES further from the
bifurcation point at $\frac{y^+-y^-}{2} = +\infty$.

Having found the location of the QES, we may re-evaluate the generalized entropy
of the island phase by substituting eq.~\eqref{eq:buryTheirParentsStrife} into
eq.~\eqref{eq:aPairOfStarCrossedLovers}, obtaining
\beqa
    \left[
      \frac{\area(\RT)}{4 G_\bulk} + \phibare_\QES
    \right]_{\island}
    &=& 2\left( \phinew_0 + \frac{2\pi\phi_r}{\beta}\right)
    + \frac{2c}{3} \log\left( \frac{\beta}{\pi \cutoffbdy}
    \right) + \frac{4\pi c \bAlm}{3\beta}
    - 
    \frac{c \cutoffbrn^2}{6 \LJT^2}
    \left( 1 - e^{-\frac{4\pi\bAlm}{\beta}}\right)
    \nonumber\\
    &&\quad- \frac{c^2\beta}{18 \pi \phi_r}
    e^{-\frac{4\pi \bAlm}{\beta}}
    + c\left[
      O\left({\cutoffbrn^4}/{\LJT^4}\right)
      + O\left( {c^2\beta^2}/{\phi_r^2} \right)
    \right]\,.
      \label{eq:theFearfulPassage}
\eeqa
(We have also dropped terms of order
$\frac{c^2 \beta \cutoffbrn^2}{\phi_r \LJT^2}$
as these are inherently smaller than either the ${c\cutoffbrn^4}/{\LJT^4}$ or
${c^3 \beta^2}/{\phi_r^2}$ corrections.) The first line simply evaluates
the generalized entropy, given in eq.~\eqref{eq:aPairOfStarCrossedLovers}, at
the bifurcation surface, \ie{}taking $a\to+\infty$. In particular, we recognize
the first term as giving the Bekenstein-Hawking result for the course-grained
entropy of two black holes
\begin{align}
  2S_\BH
  =&
     \ 2\left( \phinew_0 + \frac{2\pi\phi_r}{\beta}\right)
     \,.
     \label{eq:ofTheirDeathMarkdLove}
\end{align}
This classical contribution dominates eq.~\eqref{eq:theFearfulPassage} in the
limit
$S_\BH \gg c$
and corresponds
to eq.~(30) in \cite{Almheiri:2019yqk}. The other terms on the first line of
eq.~\eqref{eq:theFearfulPassage} evaluate the von Neumann entropy, given in
eq.~\eqref{eq:breakToNewMutiny}, after re-absorbing the UV divergence on the
brane into $\phinew_0$. Specifically, the second term gives the UV
contribution from the entangling surface on the asymptotic boundary (also
appearing in the no-island phase in eq.~\eqref{eq:hawking2d}), and the third and
fourth terms give finite contributions to the renormalized entropy including a
$\cutoffbrn^2/\LJT^2$
correction. Moving to the second line in
eq.~\eqref{eq:theFearfulPassage}, we have a correction due to the displacement
of the QES location $a$ from the bifurcation point. Here, the dilaton and von
Neumann components of generalized entropy both receive contributions at order
$\frac{\phi_r}{\beta} \cdot \frac{c^2 \beta^2}{\phi_r^2} \sim
\frac{c^2\beta}{\phi_r}$
. Note that there are no dilaton corrections at
orders
$\frac{\phi_r}{\beta}\cdot\frac{c\beta}{\phi_r}$
and
$\frac{\phi_r}{\beta} \cdot \frac{c \beta \cutoffbrn^2}{\phi_r \LJT^2}$
because
the bifurcation point extremizes the dilaton profile\footnote{It is
  helpful to consider the coordinate $\rhoDom = \sqrt{\rho^2-1}$, in terms of
  which eq.~\eqref{eq:dilatonprof} reads
  $\phiren = \phinew_0 + \frac{2\pi\phi_r}{\beta}
  \sqrt{1+\rhoDom^2}
  $
  and the brane metric $\LJT^2 ds^2_{\AdS_2} =
  \LJT^2\left( -\rhoDom^2 d\tau^2 + \frac{d\rhoDom^2}{\rhoDom^2+1} \right)$,
  near the horizon $\rhoDom=0$,
  resembles the standard flat metric $-\rhoDom^2 d\tau^2 + d\rhoDom^2$ in polar coordinates with $\rhoDom$ the
  usual radial coordinate. The dilaton and the von Neumann entropy in eq.~\eqref{eq:aPairOfStarCrossedLovers} should then have an
 expansion in terms of non-negative integer powers of $\rhoDom_\QES$.
  Eq.~\eqref{eq:buryTheirParentsStrife} gives
  the first corrections to $\rhoDom_\QES=0$ at orders
  $\frac{c\beta}{\phi_r}$
  and
  $\frac{c\beta \cutoffbrn^2}{\phi_r \LJT^2}$
  , leading to the corrections mentioned in the main text.}. The order
$\cutoffbrn^2/\LJT^2$
correction in the QES location given in eq.~\eqref{eq:buryTheirParentsStrife}
is not visible at the order shown in eq.~\eqref{eq:theFearfulPassage}.

\subsection{Page curve}

Collecting together the results of the previous subsection, we have two phases.
At early times, we have the no-island phase, with generalized entropy given by
eq.~\eqref{eq:hawking2d}. Over time, this entropy grows at a rate proportional
to the temperature $1/\beta$ and the number $c$ of matter degrees of freedom
participating in Hawking radiation, as emphasized in eq.~\eqref{eq:hawking2d2}.
This growth, however is capped off by an island phase, where quantum extremal
surfaces on the brane just outside the black hole horizon surround an island,
containing a portion of the black hole interior, now belonging to the
entanglement wedge of the bath. In this latter phase, generalized entropy is
given by the constant value written in eq.~\eqref{eq:theFearfulPassage} which is
dominated by double the Bekenstein-Hawking black hole entropy, as given in
eq.~\eqref{eq:ofTheirDeathMarkdLove}. Viewing
eq.~\eqref{eq:ofTheirDeathMarkdLove} as the course-grained entropy for the two
sides of the black hole, this is precisely the expected maximal entropy of the
system.

\begin{figure}[t]
	\def\svgwidth{.4\linewidth} \centering{ 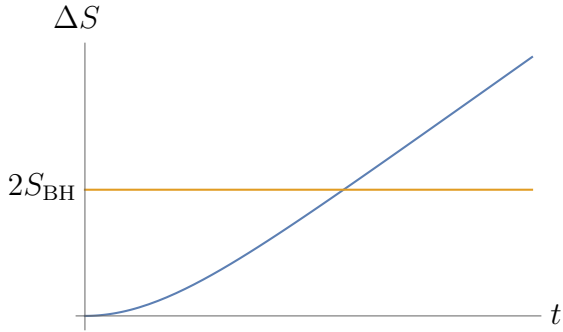 }
	\caption{Page curve for the equilibration of our topological black hole in
    $d=2$. We plot the entropy $\Delta S=S(t)-S(0)$ of the subregion on the CFT
    which is associated to the radiation, where we subtract the value of the
    entropy at $t=0$. 
    }
	\label{fig:Page-curve}
\end{figure}

To find the Page time $\tau_P=2\pi t_P/\beta$ marking the transition between
the two phases, we equate the corresponding generalized entropies given
in eqs.~\eqref{eq:hawking2d2} and \eqref{eq:theFearfulPassage}:
\beqa
  \tau_P
  = \frac{2\pi t_P}{\beta}
  &=& 
     \frac{3}{c}\left( \phinew_0 + \frac{2\pi\phi_r}{\beta} \right)
     + \log(2) + \frac{2\pi \bAlm}{\beta}
     - 
     \frac{\cutoffbrn^2}{4 \LJT^2}
     \left( 1-e^{-\frac{4\pi \bAlm}{\beta}} \right)
     - 
     \frac{c \beta e^{-\frac{4\pi\bAlm}{\beta}}}{12 \pi \phi_r}
  \nonumber\\
   &&\qquad\qquad\quad+ 
     O\left({\cutoffbrn^4}/{\LJT^4}\right)
     + 
     O\left( {c^2 \beta^2}/{\phi_r^2} \right)
     \,.
\eeqa
Overall, we recover a Page curve, with entropy growing linearly in a no-island
phase up to the Page time, and saturating to a constant maximal value in an
island phase after the Page time. In figure \ref{fig:Page-curve}, we plot the
Page curve after subtracting off the initial entropy (which includes the UV
divergences from the asymptotic boundary).

%% file: images/JT.pdf_tex
\begingroup%
  \makeatletter%
  \providecommand\color[2][]{%
    \errmessage{(Inkscape) Color is used for the text in Inkscape, but the package 'color.sty' is not loaded}%
    \renewcommand\color[2][]{}%
  }%
  \providecommand\transparent[1]{%
    \errmessage{(Inkscape) Transparency is used (non-zero) for the text in Inkscape, but the package 'transparent.sty' is not loaded}%
    \renewcommand\transparent[1]{}%
  }%
  \providecommand\rotatebox[2]{#2}%
  \newcommand*\fsize{\dimexpr\f@size pt\relax}%
  \newcommand*\lineheight[1]{\fontsize{\fsize}{#1\fsize}\selectfont}%
  \ifx\svgwidth\undefined%
    \setlength{\unitlength}{466.78511973bp}%
    \ifx\svgscale\undefined%
      \relax%
    \else%
      \setlength{\unitlength}{\unitlength * \real{\svgscale}}%
    \fi%
  \else%
    \setlength{\unitlength}{\svgwidth}%
  \fi%
  \global\let\svgwidth\undefined%
  \global\let\svgscale\undefined%
  \makeatother%
  \begin{picture}(1,0.62014768)%
    \lineheight{1}%
    \setlength\tabcolsep{0pt}%
    \put(0,0){\includegraphics[width=\unitlength,page=1]{images/JT.pdf}}%
    \put(0.76413882,0.09829839){\color[rgb]{0,0,0}\makebox(0,0)[lt]{\lineheight{1.25}\smash{\begin{tabular}[t]{l}$\epsilon_1$\end{tabular}}}}%
    \put(0.77257204,0.49501198){\color[rgb]{0,0,0}\makebox(0,0)[lt]{\lineheight{1.25}\smash{\begin{tabular}[t]{l}$\epsilon_2$\end{tabular}}}}%
    \put(0,0){\includegraphics[width=\unitlength,page=2]{images/JT.pdf}}%
    \put(0.86558684,0.30558311){\color[rgb]{0,0,0}\makebox(0,0)[lt]{\lineheight{1.25}\smash{\begin{tabular}[t]{l}$\varphi$\end{tabular}}}}%
    \put(0.81960974,0.18970013){\color[rgb]{0,0,0}\makebox(0,0)[lt]{\lineheight{1.25}\smash{\begin{tabular}[t]{l}$\rr=\rr_{\text{UV}}$\end{tabular}}}}%
    \put(0.8217274,0.14751176){\color[rgb]{0,0,0}\makebox(0,0)[lt]{\lineheight{1.25}\smash{\begin{tabular}[t]{l}$\text{UV cutoff}$\end{tabular}}}}%
    \put(0,0){\includegraphics[width=\unitlength,page=3]{images/JT.pdf}}%
    \put(0.76037123,0.5449541){\color[rgb]{0,0,0}\makebox(0,0)[lt]{\lineheight{1.25}\smash{\begin{tabular}[t]{l}$\varphi_\QES$\end{tabular}}}}%
    \put(0.68332149,0.5809776){\color[rgb]{0,0,0}\makebox(0,0)[lt]{\lineheight{1.25}\smash{\begin{tabular}[t]{l}$\varphi_2$\end{tabular}}}}%
    \put(0,0){\includegraphics[width=\unitlength,page=4]{images/JT.pdf}}%
    \put(0.71252824,0.02073216){\color[rgb]{0,0,0}\makebox(0,0)[lt]{\lineheight{1.25}\smash{\begin{tabular}[t]{l}$\varphi_1=-\varphi_b$\end{tabular}}}}%
    \put(0,0){\includegraphics[width=\unitlength,page=5]{images/JT.pdf}}%
  \end{picture}%
\endgroup%

%% file: images/Pagecurve2d.pdf_tex
\begingroup%
  \makeatletter%
  \providecommand\color[2][]{%
    \errmessage{(Inkscape) Color is used for the text in Inkscape, but the package 'color.sty' is not loaded}%
    \renewcommand\color[2][]{}%
  }%
  \providecommand\transparent[1]{%
    \errmessage{(Inkscape) Transparency is used (non-zero) for the text in Inkscape, but the package 'transparent.sty' is not loaded}%
    \renewcommand\transparent[1]{}%
  }%
  \providecommand\rotatebox[2]{#2}%
  \ifx\svgwidth\undefined%
    \setlength{\unitlength}{720bp}%
    \ifx\svgscale\undefined%
      \relax%
    \else%
      \setlength{\unitlength}{\unitlength * \real{\svgscale}}%
    \fi%
  \else%
    \setlength{\unitlength}{\svgwidth}%
  \fi%
  \global\let\svgwidth\undefined%
  \global\let\svgscale\undefined%
  \makeatother%
  \begin{picture}(1,0.61944444)%
    \put(0,0){\includegraphics[width=\unitlength,page=1]{images/Pagecurve2d.pdf}}%
    \put(-0.0462963,0.64969135){\color[rgb]{0,0,0}\makebox(0,0)[lb]{\smash{$\Delta S$}}}%
    \put(-0.14351854,0.29166666){\color[rgb]{0,0,0}\makebox(0,0)[lb]{\smash{$2S_{\text{BH}}$}}}%
    \put(1.01697525,0.01543209){\color[rgb]{0,0,0}\makebox(0,0)[lb]{\smash{$t$}}}%
  \end{picture}%
\endgroup%

%% file: sections/07_discussion.tex
In this paper, we applied the framework introduced in \cite{Chen:2020uac}, which
uses Randall-Sundrum plus DGP gravity, to extend the discussion of quantum
extremal islands in \cite{Almheiri:2019yqk} to higher dimensional black holes.
As reviewed in section \ref{sec:RS}, this setup precisely realizes the three
different perspectives of the holographic system described in
\cite{Almheiri:2019hni}. From the boundary perspective, the system is described
in terms of the $d$-dimensional boundary CFT coupled to a conformal defect. The
usual holographic dictionary then yields the bulk perspective, where the dual
description is Einstein gravity in a $(d+1)$-dimensional AdS bulk spacetime
bi-partitioned by a $d$-dimensional brane. The brane perspective is an
intermediate characterization of this system given by the $d$-dimensional
effective theory induced by the bulk theory on the asymptotic boundary and the
brane. That is, in this description, the boundary CFT spans the asymptotic
boundary, which is non-gravitational, and the brane which supports a
gravitational theory by the usual Randall-Sundrum mechanism.

We have considered the vacuum state of the system with respect to global time,
which simplifies the bulk geometry to be pure AdS. However, as discussed in
sections \ref{sec:nonextremal} and \ref{sec:numerics}, by viewing this setup in
AdS-Rindler coordinates, the global vacuum can be re-interpreted as in terms of
a massless hyperbolic black hole. This induces a similar description of the
brane geometry as a black hole of one lower dimension. The `two' asymptotic
boundaries then play the role of bath CFTs in equilibrium with the black hole on
the brane at a finite temperature $T=\frac1{2\pi R}$. Similarly, as explained in
section \ref{sec:extremal}, viewing our setup in Poincar\'e coordinates, we have
an extremal horizon in the bulk and on the brane. The latter was coupled to a
(single) $T=0$ bath CFT on the asymptotic boundary.

While islands have been numerically studied previously in
\cite{Almheiri:2019psy}, our approach provides a relatively simple setting in
which analytic calculations are possible. In particular, the doubly-holographic
nature of our model reduces the entropy calculations involving islands in the
presence of massless hyperbolic, or extremal black holes of arbitrary dimension
to holographic entanglement entropy calculations in (locally) pure AdS in one
dimension higher. From the $d$-dimensional brane perspective, when computing the
entropy of a boundary region $\bdyReg$ in the island phase, a quantum extremal
surface $\RTbrn$ marks the boundary of an island on the brane stretching to the
horizon; this island belongs to the entanglement wedge of the bath region
$\bdyReg$. From the bulk perspective, the RT surface of $\bdyReg$ runs into the
bulk from its anchoring surface $\RTbdy=\partial\bdyReg$ and intersects the
brane at $\RTbrn$. As noted in \cite{Almheiri:2019psy}, the entanglement wedge
of $\RTbrn$ stretches through the bulk and is manifestly connected to the island
on the brane in this higher-dimensional picture, despite the apparent
disconnection in the effective $d$-dimensional theory. To determine the RT
surface in an island phase, we must not only extremize the area functional
locally within the bulk, but also extremize with respect to the intersection
of the RT surface and the brane. Since the deep bulk (IR) and near-brane (UV)
contributions (further modified by DGP contributions) to the RT area,
respectively, can be interpreted as renormalized von Neumann and gravitational
Wald-Dong entropies \cite{Chen:2020uac}, this bulk calculation is equivalent to
the island prescription of extremizing generalized entropy over candidate
quantum extremal surfaces.

The most striking difference between our holographic construction
and the two-dimensional model of \cite{Almheiri:2019yqk} is that, as detailed in section 
\ref{app:Page2d},  JT gravity does not appear automatically but has to be added by hand to
the brane theory for $d=2$, in analogy to the DGP terms in higher dimensions. 
However, this may be contrasted with the induced gravity on the branes in
higher dimensions, where adding a DGP term provides finer control over the model, but is not
strictly necessary for interpreting the brane perspective as an effective CFT
coupled to gravity. Having added JT gravity as a DGP term, we showed in section
\ref{app:Page2d} that applying the RT formula in the $\AdS_3$ bulk and including
the DGP entropy, as in the $d=2$ analogue of eq.~\eqref{eq:island2}, correctly
reproduces the results of \cite{Almheiri:2019yqk} at leading order in an expansion in terms of 
small brane angles, \ie $\braneAngle\ll1$. A finite
$\braneAngle$ imposes a finite UV cutoff in the effective brane theory, as shown in eq.~\reef{bored24}, and therefore
subleading corrections to entropy formulas appear in the island phase -- see eq.~\reef{eq:theFearfulPassage}. 
Of course, with a finite UV cutoff, we would not, for instance, expect the holographic entropy to precisely
satisfy the CFT transformation rules of the entanglement entropy used by \cite{Almheiri:2019yqk}
in deriving their results \cite{Chen:2020uac}. These
corrections have the effect of pushing the QES slightly further from the horizon,
lowering the entropy of the island phase, and shifting the Page transition to an
earlier time.

As discussed extensively in \cite{Chen:2020uac}, our braneworld construction
clarifies further conceptual puzzles that appeared early discussions of quantum
extremal islands in a holographic framework, \eg \cite{Almheiri:2019hni,
  Almheiri:2019yqk, Chen:2019uhq}. One particularly confusing feature of the island rule is the (implicit) appearance of the
entanglement entropy of the QFT degrees of freedom in the region $\bdyReg$ on both sides of eq.~\eqref{eq:islandformula}. Our model puts the explanation of this fact given in \cite{Almheiri:2019yqk} on solid footing. The entanglement entropy in the left hand side of eq.~\eqref{eq:islandformula} computes the full entanglement entropy in the UV complete picture (the boundary perspective), while the entropy on the right hand side is to be interpreted in an effective, semiclassical theory (our brane perspective). In partiular, as noted in section \ref{sec:RS}, the interpretation of the brane perspective as $d$-dimensional Randall-Sundrum gravity coupled to a CFT only holds for the low energy physics at scales longer than the short distance cutoff $\tilde \delta \simeq L$. At shorter distance scales, gravity is no longer localized to the brane. In contrast, the boundary
perspective or the bulk perspective gives a complete description of quantum
state.\footnote{By the standard rules of the AdS/CFT correspondence, the
  boundary and bulk perspectives give an equivalent descriptions of the physical
  phenomena.}

\subsection*{Non-extremal black holes in higher dimensions}
As noted above, in section \ref{sec:nonextremal}, we considered AdS-Rindler
coordinates in the bulk, providing a description of the pure AdS spacetime as a
two-sided massless non-extremal black hole. A similar black hole geometry is
induced on the brane, coupled to and in equilibrium with bath regions on the
asymptotic boundary in both Rindler wedges. We considered the entropy of bath
regions $\bdyReg$ complementary to belts centered around the defects in the two
Rindler wedges. This setup, from the perspective of the effective theory on the
brane and asymptotic baths, is analogous to the two-dimensional setup at finite
temperature considered in \cite{Almheiri:2019yqk}.

We find, in particular, that the information paradox for eternal black holes and
its resolution studied in \cite{Almheiri:2019yqk} makes an expected
re-appearance in higher dimensions, as reviewed in section
\ref{sec:islands_for_non_extremal}. Again, this information paradox is resolved
by the appearance of a quantum extremal island when a second quantum extremal
surface minimizes the generalized entropy in the island rule
\eqref{eq:islandformula}. Our holographic construction translates this
competition between quantum extremal surfaces to the usual competition between
different possible RT surfaces in the holographic formula \eqref{eq:island2}. In
particular, at late times, the minimal RT entropy is provided by a second
extremal surface with components which cross the brane, as illustrated in figure
\ref{fig:RTPhases_intro}. From the brane perspective, the intersection of this
RT surface with the brane becomes the quantum extremal surfaces bounding the
island in the black hole background. The island belongs to the entanglement
wedge of the bath region $\bdyReg$. Without the appearance of islands, the
entropy of bath subregions would grow ad-infinitum. With the islands however,
the ever-growing entropy of the no-island phase is eventually capped off by the
constant finite entropy of this island phase at late times.
Further, our higher-dimensional discussion provides a simple explanation for the
saturation of entropy: the connected pieces of the RT surface in the island
phase are isolated to individual Rindler wedges and are thus invariant under
time translation (\ie forward boosts in both wedges).

Recall that the global state is pure, \ie from the boundary perspective, it is a
thermofield double state of two copies of the boundary CFT plus conformal
defect. Hence the entropy of $\bdyReg$ is identical to that of its complement
$\overline\bdyReg$. This gives a useful alternative view of the evolution of the
entropy. The region $\overline\bdyReg$ consists of a belt region centered on the
conformal defect in the two bath regions. Hence from this point of view, we are
considering the entanglement entropy of two isolated boundary regions $A$ and
$B$ on either side of the corresponding eternal black hole in the bulk. This is
essentially the same system studied in \cite{Hartman:2013qma}, except that here
the spatial sections of the bath geometry are hyperbolic in the present case. As
in \cite{Hartman:2013qma}, the entropy grows at early times but then quickly
thermalizes. In this case, the growth of the entropy stops, because it is
bounded by subadditivity, \ie $S(A\cup B)\le S(A) + S(B)$. In fact, for the
holographic system, the late time entropy saturates this inequality which erases
the mutual information between two boundary subregions. The primary difference
between the framework studied in \cite{Hartman:2013qma} and our setup, is the
addition of a backreacting brane which creates extra spacetime geometry for the
RT surfaces to traverse in this late-time island phase and so delays the onset
of this phase where the entropy is saturated. From the boundary perspective,
this longer thermalization time relative to \cite{Hartman:2013qma} can be
understood as a consequence of the large number of degrees of freedom introduced
by the conformal defect.

Further as in \cite{Almheiri:2019yqk}, we find that the island extends outside the event horizon, \ie the quantum
extremal surfaces appear outside of the horizon. If we focus on the entropy of
$\overline\bdyReg$ as above, this feature again has a simple explanation in our holographic setup,
in terms of entanglement wedge nesting. Recall in the
island phase, the individual components of the RT surface yield the entropy of
the individual belt regions on the boundary of either Rindler wedge. Since these
belts are subregions of the full hyperbolic slice on which the corresponding CFT
resides, the RT surface must remain within the corresponding Rindler wedge. That
is, the bifurcation surface of the Rindler horizon in the bulk is the RT surface
corresponding to either of the copies of the CFT in the TFD state
\cite{Casini:2011kv}, and the Rindler wedge is the corresponding entanglement
wedge. Hence, by entanglement wedge nesting \cite{Wall:2012uf,Headrick:2013zda},
the RT surface and entanglement wedge for any subregion of $H_{d-1}$ on the
boundary must lie within the corresponding Rindler wedge. Finally it was
straightforward to see from eq.~\reef{eq:donTTellMe} that the horizon on the
brane is precisely the intersection of the Rindler horizon in the bulk with the
brane. Hence the quantum extremal surface on the brane, \ie the intersection of
RT surface with the brane, must lie outside of the black hole horizon. This also
means that if we consider regions $\bdyReg$ far away from the defect, the RT
surface will pass close to the horizon. Thus, analogously to the situation
discussed in \cite{Almheiri:2019yqk}, information about the horizon seems to be
contained in the entanglement of CFT regions of the bath which are furthest from
the black hole.

\subsection*{Extremal black holes in higher dimensions}

In section \ref{sec:extremal}, by taking a Poincar\'e patch of the bulk, we
considered an extremal black hole on the brane coupled to a (single) bath CFT in
a flat background. As in \cite{Almheiri:2019yqk}, we calculated the entanglement
entropy for a bath region $\bdyReg$ which corresponded to points greater than
some distance $b$ from the conformal defect. In the case of extremal black holes, we
did not find a transition as the system was time evolved, but instead found that
the appearance of an island is linked to the choice of brane angle
$\theta_\mt{B}$ (or brane tension) and the DGP coupling.

Due to the scale invariance of Poincar\'e coordinates, it is clear that as we
push the entangling surface out in the bath region, \ie increase $b$, we
proportionately reduce the size of the island. Again, this behaviour reproduces
the intuition suggested in \cite{Almheiri:2019yqk} that the region near the
extremal horizon deep in the gravitating region (our brane) is can be contained
within the far-away portion of the bath. Actually, our higher-dimensional
picture shows that these regions are not far from each other at all --- they are
both close to the spatial infinity of the Poincar\'e patch which corresponds to
a single point in the global frame. In the other extreme $b\to 0$, we find that
regions of the brane arbitrarily close to the asymptotic boundary can be
recovered by portions of the bath sufficiently close to the defect. This is in
contrast to the two dimensional JT model, where a maximum island size exists.

Interestingly, a further qualitative deviation from the two-dimensional case is
seen at small brane angles $\theta$. Recall that, in the two-dimensional JT
model, the island phase is always dominant for belt geometries in the extremal
case \cite{Almheiri:2019yqk}. In contrast, we have found in $d>3$ that islands
cease to exist for $\theta_\mt{B}$ below some critical $\theta_c>0$. As
$\theta_\mt{B}$ approaches $\theta_c$ from above, the quantum extremal surface
of the island phase runs off infinity (\ie towards the extremal horizon). For
$\theta_\mt{B}<\theta_c$, no quantum extremal surface exists on the brane and
the bulk RT surface is simply given by two planes on either side of the brane
running straight into the bulk. Since the area of these latter surfaces is IR
finite in $d>2$, their candidacy for RT surfaces must be considered even when
the alternative island-phase surfaces exist. In fact, we find that $\theta_c$ is
precisely the angle at which the entropies of the no-island-type and island-type
surfaces match -- above this angle, the island-type surfaces remain favourable
as RT surfaces. The relevance of small $\theta_\mt{B}$ (and in particular
$\theta_\mt{B} < \theta_c$) is that in this limit, the effective theory on the
brane is described by Einstein gravity with small higher curvature corrections,
which is the most interesting parameter regime. While the lack of islands for
$\theta_\mt{B}<\theta_c$ is strikingly different from the two-dimensional case,
we remark that, in the extremal case, islands are not required from an
information-theoretic standpoint and their absence should perhaps not be
terribly surprising. This is to be contrasted with the non-extremal case, where
islands are necessary, at all brane angles, to tame the otherwise unbounded
growth of black hole entropy at late times and avoid the information paradox.

Of course, an interesting question may be to examine how varying the geometry of
the entangling surface affects the appearance of quantum extremal islands at
$T=0$. For example, rather than belt geometries, one might consider spherical
regions bisected by the conformal defect.

\subsection*{Not an ensemble}
In order to derive the island formula, a crucial ingredient was the appearance of wormholes in the
replica trick. In the two-dimensional models involving
JT gravity studied so far \cite{Almheiri:2019qdq,Penington:2019kki}, the
existence of wormholes follows from the fact that JT gravity is defined by
averaging over an ensemble of Hamiltonians. For example, JT gravity emerges as
the low energy effective description of the SYK model
\cite{Maldacena:2016hyu,Sachdev:1992fk,Sachdev:2010um,Ktalks}, or has a
definition in terms of a matrix model \cite{Saad:2019lba}.

On the contrary, our construction relies only on the standard holographic rules
of the AdS/CFT correspondence where there is no such averaging of the couplings
in the boundary theory. This is in line with the general expectations for higher
dimensional holography. This lack of averaging characterizes the UV-complete
description of the system, \ie the boundary perspective. Nonetheless, quantum
extremal islands appear in the effective description of the brane perspective
and once again one likes to understand them as remnants of replica wormholes in
the limit $n\to 1$ \cite{Chen:2020uac}. One might then wonder why -- despite the
absence of ensemble averaging -- replica wormholes should appear and connect the
gravitating region in different copies of replica trick calculations.

In fact, this is not a problem, since the different effective gravity theories in the brane picture are UV completed by a single theory of gravity in the bulk and so it is natural to consider geometries connecting the branes, \ie replica wormholes in the
effective theory. In fact considering Renyi entropy calculations in the boundary theory, one sees that the corresponding bulk geometry induces connections between the different copies of the brane theories, \ie replica wormholes on the brane \cite{Chen:2020uac}. This becomes particularly clear in our setup where the brane lives in the bulk and does not serve as a boundary of spacetime. We emphasize that here this discussion implicitly relies on the standard derivation of the RT prescription for holographic entanglement entropy
\cite{Lewkowycz:2013nqa,Dong:2016hjy} in the bulk perspective, where again we
assume that there is no ensemble averaging.\footnote{Ref.~\cite{VanRaamsdonk:2020tlr} formulates a point of view where integrating out the bath CFT generates an averaging over couplings in the theory of the conformal defect.} 

Following the logic of \cite{Marolf:2020xie}, one might be tempted to turn the logic around and, given the appearance of
wormholes in the brane description of our model, conclude that there is some form of
ensemble averaging in the dual boundary theory. However,
this line of argument implicitly assumes a precise equivalence between the
boundary theory and the `bulk' gravity theory (containing wormholes). We stress
that this equivalence does not hold in our construction. Rather the
gravitational theory on the brane is an effective theory and so the arguments of
\cite{Marolf:2020xie} do not extend to this situation. Instead, in our situation
replica wormholes appear, but wormholes connecting independent instances of the
boundary theory do not play a role. For example, this implies that higher powers of the
partition function of the boundary CFT with a conformal defect will still factorize.

Nonetheless, this issue is certainly worth further examination since in two
dimensions, replica wormholes have now been shown to play an important role in a
variety of situations, \eg calculations of Renyi entropies \cite{Penington:2019kki,
  Stanford:2020wkf}, the spectral form factor
\cite{Cotler:2016fpe,Saad:2019lba}, correlation functions \cite{Mal01,
  Saad:2019pqd}, and overlap of black hole microstate wavefunctions
\cite{Penington:2019kki, Stanford:2020wkf}. Apart from Renyi entropies, it is not clear how to
reproduce these effects in our construction, or in higher dimensions more
generally. Furthermore, it was suggested in \cite{Maldacena:2004rf,
  Penington:2019kki} that in non-averaged theories wormholes might appear as a
result of some diagonal approximation. To obtain a full quantum gravitational
answer, additional off-diagonal terms need to be added. Given that we have a
system, where wormholes appear in an approximate formulation, while at the same
time having some control over a UV complete description, one might hope that
studying our system will give an idea of how this suggestion might be realized.

\subsection*{Future directions}

Having produced a setup in which quantum extremal islands can be studied with
relative ease, some possible avenues of further investigation were suggested
above, but a number of other possible extensions to the present work also come
to mind.

For example, one may consider information-theoretic questions similar to those
raised in \cite{Almheiri:2019yqk}. There, the authors investigated whether a
protocol can be implemented to retrieve information from the island. In
particular, the entanglement wedge of the complete left system plus an interval
of the right bath contains an island that naively appears causally disconnected
from the left and the right bath interval. However, by acting with operators in
the left and right baths, it was argued that sufficient negative null energy can
be generated to pull information from this region into the left exterior, to be
picked up by the left defect and bath. One could try to reproduce this protocol
in our higher-dimensional setup using insertions of operators on the left and
right asymptotic boundaries. The negative null energy produced would then shift
the bulk horizon and hence the induced horizon on the brane.

Recall that above, we described how in the present discussion the appearance of
the quantum extremal surfaces outside of the horizon was a simple result of the
nesting of entanglement wedges from the bulk perspective. However, another
question raised by \cite{Almheiri:2019yqk} is whether this protrusion of islands
outside the horizon violates causality. In particular, the portion of the island
of the baths outside the horizon appears to be causally connected to the
defects. Naively, this appears to allow communication between the baths and
defects even if the coupling between these systems is severed. The resolution of
this paradox comes from noting that a splitting quench between the defect and
bath systems would inevitably create a positive energy shock causing an outward
shift of the horizon. It was argued in \cite{Almheiri:2019yqk}, using a JT
version of the quantum focusing conjecture \cite{Bousso:2015mna,Koeller:2015qmn}, that this shift would have the final
event horizon swallow the island, preventing post-quench communication between
the bath and defect. It would be interesting to re-create this problem in our
setup to probe the quantum focusing conjecture in higher dimensions. From the
bulk perspective, a splitting quench would be implemented by a bulk
end-of-the-world brane anchored asymptotically on the splitting surface
\cite{Shimaji:2018czt}. In $d=2$, the splitting surface on the asymptotic
boundary can be obtained by a conformal transformation from a full plane; in the
bulk, the end-of-the-world brane can similarly be obtained by a diffeomorphism
from a planar brane in pure AdS. In $d>3$, however, the calculations will become
more complicated, \eg the end-of-the-world brane will, in general, backreact on
the geometry such that the bulk is no longer locally pure AdS.

Returning to the issue of extracting information from the island, entanglement
wedge reconstruction
\cite{EW1,EW2,EW3,Jafferis:2015del,Dong:2016eik,Faulkner:2017vdd,Cotler:2017erl,Chen:2019iro}
allows us to recover information about the island with data from the boundary
CFT in the corresponding boundary subregion. One interesting question would be
to evaluate the expectation value of various CFT operators in the island, \eg
reconstructing $\langle T_{ij}\rangle$ in the vicinity of the
horizon.\footnote{We thank Ahmed Almheiri for raising this question.} The latter
is particular interesting because while the appearance of quantum extremal
islands pointed out a new resolution of the information paradox, this does not
directly address the issue of firewalls \cite{Almheiri:2012rt,Almheiri:2013hfa}.
Here asking if the black hole horizon develops a firewall in the late time phase
of the Page curve can be addressed by evaluating $\langle T_{ij}\rangle$ on the
horizon. While a direct boundary reconstruction of the latter remains to be
done, we are confident that no singularities arise in our framework. The reason
is that in the bulk, the system is in the vacuum state and we are simply
examining this state from a Rindler frame of reference. Hence in fact, we expect
that $\langle T_{ij}\rangle=0$ on the horizon and throughout the black hole
solution on the brane.\footnote{The vanishing of the stress tensor on the brane is an essential feature of our construction as the AdS$_d$
brane geometry must be a solution of the corresponding gravitational equations. That is,
the CFT on the brane cannot provide a source in these equations (at least to leading order for large
$c_\mt{T}$) otherwise the geometry would deviate from AdS space. Recall that while the brane CFT is in
its vacuum state, the bath CFT is coupled to the brane along an accelerating trajectory -- see discussion
under eq.~\reef{eq:donTTellMe}. This acceleration allows the bath CFT to achieve equilibrium
at a finite temperature.} 

This is related to the fact that in the present paper, for the sake of
simplicity, we have chosen to work with a bulk that is pure AdS, \ie the
temperature was tuned to $T=\frac1{2\pi R}$. The Rindler horizon in this
geometry consequently corresponds to a massless hyperbolic black hole. An
obvious extension would then be to consider massive black holes. Again,
calculations will be made difficult by the fact that the brane and bulk equations
of motion must be solved simultaneously with the former back-reacting on the
latter. In particular, the equilibrium configuration will now involve
excitations of the CFT on the brane, \ie the effective Einstein equations on the
brane will be sourced by the stress tensor of the boundary CFT residing there.

Yet another direction to take would be to consider our setup from the
perspective of tensor networks and error correction codes \cite{Swingle:2009bg,
  Pastawski:2015qua, HayNez16, Dong:2016eik}. For instance, as noted in
\cite{Hartman:2013qma}, the MERA-like tensor network constructing the
time-evolved CFT thermofield double state on the asymptotic boundary shares a
similar geometry to codimension-one bulk spatial slices stretching through the
bulk wormhole. One might then be motivated, as in \cite{Pastawski:2015qua}, to
view these spatial slices as supporting tensor networks implementing quantum
error correction codes between the bulk and boundary. It would be interesting to
see what such a network would tell us about the effective theory (see \eg
\cite{Czech:2016nxc, Dong:2020uxp}) on the brane and how information on the
brane is ultimately encoded in the asymptotic CFT and defect theory. On a
related note, one might also study the complexity of these brane configurations,
for example, using the higher-dimensional bulk to probe holographic complexity
conjectures \cite{Stanford:2014jda,Susskind:2014moa,Brown:2015bva,
  Brown:2015lvg}, \eg see \cite{prep33}.

Above, we emphasized the effective character of the gravitational theory on the
brane with the appearance of a short distance cutoff in Randall-Sundrum gravity.
However, as discussed in \cite{Chen:2020uac}, the brane perspective also
provides an effective description of the coupling of the bath CFT to the
conformal defect. In particular, it only accounts for the couplings localized at
the defect, which dominate at low energies, but ignores the subtle `nonlocal'
couplings, which can seen as coming through the AdS$_{d+1}$ geometry with the
bulk description. Given the simplicity of our construction, it may provide a
useful framework in which to further understand these nonlocal couplings, which
implicitly provide subtle correlations between the island degrees of freedom and
those in the bath CFT \cite{Chen:2020uac,domino}.

Lastly, in order to explain the fast growth of entanglement at early times for
large regions, in section \ref{sec:page_curve_in_d_gtr_two} we computed bounds
on entanglement growth in hyperbolic space. While they display the expected
qualitative behavior, they are not particulary tight. Instead, the difference
between bounds and numerical data becomes bigger as $\chi_\Sigma$ grows. It
would be interesting to improve these bounds.